\documentclass[fleqn,usenatbib]{mnras}

\IfFileExists{newtxtext.sty}{%
\usepackage{newtxtext,newtxmath}%
}{%
\usepackage{mathptmx} 
}

\usepackage[T1]{fontenc}
\usepackage{textcomp} 

\DeclareRobustCommand{\VAN}[3]{#2}
\let\VANthebibliography\thebibliography
\def\thebibliography{\DeclareRobustCommand{\VAN}[3]{##3}\VANthebibliography}

\usepackage{graphicx}	
\usepackage{amsmath}	
\usepackage[version=4]{mhchem}
\usepackage{booktabs}
\usepackage{float}
\usepackage{subcaption}
\usepackage{xspace}
\usepackage{xparse}
\usepackage{expl3}
\usepackage{makecell} 
\usepackage{lineno}

\newcommand{\msun}{\ensuremath{M_{\odot}}\xspace}

\newcommand{\muv}{\ensuremath{M_{\mathrm{UV}}}\xspace}

\newcommand{\Te}{\ensuremath{T_{\mathrm{e}}}\xspace}
\newcommand{\Telow}{\ensuremath{T_{\mathrm{e,low}}}\xspace}
\newcommand{\Teinter}{\ensuremath{T_{\mathrm{e,inter}}}\xspace}
\newcommand{\Tehigh}{\ensuremath{T_{\mathrm{e,high}}}\xspace}
\newcommand{\den}{\ensuremath{n_{\mathrm{e}}}\xspace}
\newcommand{\denlow}{\ensuremath{n_{\mathrm{e,low}}}\xspace}
\newcommand{\deninter}{\ensuremath{n_{\mathrm{e,inter}}}\xspace}
\newcommand{\denhigh}{\ensuremath{n_{\mathrm{e,high}}}\xspace}
\newcommand{\dencrit}{\ensuremath{n_{\mathrm{e,crit}}}\xspace}
\newcommand{\snr}{\ensuremath{\mathrm{S/N}}\xspace}
\newcommand{\angstrom}{\ensuremath{\text{\AA}}\xspace}

\newcommand{\kmps}{\ensuremath{\mathrm{km\,s^{-1}}}\xspace}

\renewcommand{\micron}{\ensuremath{\mu \mathrm{m}}\xspace}

\newcommand{\jwst}{\emph{JWST}\xspace}
\newcommand{\msaexp}{\texttt{msaexp}\xspace}

\newcommand{\pyneb}{\texttt{PyNeb}\xspace}

\newcommand{\sersic}{S\'{e}rsic\xspace}

\newcommand{\oii}{\ensuremath{[\text{O\,}\textsc{ii}]}\xspace}
\newcommand{\oiiwavel}{\ensuremath{[\text{O\,}\textsc{ii}]\,\lambda\lambda3727,3729}\xspace}
\newcommand{\oiii}{\ensuremath{\text{O\,}\textsc{iii}}\xspace}
\newcommand{\oiiiwaveluv}{\ensuremath{\text{O\,}\textsc{iii}]\,\lambda\lambda1660,1666}\xspace}
\newcommand{\oiiiwavelopt}{\ensuremath{[\text{O\,}\textsc{iii}]\,\lambda\lambda4959,5007}\xspace}
\newcommand{\oiv}{\ensuremath{\text{O\,}\textsc{iv}}\xspace}

\newcommand{\ha}{\ensuremath{\mathrm{H}\alpha}\xspace}
\newcommand{\hb}{\ensuremath{\mathrm{H}\beta}\xspace}

\newcommand{\hii}{\ensuremath{\mathrm{H}\,\text{\textsc{ii}}}\xspace} 
\newcommand{\hei}{\ensuremath{\mathrm{He}\,\text{\textsc{i}}}\xspace}
\newcommand{\heii}{\ensuremath{\mathrm{He}\,\text{\textsc{ii}}}\xspace} 

\newcommand{\heiiwavelopt}{\ensuremath{\mathrm{He}\,\text{\textsc{ii}}\,\lambda4687}\xspace}

\newcommand{\ariii}{[\ensuremath{\mathrm{Ar}\,\text{\textsc{iii}}]}\xspace} 
\newcommand{\ariiiwavellong}{\ensuremath{[\mathrm{Ar}\,\text{\textsc{iii}}]\,\lambda\lambda7135,7751}\xspace}

\newcommand{\ariv}{\ensuremath{[\mathrm{Ar}\,\text{\textsc{iv}}]}\xspace}
\newcommand{\neiii}{\ensuremath{[\mathrm{Ne}\,\text{\textsc{iii}}]}\xspace}

\newcommand{\ciii}{\ensuremath{\mathrm{C\,\textsc{iii}]}}\xspace}
\newcommand{\ciiiwavel}{\ensuremath{\mathrm{C\,\textsc{iii}]\,\lambda\lambda1907,1909}}\xspace}
\newcommand{\civ}{\ensuremath{\mathrm{C\,\textsc{iv}}}\xspace}

\newcommand{\civwavelopt}{\ensuremath{\mathrm{C}\,\text{\textsc{iv}}\,\lambda5808}\xspace}
\newcommand{\siiwavel}{\ensuremath{\mathrm{[S\,\textsc{ii}]\,\lambda\lambda6716,6732}}\xspace}

\newcommand{\siiiiwavel}{\ensuremath{\mathrm{[Si\,\textsc{iii}]\,\lambda\lambda1883,1892}}\xspace}

\newcommand{\nii}{\ensuremath{\mathrm{N\,\textsc{ii}}}\xspace}
\newcommand{\niiwavel}{\ensuremath{\mathrm{[N\,\textsc{ii}]\,\lambda\lambda6548,6549}}\xspace}
\newcommand{\niii}{\ensuremath{\mathrm{N\,\textsc{iii}}}\xspace}
\newcommand{\niiiwavel}{\ensuremath{\mathrm{N\,\textsc{iii}]\,\lambda1750}}\xspace}
\newcommand{\niiiwavelopt}{\ensuremath{\mathrm{N\,\textsc{iii}\,\lambda4642}}\xspace}
\newcommand{\niv}{\ensuremath{\mathrm{N\,\textsc{iv}]}}\xspace}
\newcommand{\nivwavel}{\ensuremath{\mathrm{N\,\textsc{iv}]\,\lambda\lambda1483,1487}}\xspace}
\newcommand{\nv}{\ensuremath{\mathrm{N\,\textsc{v}}}\xspace}

\newcommand{\CO}{\ensuremath{\mathrm{C/O}}\xspace}
\newcommand{\NO}{\ensuremath{\mathrm{N/O}}\xspace}
\newcommand{\NH}{\ensuremath{\mathrm{N/H}}\xspace}

\newcommand{\OH}{\ensuremath{\mathrm{O/H}}\xspace}
\newcommand{\logOH}{\ensuremath{12+\log(\mathrm{O/H})}\xspace}
\newcommand{\logNO}{\ensuremath{\log(\mathrm{N/O})}\xspace}

\newcommand{\OAr}{\ensuremath{\mathrm{O/Ar}}\xspace}
\newcommand{\ArO}{\ensuremath{\mathrm{Ar/O}}\xspace}
\newcommand{\ArH}{\ensuremath{\mathrm{Ar/H}}\xspace}

\newcommand{\NeO}{\ensuremath{\mathrm{Ne/O}}\xspace}
\newcommand{\NeH}{\ensuremath{\mathrm{Ne/H}}\xspace}
\newcommand{\NeC}{\ensuremath{\mathrm{Ne/C}}\xspace}

\xspaceaddexceptions{]}

\renewcommand{\ion}[2]{\ensuremath{\mathrm{#1}^{#2}}\xspace}

\newcommand{\op}{\ensuremath{\mathrm{O}^{+}}\xspace}       
\newcommand{\opp}{\ensuremath{\mathrm{O}^{2+}}\xspace}     
\newcommand{\oppp}{\ensuremath{\mathrm{O}^{3+}}\xspace}     
\newcommand{\np}{\ensuremath{\mathrm{N}^{+}}\xspace}       
\newcommand{\npp}{\ensuremath{\mathrm{N}^{2+}}\xspace}     
\newcommand{\nppp}{\ensuremath{\mathrm{N}^{3+}}\xspace}    
\newcommand{\cpp}{\ensuremath{\mathrm{C}^{2+}}\xspace}     
\newcommand{\hp}{\ensuremath{\mathrm{H}^{+}}\xspace}       
\newcommand{\arpp}{\ensuremath{\mathrm{Ar}^{2+}}\xspace}   
\newcommand{\arppp}{\ensuremath{\mathrm{Ar}^{3+}}\xspace}  
\newcommand{\nepp}{\ensuremath{\mathrm{Ne}^{2+}}\xspace}   
\newcommand{\splus}{\ensuremath{\mathrm{S}^{+}}\xspace}    
\newcommand{\sipp}{\ensuremath{\mathrm{Si}^{2+}}\xspace}    

\title[JWST Nitrogen-enhanced Galaxies]{Diverse Histories and Common Origins of Nitrogen-enhanced JWST Galaxies}

\author[Rusakov et al.]{
Vadim Rusakov,$^{1}$\thanks{E-mail: rusakov124@gmail.com}
Christopher J. Conselice,$^{1}$
Thomas Harvey$^{1}$,
Jordan C. J. D'Silva$^{1}$ and
Duncan Austin$^{1}$
\\
$^{1}$ Jodrell Bank Centre for Astrophysics, University of Manchester, Oxford Road, Manchester M13 9PL, UK
}

\date{Accepted XXX. Received YYY; in original form ZZZ}

\pubyear{\the\year{}}

\begin{document}


\label{firstpage}
\pagerange{\pageref{firstpage}--\pageref{lastpage}}
\maketitle

\begin{abstract}
Early JWST spectra revealed galaxies with a strong nitrogen excess, challenging galactic chemical evolution models. Using public JWST surveys, we construct a sample of 76 N/O-enhanced galaxies (NOEGs) at $4<z<8.5$, the largest at high redshift to date. The NOEG fraction rises from $\sim$3\% to $\sim$18\% between $z\sim4$ and 7---well above the $\sim$2\% measured locally---potentially driven by burstier, cluster-dominated star formation. Stacked spectra of the most nitrogen-rich galaxies show signatures of low-metallicity Wolf-Rayet (WR) stars, a likely source of primary nitrogen within the first few Myr of a starburst, with UV and optical continua dominated by young stellar emission and Balmer jumps evident in some cases. Many NOEGs also exhibit ionised outflows: 40\% show secondary [\oiii] and \ha components, while stacked spectra of the remainder reveal a broadened, offset \ha without forbidden-line counterparts, suggesting dust-attenuated or faded outflows. The continuum in the latter shows a weak Balmer break, indicating these galaxies are past their most recent starburst. This suggests that outflows dilute gas metallicity after the first few Myr of the initial enrichment and enable renewed \NO enhancement driven by low-metallicity Asymptotic Giant Branch (AGB) stars. We conclude that NOEGs are caught briefly after a recent burst: either within \(\sim\)\,10~Myr, when WR winds drive nitrogen enrichment, or after 30--40~Myr, when AGB winds take over---following an outflow driven by radiative or supernova feedback, consistent with recent chemical evolution models.

\end{abstract}




\section{Introduction}

Recent observations with James Webb Space Telescope (JWST) Near-infrared Spectrograph (NIRSpec) have unveiled multiple chemically unusual systems with strong nitrogen emission lines in the first 1.5 Gyr of the universe (e.g., \citealp{Cameron2023_GNz11,Senchyna2024}).  Although the gas-phase metallicity of galaxies clearly decreases with redshift (e.g., \citealp{Maiolino2008,Sanders2021,Nakajima2023,Langeroodi2023,Curti2024_MZR,Heintz2023}), the interstellar medium (ISM) appears to be surprisingly enriched with nitrogen fused in stars.  In particular, these galaxies exhibit super-solar nitrogen abundance at low overall metallicity.  This is not expected, or explained, in models of simple galactic chemical evolution (GCE; e.g., \citealp{Kobayashi2020,Chiappini1997}), where nitrogen abundance in the ISM is expected to gradually accumulate in proportion to metallicity via stellar yields as galaxies age.

It is expected that the enrichment of nitrogen in the ISM is dominated by intermediate-mass asymptotic giant branch (AGB) stars, with small contributions of core-collapse supernovae \citep[CCSN;][]{Henry2000,Kobayashi2020}.  A small amount of primary nitrogen forms in the H-burning shell of massive stars during the CNO cycle \citep{MeynetMaeder2002,Meynet2005,LimongiChieffi2018}. In fact, \(^{14}\mathrm{N}\) comprises the bulk of the CNO cycle products, as it is accumulated at the main bottleneck reaction, $^{14}\mathrm{N}(p,\gamma)^{15}\mathrm{O}$ \citep{Clayton1983}.  Some of this nitrogen is not reused during the He-burning and eventually escapes into the ISM via stellar winds or after a CCSN.  However, as galaxies age, the ISM becomes pre-enriched with the CNO products and the secondary production channel starts at high metallicity.  This allows stars to fuse even more \(^{14}\mathrm{N}\) via the hot-bottom burning in AGB stars from seed C and O material dredged up from the core to the H-burning shell \citep{RenziniVoli1981,KarakasLugaro2016,Ventura2013}.  These products, together with the primary nitrogen, are expelled by AGB winds, raising the overall nitrogen abundance at high metallicity.  The same channels lead to galactic enrichment of C and O elements, with oxygen primarily dominated by CCSN and carbon produced almost equally by both CCSN and AGB stars \citep{Kobayashi2020}.

The delay times between different CNO enrichment channels result in an observationally well-established trend between the abundance ratios \NO (or \CO) and the metallicity proxy \OH \citep{Henry2000,Vincenzo2016}.  In low-metallicity galaxies with \(\logOH \lesssim 8.0\), \NO is sub-solar and forms a plateau at \(\logNO \approx-1.7\) to \(-1.5\) \citep{Nicholls2017}.  For example, this is seen in metal-poor halo stars \citep{Spite2005}, \hii regions in nearby dwarf and late-type galaxies \citep{Garnett1990,VilaCostas1993,Berg2012}.  After a delay of 100--500 Myr, depending on the AGB star mass, the ``secondary'' \NO increases linearly at \(\logOH \gtrsim 8.0\).  This stage is observed in Blue Compact Galaxies \citep{Izotov1999} and \hii regions in early-type galaxies \citep{VilaCostas1993,Belfiore2017}.  This trend also extends to metal-poor Damped Ly-\(\alpha\) (DLA) systems at the cosmic noon \citep{Pettini1995,Pettini2008}.  Despite the ubiquity of this trend, it is often associated with a large scatter, not completely explained by the measurement uncertainties (0.3 dex for MW Halo stars in \citealp{Spite2005}; \(\sim1.0\) dex in DLAs in \citealp{Lu1998,Pettini2002}).  This dispersion likely originates from differences in timing of the release of N (especially at low metallicity) and partly from unaccounted ionisation corrections \citep{Izotov1999}.  It is noteworthy that \CO follows a similar abundance scaling trend as \NO \citep{Nicholls2017,Berg2019}, as may be expected from their common enrichment pathways. 

However, recent JWST/NIRSpec observations of high-redshift galaxies revealed several metal-poor systems with nitrogen significantly enhanced in comparison to nearby galaxies (hereafter N/O-enhanced galaxies, NOEGs).  For example, \citet{Bunker2023,Cameron2023_GNz11} identified strong nitrogen lines in the galaxy GN-z11 at \(z=10.6\), including \niii]\,\(\lambda\)1747-1754 and \niv\,\(\lambda\lambda\)1483,1487 multiplets and super-solar \NO at \(\logOH=7.91\).  Later, around twenty similar galaxies were identified at \(z>5\) with enhanced, often supersolar nitrogen abundance \citep{Curti2025,Naidu2026_CosmicMiracle,Topping2025_ElectronDensities,Topping2025_CIV_N_emitters,Zhang2026,MarquesChaves2024,Castellano2024,Labbe2024,Topping2024_Nenrichment_20pc,Schaerer2024,NavarroCarrera2024,Stiavelli2025,Yanagisawa2024,Napolitano2025,ArellanoCordova2024b}, reaching \(\logNO = 0.0-0.5\) in the most extreme cases, such as GS-3073 \citep{Ji2024_GS3073,Uebler2023_GS3073}; UNCOVER-45924 \citep{Ji2026,Labbe2024}, or CEERS-01019 \citep{Isobe2023}.  Their locus of CNO abundance ratios, with broadly super-solar \NO and sub-solar \CO and \OH, coincides with the abundance anomalies of globular clusters (GCs; \citealp{Ji2026,Senchyna2024,Gratton2019,BastianLardo2018}).  Recent studies of the UV lines \citep{Morel2026} and optical nitrogen lines \citep{Cataldi2026}, as well as stacked spectra \citep{Hayes2025,Isobe2025}, observed with JWST at \(3<z<11\) and \(1<z<6\) confirm that NOEGs represent a fraction of the total galaxy population that increases with redshift and are not a haphazard collection of serendipitous objects.

The chemical enrichment mechanisms of NOEGs and their number statistics are not established yet, although progress was made in recent studies.  For example, deep observations of the lensed galaxy RXCJ2248 \citep{Berg2026} revealed evidence of Wolf-Rayet (WR) nitrogen-type (WN) stars which are sources of primary nitrogen. The chemical evolution models of \citet{Watanabe2026,Bhattacharya2026,Berg2026,KobayashiFerrara2024} indicated that the WN stars are capable of briefly producing \NO excess in starburst galaxies before becoming diluted by CCSN yields.  On the other hand, high-resolution cosmological simulations show that the excess nitrogen may originate from AGB stars after a starburst-driven outflow shuts down star formation and resets the gas metallicity lowering \OH \citep{McClymont2026_MysteryNO}. Finally, more exotic theoretical scenarios involving supermassive stars have been proposed to explain the peculiar chemical signatures \citep{Ebihara2026,Nandal2024a,Charbonnel2023,NageleUmeda2023,Gieles2018}. It is yet unclear if one mechanism or their combination is responsible for the production of excess nitrogen.  This is another issue we investigate in this paper.

To assess the role of different suggested mechanisms and their enrichment timescales, we perform a joint analysis of the CNO and alpha element abundance patterns in NOEGs, their galaxy properties and nebular line emission. In this work, we present a sample of 134 galaxies with \NO and metallicities at redshift \(4<z<8.5\), of which 76 are NOEGs, identified in multiple surveys in archival JWST data. We search their spectra for WR features and outflow signatures to produce a coherent picture of the origin of nitrogen excess in most NOEGs. Finally, we estimate the number density of NOEGs at \(4<z<7\) and their incidence at these epochs.

The summary outline of this paper is as follows. In Section~\ref{sec:data}, we describe observations, our sample selections and spectroscopic analysis. In Section~\ref{sec:properties}, we describe our methodology for measuring chemical abundances and deriving physical properties of galaxies. We demonstrate abundance measurements and the number densities of our sample in Section~\ref{sec:results}.  The discussion of WR and AGB contributions and our interpretations of the \NO enrichment evolution, as well as conclusions can be found in Sections~\ref{sec:discussion} and \ref{sec:conclusion}.

Throughout this work we use solar abundances of different elements \(A(X)=12 + \log({\rm X/H})\) from \cite{Asplund2021}: \(A({\rm He})=10.914\), \(A({\rm C})=8.46\), \(A({\rm N})=7.83\), \(A({\rm O})=8.69\), \(A({\rm Ar})=6.38\), \(A({\rm Ne})=8.06\). Our NOEGs are defined as galaxies with \(\logNO>-1.1\) and \(\logOH<8.2\), however we often compare galaxies from different regions of the \NO--\OH space and explicitly mention those selections where they apply. We report limits for fluxes and quantities derived from fluxes measured with the signal-to-noise \(2 < \snr < 3\), and upper limits at a \(2\sigma\) level, unless otherwise stated.  Line flux \snr values are calculated using a matched-filter method as described in \S\ref{sec:data_spec_lines}.  Our magnitude notation assumes the AB magnitude scale \citep{Oke1983}. We adopt a flat \(\Lambda\)CDM cosmology with \(\Omega_m = 0.3\),  \(\Omega_{\Lambda} = 0.7\), and \(H_0 = 68~{\rm km\,s^{-1}\,Mpc^{-1}}\) from \cite{PlanckCollaboration2020}.

\section{Observations \& Sample Selection}
\label{sec:data}

\begin{figure*}
\begin{center}
   \includegraphics[angle=0,width=1\textwidth]{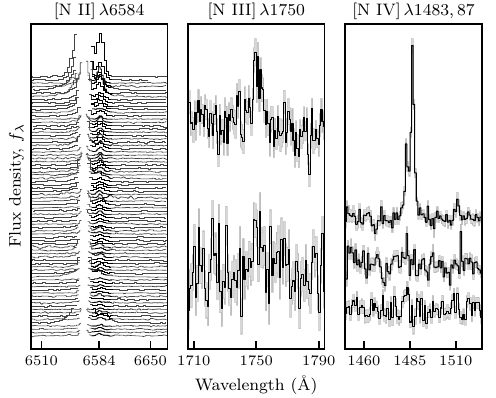}
\end{center}
\caption{A selection of spectra with optical and UV nitrogen lines: \niiwavel (left; showing 80 highest-\snr spectra), \niiiwavel (centre), \nivwavel (right). The spectra are sorted from top to bottom in the order of decreasing line signal-to-noise. Each spectrum is normalised by its median scatter in the continuum-only region. In the left panel, we masked out \ha line peaks to accentuate the weaker [\nii] emission.
}
\label{fig:sample_lines}
\end{figure*}

\begin{figure*}
\begin{center}
    \includegraphics[angle=0,width=\textwidth]{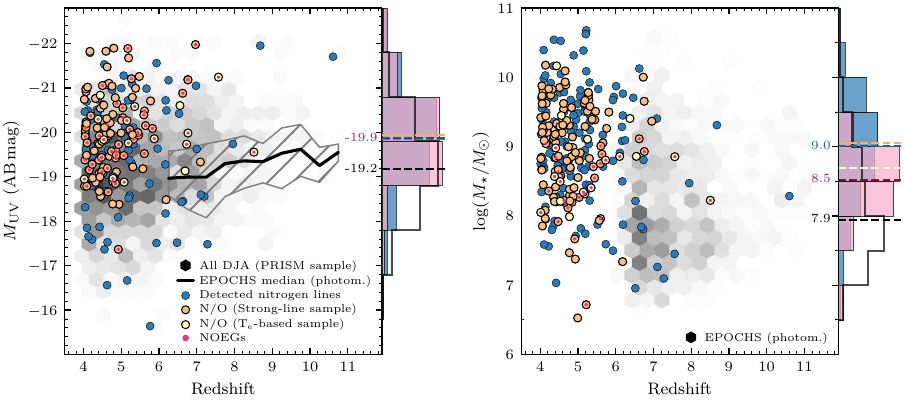}
\end{center}
\caption{Comparisons between absolute UV magnitude and stellar mass properties of our spectroscopic selections (including NOEGs), with star-forming galaxies at similar redshifts. \textit{Left} panel shows \muv for: all PRISM spectra in DJA at \(z>4\) (2D hex-bins); median \muv per redshift for galaxies in the photometric EPOCHS catalog (black solid line; \citealp{Conselice2025}); our sample of spectra with a detected nitrogen line; \NO samples based on strong-line calibrations and \Te-based. We show projected histograms of \muv for some samples and indicate their medians using dashed lines. The distribution shown here is limited to the galaxies with PRISM spectra to measure \muv and therefore shows a fraction of our sample (213\,/\,400 objects). Our sample (with nitrogen lines) has a median \(\muv=-19.9\) (dashed lines)---which is brighter than the total DJA distribution with \(\muv=-19.2\)---captures most of the magnitude range between \(4.0 < z < 8.5\). \textit{Right} panel similarly shows stellar mass estimates for: the EPOCHS catalog that covers redshift \(z>6\) (black density plot); and the same subsamples as in the left panel. The stellar masses of our spectroscopic sample are on average greater than the typical photometric masses, and are therefore likely biased to brighter systems where the weak nitrogen lines can be detected.
}
\label{fig:samples_MUV_redshift}
\end{figure*}

\subsection{Spectroscopic Data}
\label{sec:data_spec}

We use publicly available spectra from multiple NIRSpec/MSA (micro-shutter assembly) programs reduced and published on The Dawn JWST Archive\footnote{Available at: \url{https://dawn-cph.github.io/dja/}} (DJA; \citealp{Heintz2025_PRIMAL,BrammerValentino_DJAv4}).  We use \texttt{v4} reductions of \(R=1000\) and \(R=2700\) spectra \citep{Pollock2026,Valentino2025} with \({\rm grade}=3\) redshifts, which were estimated using \msaexp\footnote{\url{https://github.com/gBrammer/msaexp}} \citep{Brammer2023_msaexp} and verified by eye (G. Brammer).

Our spectroscopic fitting tests show that the extracted 1D spectra tend to overestimate uncertainties in the flux density, \(f_{\lambda}\) (unlike in older \texttt{v3} reductions). In particular, we find that the median reduced chi-square, \(\chi^2_{\nu}\), of our best-fit continuum and line emission is consistently below unity for all gratings (G140, G235, G395). We correct for this bias by scaling spectroscopic uncertainties by a factor of \(\sqrt{\chi^2_{\nu}}\). These factors are listed in Table~\ref{tab:spectra_uncertainties}. As they are calculated for a distribution of spectra with different levels of noise, we verify that the bias in \(\chi^2_{\nu}\) statistic does not depend on the signal to noise of the continuum in the spectra and, therefore, we apply these flat corrections globally to all reductions before constructing our final line flux catalogue.

\begin{table}
    \caption{Analysis of uncertainties in DJA v4 spectra based on best-fit continuum and line emission. The columns show: JWST/NIRSpec grating; median reduced chi-square, \(\chi^2_{\nu}\); uncertainty correction factor, \(\sqrt{\chi^2_{\nu}}\); number of spectra \(N(\text{spectra})\) and number of line fits \(N(\chi^2_{\nu})\), based on which the correction was calculated.
    }
    \label{tab:spectra_uncertainties}
    \begin{center}
    \begin{tabular}{ccccc}
    \toprule \toprule
        Grating & Median \(\chi^2_{\nu}\) & \(\sqrt{\chi^2_{\nu}}\) & \(N({\rm spectra})\) & \(N(\chi^2_{\nu}\)) \\
    \midrule  
        G140 & $0.757$ & $0.870$ & 738 & 1490 \\
        G235 & $0.798$ & $0.893$ & 1261 & 3554 \\
        G395 & $0.839$ & $0.916$ & 3577 & 9007 \\
    \bottomrule
    \end{tabular}
    \end{center}
\end{table}

DJA reductions include path-loss corrections for source profiles outside of the slitlet by modelling each source using a symmetric Gaussian \citep{deGraaff2025_rubies}. According to DJA\footnote{\url{https://dawn-cph.github.io/dja/blog/2025/05/01/nirspec-merged-table-v4/}}, all available grating spectra are cross-calibrated to PRISM spectra to correct for second-order deviations from different overlapping orders (as demonstrated in \citealp{Ito2026}, Appendix C), although some residual features remain. 

We experimented with performing our second-order corrections on top of the DJA products. First, we corrected for source position mismatches between NIRSpec/MSA observations by convolving all grating spectra through Near-infrared camera's (NIRCam) photometric bands and scaling grating spectra to PRISM (Sec.~\ref{sec:data_phot}). In addition, we scaled the resulting spectra to NIRCam photometry from DJA catalogs. We tested using polynomials of first, second and third order for these corrections and compared differences between fluxes of lines with \snr\(>5\) in overlapping gratings. The average fractional error on the line fluxes in overlapping DJA gratings in our sample is zero with a standard deviation of 10\%. Our added corrections did not improve on the existing DJA data, but introduced additional calibration scatter, indicating that the original DJA corrections were already sufficient. Therefore, we used DJA data products directly with their path-loss and other second order corrections. By repeating the analysis with our additional scaling, we verified that the final abundance ratios agree very well within the uncertainties and our results did not change. We note that we do scale our PRISM spectra to NIRCam photometry for spectral energy distribution (SED) modelling (\S\ref{sec:sed_fitting}).

\subsection{Spectroscopic Sample}
\label{sec:data_spec_sample}

Our sample includes observations that were conducted before December 2025.  We choose to use \(R=1000\) and \(R=2700\) spectra to resolve the auroral line \([\oiii]\,\lambda 4363\), the \(\ha+[\nii]\) complex, and various line doublets, such as \oiiwavel or \siiwavel, and other otherwise blended UV and optical lines. These observations include all possible combinations of gratings and filters with the total wavelength coverage \(0.6 - 5.5~\micron\).

The final line flux sample includes only galaxies where any of the nitrogen emission lines \nivwavel, \niiiwavel, \niiwavel has an integrated flux signal-to-noise \(\snr>3\).  We keep only objects at \(z>4\), to concentrate on high-redshift sources, and to observe rest-UV and optical nitrogen lines simultaneously at \(4.0<z<7.4\), where available.  Where multiple JWST programs have observed an emission line, we keep the highest \snr measurement. By starting with 3,014 unique objects at \(z>4\) in DJA, we construct our sample of 400 objects with at least a single nitrogen line detection that we later use for abundance measurements and for identifying NOEGs. Our nitrogen-line sample includes data from multiple spectroscopic JWST programs which we list in the Data Acknowledgements section. We demonstrate some of these spectra with the strongest nitrogen emission lines in Figure~\ref{fig:sample_lines}.

To put our sample in a broader context, we compare its distribution of UV magnitudes, \muv, and stellar masses, \(\log{M_\star}\) with wider samples of galaxies at the same epochs in Figure~\ref{fig:samples_MUV_redshift}. In the left panel we overplot a black \muv histogram of all objects in DJA (3,014), measured from low-resolution PRISM spectra matching our medium and high-resolution sample\footnote{We measured \muv by placing a top-hat filter between 1350 and 1800 \angstrom in the rest-frame of PRISM spectra.}, and a median photometric \muv in narrow redshift bins (black solid line) from the EPOCHS catalog \citep{Conselice2025}, which includes most JWST public fields at \(z>6\), spanning 214 arcmin\(^2\). Although our sample with nitrogen lines (blue circles), with a median \(\muv=-19.9\) (indicated using dashed lines), is brighter than the total DJA distribution with \(\muv=-19.2\), it still captures most of the magnitude range between \(4.0 < z < 8.5\). In the right panel we compare our sample (see \S\ref{sec:sed_fitting}) with the EPOCHS stellar masses (black density plot; \citealp{Harvey2025_epochs4,Conselice2025}). Although galaxies with nitrogen lines sample most of the range between \(7\lesssim \log{{M}_\star} \lesssim11\), their typical stellar mass of \(10^9~M_{\sun}\) exceeds the photometric stellar mass of \(10^{7.9}~M_{\sun}\) by 1.1 dex. These comparisons indicate a likely bias in our selection, which requires brighter and more massive galaxies to detect the faint nitrogen lines.

\subsection{Emission Line Measurements}
\label{sec:data_spec_lines}

We measure emission lines following a simple routine.  All lines within the range of \(\pm5000\,{\rm km\,s^{-1}}\) of each other are modelled together using Gaussian functions.  In such a relatively narrow velocity range we model the continuum using a straight line. We convolve each emission line model with the spectroscopic resolution at the line peak. As the nominal instrument documentation underestimates its resolution \citep{DeGraaff2024a}, we scale it by a constant factor of 1.7, assuming an observation is a point source centred in the slit for all grating/filter combinations.

We correct measured line fluxes for the effects of nebular dust extinction to calculate the intrinsic values.  We apply the model from \cite{Calzetti2000} with \(R_V=4.05\) based on the Balmer line decrement \ha\,/\,\hb, assuming the intrinsic case-B recombination decrement of \(I(\ha)/I(\hb)=2.86\) \citep{OsterbrockFerland2006}. Our choice follows the evidence that nebular emission in high-redshift galaxies is consistent with the Calzetti-like shallow extinction curve (e.g., \citealp{Shivaei2026}). Of 238 Balmer decrements in our sample, 74\% are consistent with no dust extinction within 3\(\sigma\), while 23\% have a median \(A_V=1.5\) mag and 8 objects have significantly negative Balmer decrements indicating potential departure from the Case B recombination scenario (see e.g., \citealp{McClymont2025_dust}).

As broadened \ha lines from active galactic nuclei (AGN) appear in around 10\% of galaxies observed with JWST at high redshift (e.g., \citealp{Juodzbalis2026_AGN_census}), they can bias \ha and \niiwavel fluxes from the host galaxy. To find galaxies with a broadened \ha and separate it from the narrow host galaxy component, we fit this line complex using two \ha models: with and without an additional Gaussian component with a flat velocity prior between 500--2000~\kmps. We select the best model using a cut in Bayesian Information Criterion (BIC) of \(\Delta \text{BIC} > 10\). We find that in a small number of cases the model selection fails to select broad-line sources with extended wings---some AGN at high redshift are known to exhibit exponential line wings \citep{Rusakov2026}. Therefore, we convolve the broad Gaussian with a symmetric exponential function, which can fit the exponentially-extended line tails and is flexible to fit the standard Gaussian shape. With the same approach, we identify secondary components in \oiiiwavelopt lines indicative of ionised outflows, which appear in up to 20--40\% of low-mass, star-forming galaxies at \(z>4\) \citep{Carniani2024}.

Our spectroscopic fitting is optimised using the Trust Region Constrained method (\texttt{trust-constr}) in \texttt{scipy}. We minimise the chi-square loss function: $$\chi^2=\sum_i\frac{(f_{\lambda,i}-m_i)^2}{\sigma_i^2},$$ calculated using the flux density \(f_{\lambda,i}\) and model \(m_i\) at each wavelength bin \(i\)---using fast and precise \texttt{JAX}-computed analytical gradients. 

Fitted parameters \(\theta\) include the amplitude of each Gaussian \(G(\lambda; \theta)\), the host galaxy velocity dispersion \(\sigma_v\), the slope and offset of the continuum and the spectrum redshift. We calculate line fluxes by integrating the best-fit Gaussian model components in the \(\pm3 \sigma_v\) region of the line position: $$F(\theta)=\int G(\lambda; \theta)d\lambda.$$ The flux uncertainties are linearly propagated from uncertainties of best-fit model parameters \(\theta\) using their covariance and \texttt{JAX} gradients of the line flux with respect to model parameters: $$\sigma^2_F = (\nabla F)^T {\rm Cov}(\theta) (\nabla F).$$ The covariance is calculated using a Hessian: \({\rm Cov = (H/2)^{-1}}\). For lines with low \snr, the model is usually not constrained, which results in negative covariances computed from the Hessian. In these cases, we calculate the flux uncertainty conservatively using the Jacobian: \({\rm Cov = J^T  W J}\), with weights calculated from spectroscopic uncertainties, \(W=1/\sigma^2\). Finally, we calculate the \snr of the lines by weighting individual wavelength elements by the model line profile (i.e., a matched-filter \snr): $$\snr=\frac{\sum_i f_{\lambda,i} w_i}{\sum_i \sigma^2_i w_i},$$ where \(w_i=G_{{\rm norm},i}/\sigma_i^2\) and \(G_{\rm norm}\) is a Gaussian normalised by its total area in the wavelength window where the line flux is integrated.

\subsection{Spectroscopic stacking}
\label{sec:stacking}

Here we describe the procedure for stacking multiple spectra, where it is required to detect faint signatures (see discussion about outflows and Wolf-Rayet signatures in Sec.~\ref{sec:disc_WR_stars}--\ref{sec:disc_nebular_outflows_fading}). Before stacking, we normalise individual spectra by the median continuum flux in the stacking window after masking out lines. We resample spectra to a common wavelength grid in the rest frame using the flux conserving method in \texttt{specutils v2.1.0} \citep{specutils2025}. We define the common grid based on the median scalar spectroscopic resolution in a stack to minimise introducing correlated noise. When we stack PRISM spectra, whose resolution varies significantly as a function of wavelength, we define the wavelength grid by stepping in \(d\lambda = \lambda / R_\text{median}\) units, where \(R_\text{median}\) is the median resolution across all spectra at each wavelength. All stacked-spectra in this work are median-combined and uncertainties are estimated by bootstrap resampling fluxes of the individual spectra in 1,000 trials (assuming uncertainties are Gaussian) and using 16th and 84th percentiles to define the uncertainty of the stacked spectrum.

\subsection{Photometric Data}
\label{sec:data_phot}

We use photometric data matching our spectroscopic selections and flux catalogs (Sec.~\ref{sec:data_spec_sample}) for the purposes of scaling spectroscopic flux (Sec.~\ref{sec:data_spec}), modelling SEDs (Sec.~\ref{sec:sed_fitting}) and for modelling morphological profiles (Sec.~\ref{sec:properties_morphology}).  We take photometric catalogs and reduced image mosaics from DJA \citep{Valentino2023}, which includes data from several JWST programs that we list in the Data acknowledgements section.

\subsection{AGN \& LRD Identification}
\label{sec:data_agn_identification}

\begin{table*}
    {
    \caption{Summary of the ionisation zones that we assume in this work.  For each zone, we list the respective ions and the assumed electron temperature and density.
    }
    \label{tab:ionisation_zones}
\begin{center}
\begin{tabular}{ccccc}
\toprule \toprule
    Ionisation Zone & Ions & \(\Te \, {\rm [K]}\) & \(\den \, {\rm [cm^{-3}]}\) & \(\den(z)\) relations \\
\midrule  

    High (30$-$55~eV) & \opp, \nppp, \arppp & \(\Te(\opp)\) & \(\den(\nppp)\), \(\den(\arppp)\) & \(5400\times(1+z)^{1.62\pm0.12}~(^3)\) \\
    
    Intermediate (15$-$30~eV) & \npp, \cpp, \arpp, \nepp & \(0.83 \times \Te(\opp)+1700\)~K (\(^1\)) & \(\den(\cpp)\), \(\den(\sipp)\) & \(1100\times(1+z)^{1.93\pm0.08}~(^4)\) \\
    
    Low (10$-$15~eV) & \np, \op & \(0.835 \times \Te(\opp)+2640\)~K (\(^2\)) & \(\den (\op), \, \den (\splus)\) & \(54^{+31}_{-22}\times(1+z)^{1.2\pm0.04}~(^5)\) \\

\bottomrule
\end{tabular}
\end{center}
\(^{1, 2}\) \Te--\Te relations from (1) \cite{Garnett1992}, (2) \cite{Pilyugin2009}.\\
\(^{3, 4, 5}\) Where \den diagnostics are missing, we infer the density from \(\den-z\) relations from (3-4) \cite{Martinez2025}, (5) \cite{Abdurrouf2024}.

}
\end{table*}

To investigate whether NOEGs are associated with AGN incidence, as has been suggested by some studies \citep{Isobe2025}, we identify AGN in our sample. In addition to galaxies with kinematically broad emission lines, we consider Little Red Dots (LRDs)---a population of compact galaxies at high redshift that are thought to host an AGN, with broad recombination lines \citep{Matthee2024} and a distinct continuum breakpoint around the Balmer series limit \citep{Setton2025}. We treat LRDs as a subclass of AGN here. We employ three selection criteria to select these objects:
\begin{enumerate}
    
    \item Presence of a broad \ha line with \(\text{FWHM}>1000~\kmps\).

    \item V-shape continuum criteria for NIRSpec/PRISM spectra from \cite{deGraaff2026}.
    
    \item V-shape continuum criteria for NIRCam photometric colours from \cite{Kokorev2024}. 
    
\end{enumerate}
We require any of these criteria to be satisfied to select an AGN and either of criteria (2)--(3) to select an LRD. For criterion (1), all lines are modelled using a Gaussian convolved with an exponential, as described in \S\ref{sec:data_spec_lines}. For criterion (2), we fit the UV and optical continua in PRISM spectra between 0.12 and 1.1 rest-frame \micron using a broken power law:
\begin{align}
    F(\lambda) &= A_\text{UV}\left(\frac{\lambda}{\lambda_{3646}}\right)^{\beta_\text{UV}},      &\quad\text{for}\quad~\lambda < \lambda_{3646}, \\
    F(\lambda) &= A_\text{optical}\left(\frac{\lambda}{\lambda_{3646}}\right)^{\beta_\text{optical}}, &\quad\text{for}\quad~\lambda > \lambda_{3646},
\end{align}
where \(\lambda_{3646}\) is the Balmer series limit at 3646 \angstrom. For this, we mask regions \(\pm\)\,5000~\kmps around prominent emission lines: UV region with multiple lines within 1200--1900~\angstrom, Balmer series lines, \(\neiii\,\lambda\lambda3869,3967\), \oiiiwavelopt, \oiiwavel, \(\hei\,\lambda\lambda5877,6680\), \siiwavel, \(\text{Pa}\)-series lines between 9015 and 10938~\angstrom. For selection (3), we use only the colour criteria from \cite{Kokorev2024} for \(4<z<6\) and \(z>6\). Using these selections, we have identified 6 LRDs satisfying any two criteria and 3 AGN with a broad line among 400 galaxies with nitrogen lines. We further compare line ratios of NOEGs with the stellar and AGN photoionisation models and discuss incidence of AGN among NOEGs in \S\ref{sec:disc_agn}.

\section{Derivation of Physical Properties}
\label{sec:properties}

\subsection{Chemical Abundance Measurements}
\label{sec:properties_abundances}

Direct abundance measurements in high-redshift galaxies have to account for their strong degree of ionisation. Similar to extreme emission line galaxies at low redshift \citep{Berg2021}, star-forming galaxies at \(z>4\) frequently exhibit strong emission lines with equivalent widths of \(\hb+[\oiii]\) of around 1000\,\angstrom \citep{RobertsBorsani2024}. This indicates that these galaxies experience strong starbursts and are commonly characterised by ionising photons with energies 30--55~eV and \(>\)\,55~eV and ionisation parameters of \(-3 \lesssim \log U \lesssim -1\). Despite this, photoionisation modelling in \cite{Martinez2025} showed that in \(\log U>-2\) galaxies a high degree of ionisation is not expected to strongly affect the ionisation profile of oxygen, which remains mostly populated by \opp ions, with only a fraction of 1--2\% comprised of \oppp. At the same time, more than 10\% of ionised nitrogen can be populated by \nppp species, depending on metallicity. We find ten galaxies among 400 in our sample with detected \heiiwavelopt emission coming from the \(\gtrsim\)\,54~eV regions, but given the intrinsic weakness of this line and other lines (e.g. \oiv which we do not detect), we expect that more galaxies can be as highly ionised. With these considerations, we adopt a three-zone ionisation model (\(<15\)~eV, \(15-30\)~eV, \(>30\)~eV) and, where possible, use the four-zone \(\log{U}\) and ionisation correction factor (ICF) calibrations from \cite{Martinez2025,Berg2021}.

All nebular properties and ionic abundances are measured using \pyneb v.\(1.1.30\) \citep{Luridiana2015}. To estimate the electron density \den and the temperature \Te, we use the iterative-solver \texttt{Atom.getTemDen} method from this package. For measuring \den in each zone, we use the rest-UV to optical diagnostics: 
\begin{enumerate}
    \item \denlow: \oiiwavel and \siiwavel; 
    \item \deninter: \ciiiwavel and \siiiiwavel;
    \item \denhigh: \(\ariv\,\lambda\lambda4740,4711\) and \nivwavel. 
\end{enumerate}
We calculate each density at a fixed \(\Te=1.5\times10^4\)~K, as the density has only a weak dependence on \Te, \(\den \propto \Te^{1/2}\), in the \den-sensitive range. This is equivalent to a 10--20\% change in \den when varying \Te between 1.5--2.0\,\(\times10^4\)~K---well within our typical uncertainties. We define the sensitivity range for every \den diagnostic, following \cite{Martinez2025}:
$$
0.1 \times n_\mathrm{e,crit}^\mathrm{min} < \den < 10 \times n_\mathrm{e,crit}^\mathrm{max},
$$
for the minimum and maximum critical densities in a density-sensitive doublet---\(n_\mathrm{crit,min}\) and \(n_\mathrm{crit,max}\), respectively---and clip the values outside of the range to these boundaries and flag them as limits. Where we do not have density diagnostics, we infer \denlow from the best-fit density-redshift \(\den-z\) relation in \cite{Abdurrouf2024}, and \deninter and \denhigh from the relations in \cite{Martinez2025}. We summarise our per-zone \Te and \den assumptions in Table~\ref{tab:ionisation_zones}.

This zone-specific approach to calculating \den addresses challenges highlighted by \cite{Martinez2025} for nitrogen-emitting galaxies at high redshift, which on average have elevated electron densities (e.g., \citealp{Topping2025_ElectronDensities,Isobe2023_densities}). First, with the \denlow constraints we can apply the common approximation \(\np/\op=\NO\) \citep{PeimbertCostero1969} where it is applicable---below the critical electron density \den of \niiwavel and \oiiwavel lines (\(\den<\dencrit\)). Based on \denlow we estimate the emissivities of [\nii] (\(\dencrit\approx10^5\)~cm\(^{-3}\)) and \oii (\(\dencrit\approx1-5\times10^{3}\)~cm\(^{-3}\)) and avoid [\nii]\,/\,\oii overestimating \NO where \denlow approaches \dencrit. Finally, \cite{Martinez2025} showed that underestimating densities \(\denhigh\) can overestimate \(\Tehigh\) by around 1300--2300~K and result in underestimating the metallicity \OH by around 0.67~dex. This bias in \(\Tehigh\) also propagates to \NO abundances through the \Te-sensitive line ratios \niv\,/\,\oiii] and \niii]\,/\,\oiii]. We follow their suggestion to minimise this bias: we first calculate \den, which depends weakly on \Te, and then, based on the \den, we calculate electron temperatures. 

The \Te constraints in our sample are based on these diagnostic line ratios:
\begin{enumerate}
    \item \Telow \([\nii]\,\lambda5755\,/\,[\nii]\,\lambda6584\),
    \item \Tehigh \([\oiii]~\lambda 4363/[\oiii]\,\lambda5007\),
    \item \Tehigh \(\oiii]\,\lambda1666\,/\,[\oiii]~\lambda 4363\),
    \item \Tehigh \(\oiii]~\lambda 1666/[\oiii]\,\lambda5007\),
\end{enumerate}
where most \Te are derived from (ii). As this diagnostic probes the high-ionisation zone, we estimate the temperature in other zones using \Te--\Te relations: 
\begin{enumerate}
    \item \(\Telow=0.835 \times \Te(\oiii)+2640\)~K from \citet{Pilyugin2009},
    \item \(\Teinter=0.83 \times \Te(\oiii)+1700\)~K from \citet{Garnett1992}.
\end{enumerate}
The uncertainties in line fluxes, \den and \Te are propagated to the final total abundances using 1,000 Monte Carlo draws for each value.

Where the auroral lines are missing, we employ strong-line calibrations to estimate \OH and \NO, as described later in this section. The summary of atomic and collisional data that we use is provided in Table~\ref{tab:atomic_data}.

We assume that the oxygen abundance is dominated mostly by the singly and doubly ionised oxygen \citep{Martinez2025,Berg2021} and is calculated as:
\begin{equation}
    {\rm \frac{O}{H}} = \frac{\ion{O}{+}}{\ion{H}{+}} + \frac{\ion{O}{2+}}{\ion{H}{+}}.
\end{equation}

\begin{figure}
\begin{center}
   \includegraphics[angle=0,width=1\columnwidth]{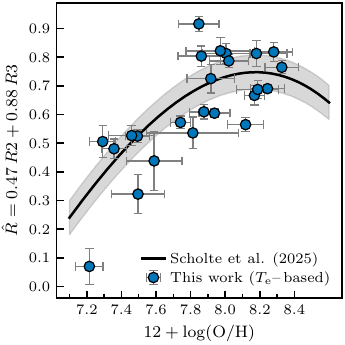}
\end{center}
\caption{Comparison of the directly measured metallicities \OH in this work with the \(\hat{R}\) calibration from \protect\cite{Scholte2025}. To show the uncertainty of the relation we use the residual scatter from calibration in their paper. The close agreement with the relation shows that \(\hat{R}\) can be used to accurately infer metallicity for our subsample of data without \Te measurements.
}
\label{fig:rhat_metallicities}
\end{figure}

Where we cannot measure \Te, we estimate \OH using the \(\hat{R}\) diagnostic introduced in \cite{Laseter2024}, where \(\hat{R}=0.47R_2 + 0.88R_3\), \(R_2=\log(\oiiwavel/\hb)\) and \(R_3=\log(\oiii\,\lambda5007/\hb)\). We use the \(\hat{R}\) calibration from \cite{Scholte2025}, which is based on a sample of local DESI (Dark Energy Spectroscopic Instrument) galaxies at \(\logOH<8.2\), and at \(\logOH \geq 8.2\) we revert to the calibration from \cite{Laseter2024} based on a smaller sample of JWST galaxies at \(2<z<9\). Being independent of the ionisation state, this relation was shown to be largely independent of redshift in \cite{Scholte2025} and therefore applicable to our sample here. Figure~\ref{fig:rhat_metallicities} shows that half of our sample has on average 0.26~dex higher metallicity than the \(\hat{R}\) calibration---the difference that disappears if we change from a three-zone \den model approach to a single-zone density model used in \cite{Scholte2025}---however, there is a good agreement on average. 

Figure~\ref{fig:rhat_metallicities} clearly shows that this diagnostic is degenerate for metallicities around \(7.8 < \logOH < 8.3\), where it turns around, and can produce large uncertainties. We resolve this degeneracy by encoding an external prior, based on the mass-metallicity relation (MZR) from \cite{Sarkar2025}. We compute the likelihood over the whole metallicity range for this diagnostic (\(6.49 < \logOH < 8.69\)) and multiply it by the MZR-based prior:
\begin{equation}
    p(x \vert \hat{R}_\text{obs}, M_\star, z) \propto \mathcal{L}(\hat{R}_\text{obs} \vert x)~ \pi(x \vert M_\star, z),
\end{equation}
where the likelihood is a Gaussian with the combination of observed and calibration scatter in the \(\hat{R}\) direction, \(\sigma^2_\text{tot} = \sigma_\text{obs}^2 + \sigma_\text{int}^2\), with \(\sigma_\text{int}=0.06\) \citep{Scholte2025}:
\begin{equation}
\mathcal{L}(\hat{R}_\text{obs} \vert x) = \mathcal{N}(x; \hat{R}_\text{obs}, \sigma_\text{tot}),
\end{equation}
and the prior is a Gaussian with the calibration scatter of \(\sigma_\text{prior}=0.16\) from \cite{Sarkar2025}:
\begin{equation}
\pi(x \vert M_\star, z) = \mathcal{N}(x; x_\text{MZR}, \sigma_\text{prior}).
\end{equation}
Finally, we also convolve the metallicity posterior with a Gaussian \(\mathcal{N}(0,\sigma_\text{int})\) to account for the calibration scatter \(\sigma_\text{int}=0.14\) in the \(\logOH\) direction \citep{Scholte2025}. We treat \OH as upper limits when \([\oiii]\,\lambda5007\) or \(\oiiwavel\) are upper limits, except for \(\oiiwavel\) upper limits at \(\log U>-2.5\)---in these cases we expect the total oxygen abundance to be dominated by \opp with \(\op/\mathrm{O}<0.3\) \citep{Martinez2025}, which is smaller than our typical uncertainty \(\sigma(\OH)/(\OH)=0.43\). Whenever \([\oiii]\,\lambda5007\) is missing, or if \(\oiiwavel\) is missing at \(\log U<-2.5\), we treat \OH as a lower limit.

To estimate the total \NO abundance ratio, we calculate individual ionic abundances in the low ionisation zone with \np/\op (from \niiwavel and \oiiwavel lines), intermediate with \npp/\opp (\niiiwavel, \oiiiwaveluv) and high with \nppp/\opp (\nivwavel, \oiiiwaveluv). Where we do not detect \oiiiwaveluv, we use \oiiiwavelopt as a fallback. As we do not detect all three lines in any galaxies, we apply ICFs to calculate the total \NO. We use metallicity- and \(\log U\)-dependent ICF values binned in \den from \cite{Martinez2025}:
\begin{align}
    \frac{\rm N}{\rm O} &= \frac{\np}{\op} \times {\rm ICF}(\np/\op), \\
    \frac{\rm N}{\rm O} &= \frac{\npp}{\opp} \times {\rm ICF}(\npp/\opp), \\
    \frac{\rm N}{\rm O} &= \frac{\nppp}{\opp} \times {\rm ICF}(\nppp/\opp).
\end{align}
For these ICFs, we use three diagnostics for calculating \(\log U\) with calibrations from \cite{Martinez2025}, in the order of priority: density-independent N43 (\(\niv\,/\,\niii]\)) probing the high-ionzation zone; \opp\,/\,\op for the intermediate zone; and O32 (\([\oiii]\,\lambda5007\,/\,\oiiwavel\)) also for the intermediate zone but more sensitive to \den. The uncertainties in \(\log U\) and ICF values include the calibration scatter \citep{Martinez2025}, and uncertainties in the diagnostics, metallicity, and electron density, where relevant. For the galaxies without detected auroral lines and direct ionic abundances, we use strong-line calibration from \cite{PerezMontero2009} to derive the total \NO from individual line ratios \([\nii]\,\lambda6584\,/\,\oii\,\lambda3729\).

When calculating the total \CO ratio, we apply the metallicity-dependent ICF from \cite{Berg2019} to the ionic ratios based on \ciiiwavel and \oiiiwaveluv lines:
\begin{equation}
    {\rm \frac{C}{O}} = \frac{\cpp}{\opp} \times {\rm ICF \left( \frac{\cpp}{\opp} \right)}. 
\end{equation}

Similarly, the ratios of alpha-element abundances are computed using the ICFs from \cite{ads21}. After calculating the ionic ratio based on \ariiiwavellong, \(\ariv\,\lambda\lambda4711,4740\), \oiiiwavelopt and \(\neiii\,\lambda3968\), \ArO is given by: 

\begin{equation}
    {\rm \frac{Ar}{O}}=\frac{\arpp}{\opp} \times {\rm ICF \left( \frac{\arpp}{\opp} \right) },
\end{equation}
or:
\begin{equation}
    {\rm \frac{Ar}{O}}=\frac{\arpp+\arppp}{\opp} \times {\rm ICF \left( \frac{\arpp+\arppp}{\opp} \right) },
\end{equation}
and \NeO:
\begin{equation}
    {\rm \frac{Ne}{O}}=\frac{\nepp}{\opp} \times {\rm ICF \left( \frac{\nepp}{\opp} \right) }.
\end{equation}

By taking advantage of rest-UV to optical \jwst coverage for galaxies at \(z>4\), we estimate \NO abundances in 171 galaxies. Our final sample consists of 23 direct-\Te \(\np/\op\) abundances from optical emission lines and 145 \NO abundances based on strong-line calibrations. We further measure \Te-based abundances \(\npp/\opp\) for 2 galaxies, and \(\nppp/\opp\) for 4 galaxies. This adds up to 174 \NO ratios, three of which are shared between any pair of ionic ratios. Of these 171 galaxies, 134 have \OH measurements spanning \(7.2 < \logOH < 8.6\). Finally, beyond this sample, we estimate limits on \NO for 24 galaxies (10 with \Te). Interestingly, only one object simultaneously has a UV multiplet and the optical nitrogen line detected, which indicates either that the nitrogen gas is highly stratified in most galaxies, and we observe the emission of the most abundant ions only, or that observed subsamples are biased, because detecting the UV lines requires deeper spectra.

Our \NO sample probes a wide range of UV magnitudes and stellar masses at redshift 4.0--8.5, but its comparison with photometric datasets shows that it is still likely biased to brighter and more massive systems (Figure~\ref{fig:samples_MUV_redshift}). The left panel in the figure shows that \NO objects (orange and light-yellow circles) probe 3.5 magnitudes in \muv, but are brighter by about 0.6 AB mag than the total DJA spectroscopic sample, and also brighter than the photometric EPOCHS sample. We find similar bias when we select NOEGs specifically (76 galaxies with \(\logNO>-1.1\) and \(\logOH<8.2\))---the panel on the right shows that NOEGs are 0.6 dex more massive than the photometrically-selected galaxies, although not as massive as the median nitrogen-line sample. Therefore, we note that our sample of NOEGs is likely biased high in stellar mass and brightness and underestimates the total number of such systems at \(4<z<8.5\).

We will make our full line flux and abundance catalogs public after publication.

\subsection{Spectro-photometric SED Fitting}
\label{sec:sed_fitting}

To estimate the stellar masses of the galaxies in our sample, we model their SEDs.  Depending on availability, we use spectroscopic and photometric data to extract the physical properties. We scale PRISM spectra to photometry. For this, we integrate the spectra through NIRCam filters and scale them to the photometric fluxes using a third-order Chebyshev polynomial.

Using \texttt{bagpipes} v.\(1.2.0\) \citep{Carnall2019_bagpipes,Carnall2018_bagpipes_spec}, we fit BPASS \(2.2.1\) \citep{Stanway2018_bpass} stellar evolutionary models by sampling their posterior distribution with the \texttt{nautilus} sampler. We use the standard stellar BPASS model which assumes a Kroupa initial mass function (IMF) with the broken power-law slopes \(\alpha_1=1.30\), \(\alpha_2=2.35\), the stellar masses from 0.1 to 300 \(M_{\odot}\) and a breakpoint at 0.5 \(M_{\odot}\) \citep{Kroupa2001}. These stellar models are processed with \texttt{CLOUDY} \citep{Ferland2017_CLOUDY} to generate the nebular emission grids. We assume a non-parametric star formation history with the continuity star formation history prior \citep{Leja2019}, which together with precise spectroscopic information is expected to recover star formation histories with minimal bias compared to other models and with representative uncertainties. We summarise the detailed parameters of our SED model in Table~\ref{tab:sed_models} and discuss that in some cases the SED model does not fully reproduce the nebular \ha emission in the spectra (Figure~\ref{fig:SFR_Ha_comparison}).

\subsection{Morphological Profiles}
\label{sec:properties_morphology}

Sizes of galaxies are measured by modelling their profiles in \(2'' \times2''\) cutouts from F444W JWST/NIRCam images using the \sersic profile:
\begin{equation}
    I(R) \propto F_{\mathrm{total}} \exp{[(R/R_e)^{1/n}-1]},
\end{equation}

\noindent
with the total flux \(F_{\mathrm{total}}\), half-light (effective) radius \(R_e\) and \sersic index \(n\). The profile is convolved with a point-spread function (PSF) in corresponding fields. We construct an empirical PSF for each field using the \texttt{psfs} module of the \texttt{EXPANSE} pipeline\footnote{\url{https://github.com/tHarvey303/EXPANSE}} from \citep{Harvey2025_EXPANSE}, which is based on the code \texttt{aperpy} \citep{Weaver2024_aperpy}. Where multiple sources are present in the cutouts (as identified using the \texttt{sep} package, \citealp{sep_package}), each is modelled with an individual \sersic profile. After visually inspecting image residuals we add sources if they are missed by \texttt{sep} (for example faint or blended sources) and refit. We mask out sources centred within 10 pixels of an image edge to avoid sampler convergence issues that occur when a significant fraction of a light profile is cut. This modelling is done with the \texttt{pysersic} package, where we use a No-U-Turn Markov Chain Monte Carlo sampler (NUTS, \citealp{NUTS_sampler}) to construct posterior distributions of model parameters. Each model with \(k\) parameters is sampled 3500 times with \(4k\) chains with the first 500 samples discarded as a burn-in sample. We demonstrate the best-fit profiles and residual images in Figure~\ref{fig:appendix_sersic_images} (Appendix~\ref{app:nircam_images}).


\begin{figure*}
\centering
    \centering
    \includegraphics[width=\linewidth]{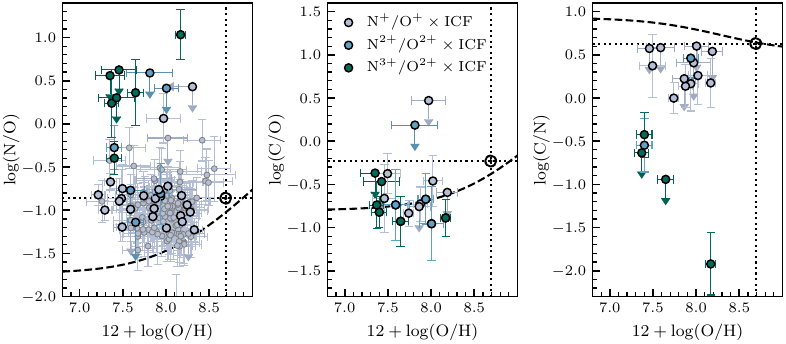}
    \caption{Ratios of total elemental abundances: \NO (left), \CO (middle), \({\rm C/N}\) (right); shown as a function of metallicity for our sample. We show \NO ratios based on ionisation-corrected \np/\,\op, \npp/\,\opp and \nppp/\,\opp abundances in different colours. Large circles with a solid black outline show \Te-based measurements, whereas faint smaller circles are measurements based on strong-line calibration (i.e., the \np/\,\op subsample). The dashed line shows the empirical relation for galactic stars from \protect\cite{Nicholls2017} and the \(\odot\) sign and dotted lines indicate the solar abundances.}
    \label{fig:CNO_OH}
\end{figure*}

\begin{figure*}
\centering
    \centering
    \includegraphics[width=0.7\linewidth]{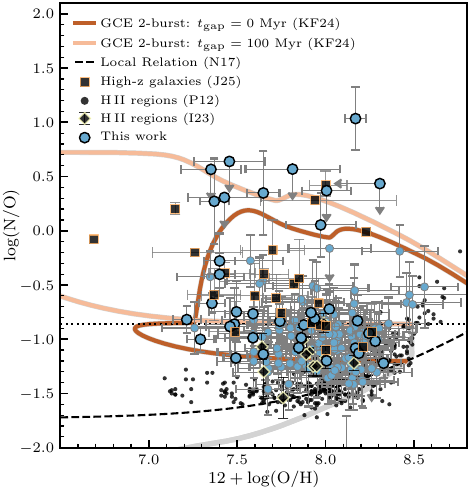}
    \caption{Comparison between \NO versus \OH ratios in our sample with literature values. Our sample is shown in blue circles. Large circles with the black outline show \Te-based abundances, whereas the smaller circles with a grey outline are derived from strong line calibrations (see \S~\ref{sec:properties_abundances}). For the high-redshift sample, we use the compilation of measurements from \protect\cite{Ji2026} (see text for individual references). The selection of local \hii regions is taken from \protect\cite{Izotov23,Pilyugin2012}. We show the local empirical fit for \NO--\OH from \protect\cite{Nicholls2017} and two GCE model tracks for a two-burst model from \protect\cite{KobayashiFerrara2024}: with (light brown) and without (dark brown) a 100-Myr gap separating two star formation episodes.}
    \label{fig:NO_all}
\end{figure*}

\section{Results}
\label{sec:results}

\subsection{CNO abundances: excess \NO and normal \CO}
\label{sec:results_NO}

\begin{figure*}
\centering
    \includegraphics[width=\linewidth]{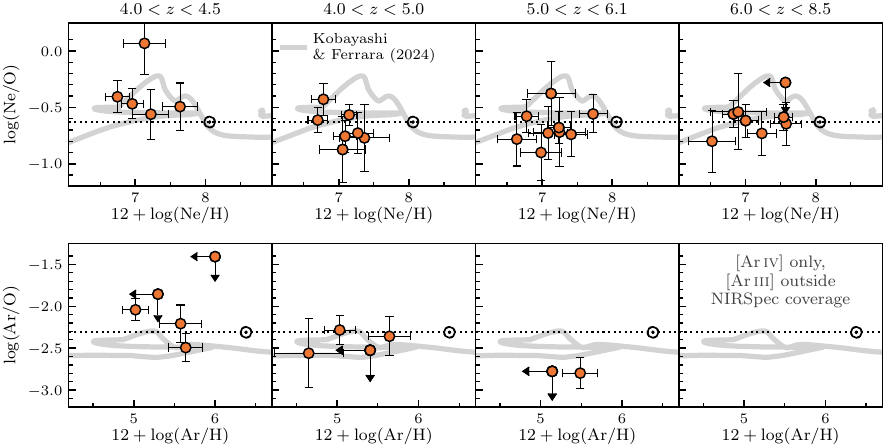}
\vspace{0.4cm}
\caption{Abundance ratios: \NeO versus \NeH (top) and \ArO versus \ArH (bottom), in bins of increasing redshift from left to right. The grey track represents a two-burst GCE model from \protect\cite{KobayashiFerrara2024} with no time gap between the bursts (see \S\ref{sec:results_NO} for description). The solar values are indicated using the black dotted lines and the \(\odot\) symbol. All galaxies on average have solar \NeO values at all epochs. They have tentatively supersolar \ArO values at \(z<4.5\) and decreasing values at \(z>4.5\) and \(z>5.0\) indicating that CCSN enrichment is active at all times, but the yields of Type Ia SNe possibly decrease at higher redshift (see text).}
\label{fig:argon_neon_NO}
\end{figure*}

Here, we define NOEGs as galaxies with metallicity range \(\logOH < 8.2\), where most galaxies are expected to be producing primary nitrogen, and \(\logNO > -1.1\), which we choose to match for consistency with some recent studies (e.g., \citealp{Cameron2026,Bhattacharya2026}), and which is around 0.4--0.6 dex greater than \NO in most local galaxies \citep{Nicholls2017}. We use this definition when we count the number of \NO-enhanced galaxies or estimate number densities. Where we deviate from this definition for comparing different subsamples with lower or higher \NO, we specify that.

According to this definition, more than half of our sample (76\,/\,134) are NOEGs (left panel in Figure~\ref{fig:CNO_OH}). The strongest \NO ratios typically arise from the ionisation-corrected \npp/\,\opp and \nppp/\,\opp abundances in the \(\gtrsim30\)~eV regions, reaching \(\logNO=0.5-1.0\), with a median of 0.35. Although the majority of the \NO sample originate from the 14--30~eV regions with \np/\,\op and \logNO between \(-1.5\) and \(0.0\), the median galaxy is still a NOEG at \(\logNO\approx-1.0\). As may be expected, the higher-ionisation regions have lower median metallicities: from \(\logOH=8.01\pm0.03\) for ratios derived from \np\,/\,\op to \(\logOH=7.67\pm0.06\) for \npp\,/\,\opp and \(7.52\pm0.06\) for \nppp\,/\,\opp. 

Galaxies with the UV-based \NO ratios must have intrinsically higher N abundances than the optical-based ones, as their differences in \OH cannot explain the difference in \NO. Typical UV-based ratios have around 0.4 dex lower abundance of \OH than the optical-based ones, which only partly explains the difference of 1--2 dex in \NO values. Therefore, it is the case that the \(\gtrsim\)\,30 eV regions in our sample have higher intrinsic \NH abundance. This suggests that either there is a different mechanism for producing nitrogen in \np and \npp, \nppp galaxies, or that the chemical enrichment history may be responsible for the difference in \NO.

Unlike nitrogen, we find that most of these galaxies are not enhanced in carbon (middle panel, Figure~\ref{fig:CNO_OH}), consistent with other studies of NOEGs (\citealp{Ji2026} and references therein). The \CO abundance closely follows the empirical low-redshift relation from \cite{Nicholls2017}. This is explicitly seen as subsolar \(\text{C/N}\) in the right panel of the figure. This suggests that the carbon enrichment mechanisms are similar to those in the local universe and are dominated by the yields of CCSN. It also suggests that there are no processes that affect relative carbon and oxygen abundances at low metallicity---any sources of carbon in these galaxies appear to produce a fixed \CO ratio. Therefore, this suggests that excess nitrogen must be produced by some mechanism distinct from the typical CNO enrichment by CCSN.

Finally, we show a comparison of our \NO sample with the individual NOEGs studied at high redshift and nitrogen measurements in local \hii regions and DLAs in Figure~\ref{fig:NO_all}. For the measurements at \(z>6\), we use a compilation from \cite{Ji2026}, based on the studies by \cite{Curti2025,Topping2025_ElectronDensities,Stiavelli2025,Zhang2026,ArellanoCordova2024b,Castellano2024,Ji2024_GS3073,Labbe2024,MarquesChaves2024,Napolitano2025,NavarroCarrera2024,Topping2024_Nenrichment_20pc,Schaerer2024,Cameron2023_GNz11,Isobe2023,Uebler2023_GS3073}. We also overplot \NO--\OH of the local \hii regions from \cite{Izotov23,Pilyugin2012}. Our sample expands the number of existing observations significantly in a similar range of metallicity and \NO as in the previous literature, especially in the high-\NO space. We note that our sample includes several sources identified previously. For these objects our \NO and \OH measurements agree within 0.2--1.5\,\(\sigma\) with the literature values despite often using different ICF values: CEERS-EGS-1746, CEERS-EGS-1665, CEERS-EGS-1477 \citep{Stiavelli2025,Sanders2024}; EXCELS-UDS-70864 \citep{ArellanoCordova2024b}; 2478-RXCJ2248 \citep{Berg2026,Topping2024_Nenrichment_20pc}.

For comparison with all measurements, we overplot the GCE model tracks from \cite{KobayashiFerrara2024} corresponding to a two-burst model with (blue line) and without (yellow line) a 100\,Myr gap between the bursts, where SFR of each burst is exponentially declining with timescales of 200\,Myr and 0.2\,Myr for the first and second bursts. Each burst follows an inflow of pristine gas with an exponential timescale \(\tau_i=1\)\,Myr (horizontal segments at around \(\logNO\approx-1.3\) and \(-0.9\)) and the \NO enrichment matching observations is produced by the WR stars. These models successfully outline the locus of measurements and demonstrate how these systems may develop over time; however, they require further development to explain all of these observations, as we discuss in \S\ref{sec:disc_WR_AGB_stars}--\ref{sec:disc_nebular_outflows_fading}. For example, the time spent by a galaxy in the main GCE loop (darker brown colour) is only around 8 Myr, during which period WR nitrogen yields become well-mixed with the CCSNe yields. Besides, we do not find any galaxies with \(\logNO<-0.5\) and \(\logOH>8.6\) where the model evolves after the WR enrichment. Therefore, these models may explain some of the youngest and most actively star-bursting systems, but not the ones that had star formation histories for longer than \(\sim10\)~Myr, which we also discuss later.

\subsection{Alpha element abundances: CCSN-dominated enrichment}
\label{sec:results_alpha_elements}

Although alpha elements like O, Ne are produced almost exclusively by CCSN, Ar has a non-negligible contribution from Type Ia SNe (\(\sim30\%\) of solar Ar in galactic chemical evolution models; \citealp{Kobayashi2020}). As a result, \(\log(\OAr)\) versus \(12+\log(\text{Ar\,/\,H})\) acts as a nebular analogue of the stellar [\(\alpha\,/\,\text{Fe}\)] versus [\(\text{Fe\,/\,H}\)] plane: CCSN produce a roughly flat \(\log(\OAr)\) with time, while delayed Type Ia enrichment lowers \OAr at fixed \(\text{Ar\,/\,H}\). This makes \OAr a useful tracer of the relative CCSN versus Type Ia contribution, and hence of star formation history, in emission-line nebulae where Fe may not be accessible \citep{Arnaboldi2022,Kobayashi2023,Bhattacharya2025_argon}. Therefore, if the excess nitrogen is produced while our galaxies are still young for Type Ia SN enrichment, this may be reflected in their lower-than-usual \ArO.

Our \NO sample has \NeO ratios consistent with solar values in every redshift bin between 4.0--8.5 (top panel, Figure~\ref{fig:argon_neon_NO}), but \ArO appears to decrease at higher redshift (bottom panel). The near-solar \NeO abundance and the lack of its redshift evolution is in agreement with \hii regions across a range of metallicities \citep{ArellanoCordova2024_CLASSY,DominguezGuzman2022,Berg2020,Croxall2016,Izotov2006}, local star-forming galaxies \citep{ArellanoCordova2024} and high-redshift galaxies \citep{Stanton2025}. \ArO in our NOEGs is tentatively supersolar at \(4.0<z<4.5\) with \((\ArO)=1.26 \pm 0.27\,(\ArO)_{\odot}\), solar at \(4.5<z<5.0\), with \((\ArO)=0.94 \pm 0.29\,(\ArO)_{\odot}\) and is subsolar at \(5.0<z<6.0\), with \((\ArO)=0.33 \pm 0.14\,(\ArO)_{\odot}\). At \(z>6.0\), \ariii shifts outside the NIRSpec wavelength range and \ariv is not detected. It is possible that the \ariv lines, which have 2--10 times lower emissivity than \ariii, depending on the density and temperature, may be undetected at higher redshift (most of our measurements are \ariii-based), although, at the same time, harder ionising radiation at higher redshift may ionise \arpp increasing the abundance of \arppp. Therefore, we interpret the overall \ArO redshift trend as a decrease in Ar abundance and Type Ia SN rates at higher redshift.

To conclude, our measurements of \NeO are consistent with expectations of CCSN enrichment, while the \ArO ratios suggest that Type Ia SNe contributions may still be delayed in all NOEGs. This suggests that chemical enrichment is dominated by the CCSN in these galaxies. Therefore, NOEGs at high redshifts must be younger than roughly the peak delay time for Type Ia SNe enrichment---between 100 Myr and 1 Gyr \citep{Maoz2014}. These findings are consistent with subsolar argon abundance of \((\ArO)=0.65 \pm 0.10\,(\ArO)_{\odot}\) as found in \cite{Stanton2025} in star-forming galaxies at \(1.8<z<5.3\). In this time window, some of the younger systems may still have retained their \(>\)15\,\(M_\odot\) stars producing primary nitrogen (see discussion in \S\ref{sec:disc_WR_stars}).

\subsection{Physical properties of NOEG hosts}
\label{sec:sed_properties}

\begin{figure*}
\centering
    \includegraphics[angle=0,width=1\textwidth]{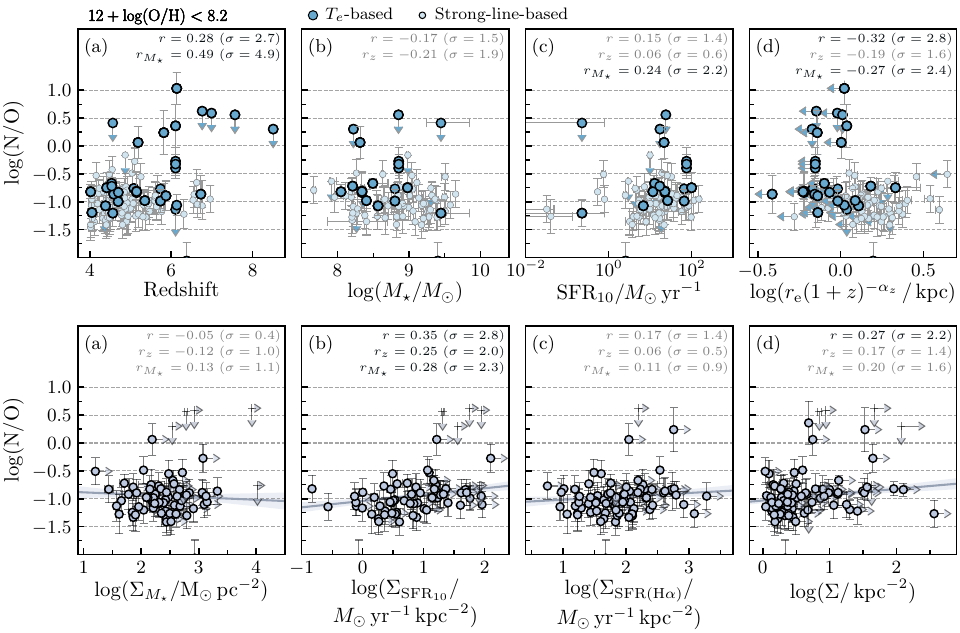}

\caption{Comparisons between \NO and physical properties of NOEGs. In the \textit{top} row, the panels show \logNO against: (a) cosmological redshift; (b) stellar mass, \(M_\star\); (c) SFR from SED fitting; and (d) the effective radius of the S\'{e}rsic profile, \(r_\text{eff}\), corrected for the redshift evolution from \protect\cite{Morishita2024_sizes}. Large blue points show the \Te-based \NO, and the light-blue faint points show strong-line-based \NO measurements. In the top-right of each panel we show the Spearman correlation coefficient \(r\) and correlation significance \(\sigma\), as well as \(r\) values after controlling for the stellar mass \(r_{M_\star}\) or redshift \(r_{M_\star,z}\) (see text). For better readability, we highlight correlation coefficients with \(\sigma \geq 2\) in bold. The correlation with redshift is the most significant and likely encodes the tentative correlations with SFR and galaxy compactness, as we discuss in the text. In the \textit{bottom} row, the panels show the surface densities of: (a) stellar mass \(M_{\star}\); (b) \(\text{SFR}_{\text{10}}\) over the last 10 Myr from best-fit SEDs; (c) \(\text{SFR}(\ha)\) sensitive to SFR in the past 3--10 Myr; and (d) inverse surface area for reference. The same correlations as in the top panel propagate to the tentative correlations with their surface densities. The solid lines are best-fit linear relations that confirm the correlation statistics (see text). We exclude AGN and LRDs (see \S\ref{sec:data_agn_identification}) from the plots of properties derived from SED fitting.
}
\label{fig:NO_sed_sersic_correlations}
\end{figure*}

\begin{figure*}
\centering
    \centering
    \includegraphics[width=1\textwidth]{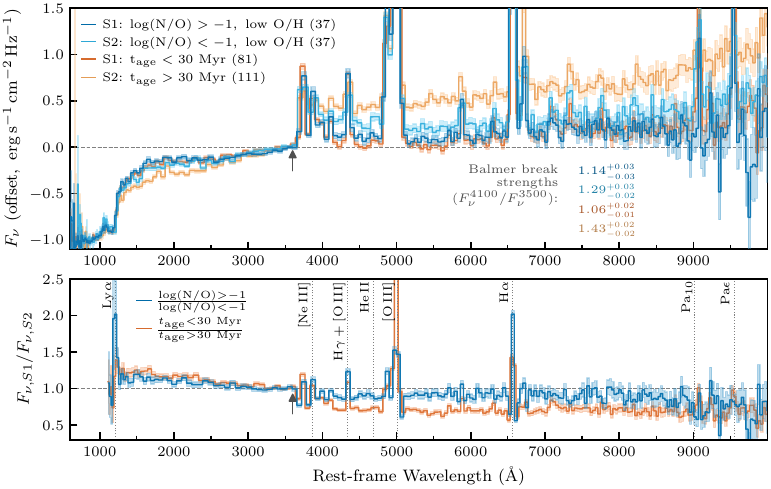}
    \caption{Comparison of median-stacked PRISM spectra for distinct groups of galaxies in our sample: galaxies with \NO above and below \(\logNO=-1\); and galaxies with the mass-weighted age younger and older than \(t_\text{age}=30\)~Myr. \textit{Top}: median-combined PRISM spectra of each galaxy group. Each stacked PRISM is normalised to the flux at the Balmer series limit at 3646 \angstrom (indicated with an arrow) and is shifted down by the flux density at this wavelength. Balmer break strengths, annotated for each median spectrum, indicate that older and lower-\NO galaxies have had a recent downturn in star formation rate, whereas the higher-\NO and younger galaxies still preserve massive stars. \textit{Bottom}: ratios of high-to-low-\NO and younger-to-older median spectra. The ratios of the complementary subsamples demonstrate that differences between high and low-\NO samples are similar to differences between younger and older galaxies in our sample: UV and optical continuum differences, as well as the ratios of nebular lines.
    }
    \label{fig:stacked_prisms}
\end{figure*}

To learn more about the origin of the nitrogen excess we also investigate whether it correlates with properties of the host galaxies. In this part of the analysis, we consider a combination of morphological and stellar population properties and demonstrate the correlations (or lack thereof) using scatter plots in Figure~\ref{fig:NO_sed_sersic_correlations}.

For NOEGs, we find that the best predictor of the \NO ratio is the redshift and galaxy compactness, whereas stellar mass and SFR have insignificant or tentative correlations (top row in Figure~\ref{fig:NO_sed_sersic_correlations}). We calculate Spearman correlation statistics considering all data points (including limits) and report three correlation coefficients between \NO and each of the physical properties: raw coefficient \(r\); redshift-controlled coefficient \(r_z\) after fitting for and subtracting a linear relationship between
\NO and physical properties with redshift; \(r_{M_\star}\) correlation coefficient after removing the stellar mass dependence; and correlation significance \(\sigma\) from a two-tailed statistical test. As a result, the correlation between \logNO and redshift is the most significant, particularly when removing the stellar mass dependence: \(r_{M_\star}\approx0.5\) with \(\sigma\approx5\). Tentatively, more compact galaxies have higher \logNO as well (\(r\approx 0.3\) with \(\sigma \approx 3\)). The sizes that we plot are corrected for redshift evolution assuming the relation from \cite{Morishita2024_sizes} with \(\alpha_z=-0.44 \pm 0.21\). The correlation with the SFR is tentative when considered at face value, and is covariant with the redshift evolution as evidenced by the low significance of the \(r_z\) correlation. Finally, the stellar mass is not correlated with \NO of NOEGs, although we verify that this correlation becomes \(r=0.47\) and is significant (\(4\sigma\)) for high-metallicity galaxies at \(\logOH>8.2\). The latter correlation is an indication of the secondary nitrogen production in more massive galaxies with low-mass, high-metallicity AGB stars dominating N-enrichment and is almost completely explained by the covariance between \logNO and \(\log(\OH)\).

The tentative size and SFR correlations propagate to the correlations with the surface densities of these properties (bottom panel in Figure~\ref{fig:NO_sed_sersic_correlations}). We calculate the surface densities by dividing the properties by the surface area \(\pi r_\text{e}^2\), where \(r_e\) is the effective S\'{e}rsic radius. As above, the stellar mass density does not differentiate between \NO values, whereas the SFR surface density and basic inverse area \(\Sigma\) (plotted for reference) correlate tentatively at \(2.9\sigma\) and \(2.4\sigma\). By fitting a linear relationship using orthogonal distance regression method \texttt{scipy.odr}, we find the slopes of these relations to be each \(\beta\approx0.12\) with \(2.5\sigma\) significance. Finally, the surface density of the SFR calculated from \ha luminosity has a similar slope, but the scatter in the data renders it insignificant. These results qualitatively agree with the findings in the JADES (JWST Advanced Deep Extragalactic Survey) survey \citep{Cameron2026}.

The SFR and SFR density trends in our NOEGs imply that the increase in \NO may be driven by young stellar populations in more compact environments at higher redshift. The correlation statistics imply that up to around 10\% of the increase in \NO can be explained by the increase in either SFR or \(r_e(1+z)^{\alpha_z}\) (inferred from \(r_{M_\star}^2\)), which likely translates into the correlation with redshift explaining around one quarter of the increase in \NO. We note that these correlations are weak in our analysis, as the large scatter in the data and the uncertainties prevent us from constraining them more confidently. It is also worth noting that our galaxy sizes are measured in the NIRCam\,/\,F444W filter, whose PSF resolution is worse by around 1.6 to 3.6 times compared to shorter-wavelength NIRCam filters and therefore our \(r_e\) upper limits may affect our analysis. Besides, this band is not necessarily tracing young stellar populations, although \cite{Morishita2024_sizes} find no significant differences between the rest-UV and rest-optical sizes of star-forming galaxies at high redshift. In Figure~\ref{fig:appendix_size_mass}, we compare our nitrogen-line sample with the size-mass relation from \cite{Morishita2024_sizes}, where the NOEGs tend to be upper limits below the rest of the nitrogen sample.

\subsection{Does high \NO trace young stars?}

\begin{figure*}
\centering
    \includegraphics[width=\linewidth]{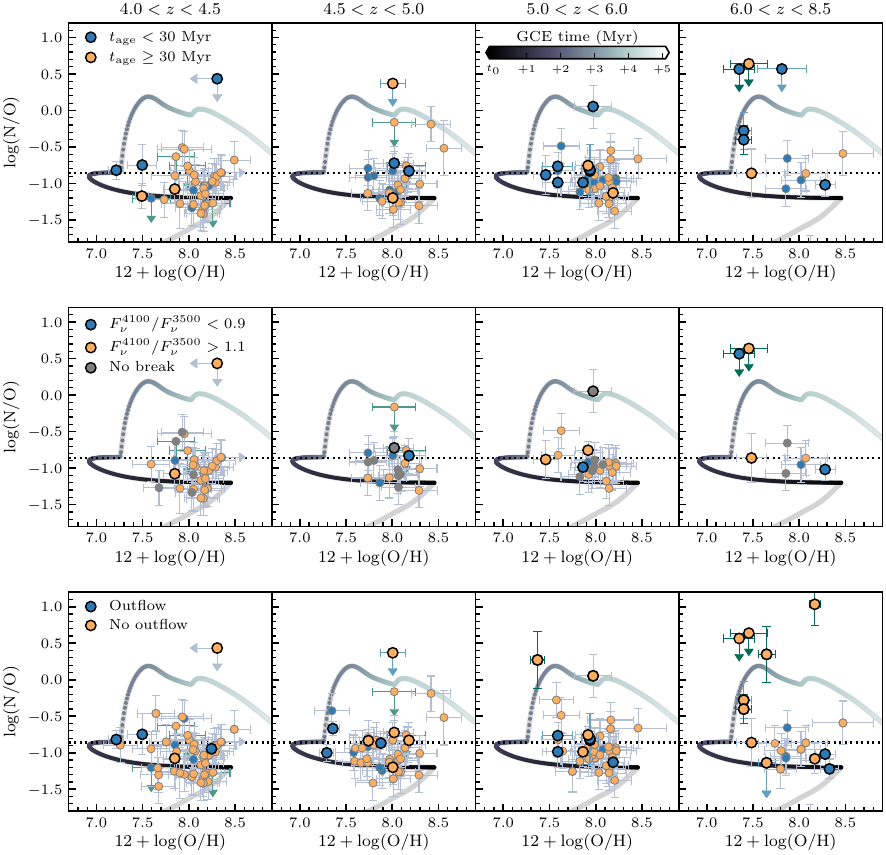}
\vspace{0.4cm} 
\caption{\NO and metallicity in bins of increasing redshift from left to right. Top: galaxies with mass-weighted age \(t_0<10~\text{Myr}\) (blue circles) and \(t_\text{age}\geq10~\text{Myr}\) (orange circles). Middle: Balmer break discontinuity, where the top-hat flux ratio at 4100 and 3500 \angstrom \(F_\nu^{4100}/F_\nu^{3500}>1.1\) (orange) indicates the presence of a break and \(F_\nu^{4100}/F_\nu^{3500}<0.9\) (blue) indicates a jump with an otherwise smooth continuum (grey) in the rest of the objects. Bottom: shows galaxies with (blue) and without (orange) a secondary broader line component detected in \oiiiwavelopt or \ha that likely indicates an outflow. The colour of uncertainty bars matches Figure~\ref{fig:CNO_OH}, where green, blue and grey correspond to \nppp, \npp and \np-based \NO ratios. The colour track represents a two-burst GCE model from \protect\cite{KobayashiFerrara2024} with no time gap between the bursts and an inflow causing the second burst starting at \(t_\text{age}\) (see colour bar). The solar values are indicated using the black dotted lines and the \(\odot\) symbol. Higher-redshift galaxies have lower mass-weighted ages on average and sample more of the example GCE track. NOEGs at higher redshift and higher \NO have a higher incidence of detectable outflows on average.}
\label{fig:NO_age_outflows_vs_models}
\end{figure*}

In the previous sections we presented several indications that NOEGs (especially at higher redshift) may have younger stellar populations: alpha-element abundances indicate decreasing Type Ia SN rates, higher \NO abundances have lower metallicities, and higher SFR and more compact sizes suggest younger star formation.

Here, we show more explicit evidence that the difference between high and low-\NO galaxies at low metallicity is that of the difference between galaxies with different star formation histories. In particular, we compare low-resolution PRISM spectra of the galaxies from our sample with \NO above and below \(\logNO=-1\) (median of our sample) to the galaxies younger and older than the mass-weighted age \(t_\text{age}=30\)~Myr and with \(\logNO>-1.3\). We median-combine these groups of spectra (see \S\ref{sec:stacking}) after normalising them by the flux density in the region 3500--3600 \angstrom in the top panel of Figure~\ref{fig:stacked_prisms}. We annotate the strength of their Balmer break, measured as the ratio of the integrated flux density \(f_{\nu}^{4100}/f_{\nu}^{3500}\) in two top-hat filters: at 3400--3600 and 4000--4200 \angstrom; and excluding \(\pm\)\,3500~\kmps regions around emission lines \oiiwavel, \(\neiii\,\lambda\lambda3869,3967\) and Balmer lines at 3970, 4103, and 4342 \angstrom. As expected for the younger galaxies with \(t_\text{age}<30\)~Myr, they have a continuous spectrum with no break, where O and B-type stars likely dominate the continuum emission. But older galaxies have a Balmer break strength of 1.4, which arises from the continuum of the less-massive A-type stars. Interestingly, the stacked spectra of galaxies with high and low \NO ratios are similar to those of younger and older galaxies, respectively---highest \NO is found in galaxies still preserving young massive stars, lower \NO have experienced a gap since the latest burst of star formation, albeit not as prolonged as that of the oldest galaxies.

We show the similarities between the two groups more explicitly by plotting the ratio of the combined spectra of high-to-low-\NO galaxies and younger-to-older galaxies in the bottom panel of Figure~\ref{fig:stacked_prisms}. Both high-\NO and younger galaxies have brighter UV continua and stronger nebular lines (annotated on the plot) than the lower-\NO and older galaxies, as well as brighter UV-to-optical continua without the Balmer discontinuity. Therefore, the key distinctions between the high and low-\NO galaxies with primary nitrogen are the presence of \(>\)\,15\,\(M_\odot\) stars in the former and the ongoing---as opposed to recent---burst of star formation.

In fact, when we split galaxies by redshift in Figure~\ref{fig:NO_age_outflows_vs_models}, we notice that individual high-\NO galaxies at higher redshift also have typically \(t_\text{age}<30\)~Myr ages (top panels) and/or lowest metallicities, some have a Balmer jump from the nebular continuum (middle panels) and no ongoing outflows, which together indicate that they experience ongoing young starbursts.

\subsection{Number density of NOEGs}
\label{sec:number_density_noegs}

\begin{figure}
\begin{center}
    \includegraphics[angle=0,width=1\columnwidth]{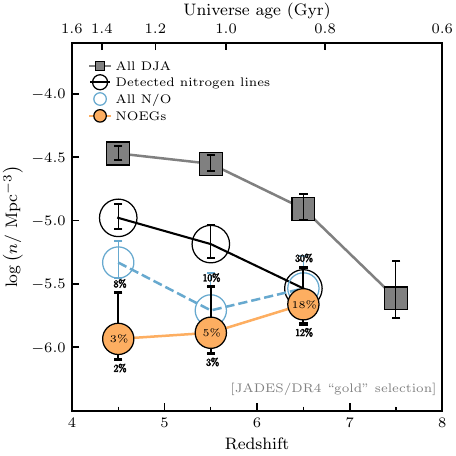}
\end{center}
\caption{Number density of NOEGs as a function of redshift compared to the total galaxy population observed with NIRSpec/MSA. In addition we plot the total \NO sample and galaxies with at least one detected nitrogen line. In order to reduce observational bias from complex selection functions of different surveys that end up in our sample we kept only objects matching the JADES DR4 ``Gold'' F444W (\(z<5.7\)) and UV-selected (\(z>5.7\)) samples \citep{Scholtz2026_jades}. Additionally, we conservatively select sources with AB magnitudes brighter than \(\text{F444W}<26\). Uncertainties are 16th and 84th percentiles of the Gamma distribution for a small number of data points \(N\) per bin (following \citealp{Gehrels1986}). Percentages indicate the NOEGs fraction of the total galaxy sample, and \% uncertainties are 16th and 84th percentiles of the beta posterior on the continuous probability \(k/n\) for \(k\) NOEGs and \(n\) total galaxies. Note, the number of NOEGs in each redshift bin is [2, 2, 3, 0].}
\label{fig:number_density_NOEGS}
\end{figure}

To explore the fraction of the galaxy population with enhanced \NO as a function of redshift we use the most observationally complete fraction of our sample. Estimating the number density of NOEGs remains challenging due to varying survey depths and complexity of selection functions in JWST surveys with NIRSpec: wavelength, colour, mass and various arbitrary selection criteria deem most of our sample inappropriate for such analysis. However, we can try to estimate a lower limit using a purely magnitude or flux-limited subsample that was observed as part of the JADES survey \citep[e.g.,][]{Eisenstein2026_jades} in GOODS fields.

We use the selection criteria of the latest ``gold'' sample from JADES Data Release 4 from \cite{Scholtz2026_jades}, which include NIRCam F444W flux at \(1.5 < z < 5.7\) and mostly HST-based rest UV magnitude at \(z > 5.7\). Additionally, to ensure that our sample is as uniform as possible we make a conservative cut at AB magnitude \(\text{F444W}<26\). This selection leaves the total of 121 objects at \(4<z<8\) among NIRSpec/MSA spectra in DJA, 32 galaxies with at least one detected nitrogen line and 7 NOEGs with \(\logNO>-1.1\) and low metallicity \(\logOH < 8.2\).

The galaxy number densities in Figure~\ref{fig:number_density_NOEGS} show that the fraction of NOEGs increases with redshift with respect to all galaxies. These number densities are calculated for the combined area of the ``gold'' sample's photometric footprint of 0.157 deg\(^2\) in GOODS-S/N fields. The number density of NOEGs (yellow circles) remains approximately constant despite the number density of all galaxies (grey squares) diminishing by \(\sim\)\,1.3, 2.5 and 5.7 times in each subsequent redshift bin. As a result, the fraction of NOEGs (shown as percentages in the figure) increases from \(3^{+5}_{-1}\%\) of all galaxies to \(18^{+12}_{-6}\%\) between redshifts \(z=4-6\) and \(6-7\). For comparison, we overplot the number densities of all galaxies with a detected nitrogen line (empty black circles) and galaxies with \(\logNO>-1.5\). They demonstrate that NOEGs likely make up most galaxies with nitrogen lines at \(z=6.5\) within uncertainties, and most galaxies with \NO in our sample become NOEGs at \(z\sim6\). As our findings in individual redshift bins are based on a small number of galaxies, their confidence is not greater than 1--2 sigma; however, assuming the completeness across all redshift bins is similar here, we can more confidently say that NOEGs become more common at high redshift. Finally, we stress that the NOEG fraction is likely a lower limit here, as our sample is brighter and has higher stellar masses than a typical galaxy found in spectroscopic or photometric surveys, as can be seen in Figure~\ref{fig:samples_MUV_redshift}.


\section{Discussion}
\label{sec:discussion}

A population of NOEGs at high redshift has been discovered recently and therefore most studies have focused on individual galaxies or highly-selective samples. Case studies have analysed the strongest nitrogen emitters identified by their UV multiplets \niiiwavel and \nivwavel  \citep[e.g.,][]{MarquesChaves2024,Castellano2024,Topping2025_CIV_N_emitters,Topping2024_Nenrichment_20pc,Schaerer2024,NavarroCarrera2024}, which are found at redshift \(z>6-8\). Several population studies identified NOEGs in low-resolution R100 spectra \citep{Isobe2023,Isobe2025,Hayes2025,Morel2026} and are therefore limited to abundances based only on the unresolved UV lines, while missing \niiwavel, and are more prone to bias due to not being able to resolve temperature-sensitive auroral lines or density-sensitive doublets. Incomplete samples, lack of a combined UV and optical-based range of elemental abundances across ionisation zones and nebular gas conditions limit our ability to conduct systematic population studies of NOEGs.

More recently, \cite{Cameron2026} measured \NO across redshifts \(1.5<z<7.0\) using R1000 spectra from the JADES survey in GOODS fields, tracing nitrogen from low to high ionisation zones. With the addition of strong-line measurements, at least 13\% of galaxies in their sample are moderately enhanced at \(-1.1<\logNO<-0.6\), with no galaxies enhanced at the level of GN-z11 \( \left( \logNO>-0.6 \right) \) at these redshifts. They identify correlations between \NO and SFR of galaxies and find that the stellar mass does not trace the \NO in NOEGs.

In this work, we take a similar approach to constructing a sample of NOEGs at \(4<z<8.5\), but using all existing archival data, including R1000 and R2700 spectra, and analyse abundances of C, N, O, Ne, Ar elements and physical properties of galaxies. This information allows us to probe \NO--\OH and \CO--\OH (Figure~\ref{fig:CNO_OH}) with the currently largest and uniform single data sample of 134 galaxies, 76 of which have \(\logNO>-1.1\) and \(\logOH<8.2\). From the \NeO and \ArO ratios (Figure~\ref{fig:argon_neon_NO}) we find that the Type Ia SN rate in our \NO sample may decrease with redshift, consistent with chemical evolution models, and that these galaxies become younger. This is further corroborated by our side-by-side comparison of galaxies with high \NO ratios and young mass-weighted ages against the low \NO ratios and old ages---galaxies with \NO excess typically exhibit nebular and stellar emission consistent with young starburst galaxies.

In this section, we discuss our results in the context of the recently proposed models to explain origins of the primary nitrogen excess. We first characterise the star-formation mode of NOEGs and assess potential AGN association. Then, we discuss whether the models invoking different star formation histories, outflow mechanisms and enrichment by intermediate-mass and massive stars explain our observations. Our ultimate goal is to understand how the excess is produced and maintained in these systems.

\subsection{Star Cluster Dominated Star Formation?}
\label{sec:disc_clusters}

A natural explanation for why NOEGs become more common toward high redshift is a growing contribution from compact, cluster-dominated star formation. The NOEG incidence among the star-forming population rises sharply across \(4<z<7\)---from \(3^{+5}_{-1}\)\% to \(18^{+12}_{-6}\)\% (\S\ref{sec:number_density_noegs})---an order of magnitude above the \(2.21 \pm 0.91\)\% measured at \(z < 0.5\) with DESI \citep{Bhattacharya2026}\footnote{We note that our work imposes a slightly higher metallicity boundary for selecting NOEGs (\(\logOH<8.2\)) than in \cite{Bhattacharya2026}, but aligning our selection to \(\logOH<8.0\) does not affect our conclusions. In both cases the \NO boundary is \(\logNO>-1.1\).}. This evolution, together with the lower metallicities, higher \NO and greater compactness of the highest-redshift systems, points to star formation increasingly dominated by dense star clusters.

The epochs of NOEGs studied here approximately correspond to the formation epochs of metal-poor globular clusters \citep{ForbesBridges2010}, which JWST is now directly resolving as bound, dense proto-globular clusters in lensed galaxies at \(z\sim\)\,6--10 \citep{Vanzella2023_Sunrise,Adamo2024_CosmicGems,Mowla2024,Adamo2020}. The high gas and star-formation densities measured in star-forming systems at increasing redshift further support this picture \citep{Martinez2025,Topping2025_ElectronDensities,Abdurrouf2024,Isobe2023_densities}.

Recently, \cite{Senchyna2024} suggested that the high \NO and dense nitrogen gas in GN-z11 signify the formation of present-day globular clusters. \cite{Ji2026} demonstrated that NOEGs and metal-poor globular clusters in the Milky Way have very similar He,\,C,\,N,\,O and Fe abundance ratios, and similarly high stellar-mass and SFR densities. This link is supported by the findings in \cite{Adamo2011,Adamo2015} that the fraction of light from bound star clusters in the UV or in the near-infrared and the efficiency of cluster formation are correlated with the surface density of SFR in galaxies. Similarly to \cite{Cameron2026}, we find tentative evidence supporting these comparisons. In our sample, \NO is most strongly associated with the compactness of galaxies and SFR. We demonstrate this in Figure~\ref{fig:NO_sed_sersic_correlations} using SFR calculated from SED fitting on a 10 Myr timescale (\S\ref{sec:sed_properties}). Our high-\NO galaxies have an SFR surface density of \(\log \Sigma_{\text{SFR}_\text{10Myr}} = 0-2 ~M_{\odot}\,\text{yr}^{-1}\,\text{kpc}^{-2}\)---similar to or higher than in the Blue Compact Dwarf (BCD) galaxies in \cite{Adamo2011}, which host bound star clusters formed in a recent burst of star formation in the past 10 Myr or so. However, we cannot estimate the efficiency of cluster formation in these systems with the typical spatial resolution of around \(>400\)~pc in the NIRCam/F444W band that we used. For example, in BCDs in \cite{Adamo2011}, the clusters contribute around 20--30\% of UV and NIR emission, and around half of the SFR of those BCD galaxies can be explained by star cluster formation. Compact, potentially nucleated star formation of this kind leads us to asking whether AGN are preferentially found among NOEGs in the next subsection.

\subsection{AGN among NOEGs}
\label{sec:disc_agn}

\begin{figure}
\begin{center}
    \includegraphics[angle=0,width=1\columnwidth]{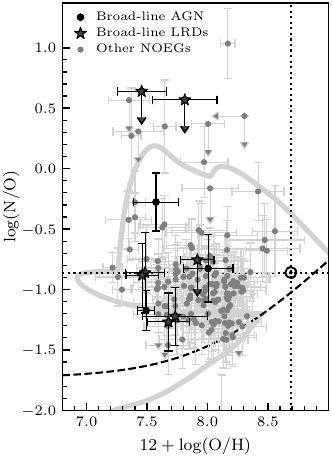}
\end{center}
\caption{NOEGs that were selected to host an AGN shown in the \NO--\OH space. The AGN selection (black circles) was based on the presence of broad lines with the width \(\text{FWHM}>1000~\kmps\) and V-shapes in the UV-to-optical PRISM spectra, characteristic of LRDs (see \S\ref{sec:data_agn_identification}). The rest of the NOEG sample is shown in grey circles. The grey track shows the two-burst model from \protect\cite{KobayashiFerrara2024} (see \S\ref{sec:results_NO}).
}
\label{fig:NO_OH_AGN}
\end{figure}

\begin{figure}
\begin{center}
    \includegraphics[angle=0,width=1\columnwidth]{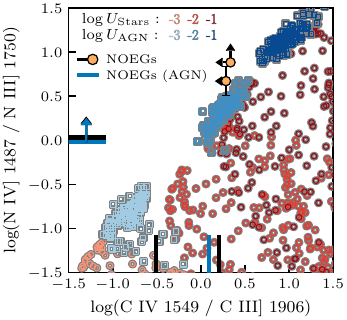}
\end{center}
\caption{Ratios of UV lines probing the ionisation parameter \(\log{U}\) in low-metallicity galaxies with \(\logNO>-1.4\) (orange circles and horizontal or vertical bars).  Blue (AGN) and red (stars) point clouds demonstrate line ratios from \texttt{CLOUDY} models and the colour code is indicated at the top left. Our constraints and limits are in the typical parameter space of AGN line ratios or in the extreme part of the stellar photoionisation space.
}
\label{fig:line_ratios_C4C3_N4N3}
\end{figure}

Strong nitrogen emission and abundance excess are associated with only 0.1\% of quasars selected in SDSS \citep{Bentz2004}, where the nitrogen abundance is estimated to be \(\sim15\) times the solar.  More recently, \cite{Isobe2025} identified the enhancement in a stack of low-resolution NIRSpec spectra of high-redshift sources with broad \ha lines, and several case studies identified \NO enhancements in high-redshift galaxies hosting a prominent AGN: GS-3073 \citep{Ji2024_GS3073,Uebler2023_GS3073}; UNCOVER-45924 \citep{Labbe2024}, or CEERS-01019 \citep{Isobe2023}; or strong high-ionisation \(\nv\,\lambda1240\) lines indicative of hard-ionizing AGN spectra \citep{Tang2025}. It is unclear whether AGN can be in part responsible for \NO enrichment, for example from an AGN-driven outflow or by maintaining hard ionising emission in NOEGs with, for example, \niv emission, as the stars age. Where nitrogen enhancements are associated with a galaxy nucleus, this may indicate chemical enrichment in a nuclear star cluster \citep{Hamann1993}. As nuclear star clusters are considered to be potential birth sites of supermassive black holes \citep{Neumayer2020}, naively, nitrogen enhancements may trace the environments of formation or early development of AGN. In the scope of this work, we only investigate the incidence of AGN among NOEGs, i.e. only their relative population importance, and leave their dedicated analysis for future work. Do any of our NOEGs host AGN?

We identify three broad-line AGN and six broad-line LRDs with V-shape continua from PRISM and photometry in our sample with \NO and \OH measurements (for classification, see \S\ref{sec:data_agn_identification}). In Figure~\ref{fig:NO_OH_AGN} we show that their host galaxies probe a wide range of values in the \NO--\OH plane, but all have low metallicity and low-to-high \NO excess. This metallicity range is similar to a prominent LRD UNCOVER-45924 with \(\logOH \approx 7.2-7.7\) (estimate from \citealp{Ji2026};\citealp{Labbe2024}) or the stacked Type I AGN with \(\logOH \approx 7.5\) \citep{Isobe2025}, and agrees with the average metallicity \(Z=0.08 Z_\odot\) of LRDs \citep{Nikopoulos2026}, although lower than some other \NO-rich AGN reported previously in the literature---e.g., GS-3073 \citep{Ji2024_GS3073,Uebler2023_GS3073} and CEERS-01019 \citep{Isobe2023} have \(\logOH>8.0\).

To expand our AGN selections, we compare high-ionisation line ratios in our galaxies with a set of \texttt{CLOUDY} models with AGN and stellar photoionisation (Figure~\ref{fig:line_ratios_C4C3_N4N3}). We constructed the AGN model grids using \texttt{synthesizer} by varying the following parameters: black hole mass \( \log{(M_{\rm BH}/M_{\odot})}\in [7.0,9.0] \), accretion rate, \(\log\lambda_{\rm Edd} \in [-1.6, 0.0]\), metallicity \(Z \in [0.0001, 0.03]\) and hydrogen density \( \log({n_{\rm H}}/{\rm cm^{-3}}) \in [2.0,5.0]\). The stellar models include ranges in stellar age \( \log{(t_{\rm age}/{\rm yr})} \in [6.0,8.0] \) and metallicity \( Z \in [0.00001, 0.04] \). Both grids also sample the ionisation parameter in the range \(\log{U} \in [-4,-1]\). 

Although we have mostly limits for our galaxies with \(\logNO>-1.4\) and \(\logOH<8.2\), they are consistent with photoionisation by AGN with \(\log U\approx-2\) or hard ionising stellar radiation, with \(\log U>-2\). Two galaxies that we classified as broad-line AGN populate the same region, suggesting that the remaining galaxies may also host an AGN. Only one \(\civ\,/\,\ciii\approx-0.5\) constraint lies in the region more typical for stellar photoionisation with \(\log U\approx-3\).

In summary, the incidence of AGN in our sample matches the expectations. Of 110 galaxies with \(\logNO>-1.4\), nine host an AGN (including LRDs) and a further 5 sources have high-ionisation line ratios consistent with AGN. This fraction is comparable to the fraction of broad-line AGN among UV-selected galaxies at \(z \gtrsim 4\) \citep{Maiolino2024,
Juodzbalis2026_AGN_census}. Therefore, there is no evidence that AGN and NOEGs are systematically associated or share common environments. In the remainder of the section we consider how \NO is produced and maintained in galaxies without assuming an AGN contribution.

\subsection{Enrichment by Wolf-Rayet and AGB stars}
\label{sec:disc_WR_AGB_stars}

As \CO versus \OH in our sample is consistent with expectations from CCSN enrichment (middle panel in Figure~\ref{fig:CNO_OH}), in agreement with previous works \citep{Berg2026,Ji2026,Cameron2023_GNz11}, excess primary \NO has to be produced by a mechanism acting on top of the expected CNO yields.

The most likely source of \NO in NOEGs is a type of star whose IMF-integrated nitrogen yield dominates ISM enrichment over other sources. Under standard stellar population assumptions, we consider chemical yields of the most likely sources: intermediate-mass AGB stars (IAGB; 4--6\,\msun), super-AGB stars (SAGB; 6.5--7.5\,\msun) and massive stars (15--120\,\msun). SAGB with \(Z=0.0001\) produce between 0.03--0.09\,\msun \(^{14}\text{N}\) per star over the TP-AGB phase \citep{Doherty2014_iii}, whereas IAGB stars with \(Z=0.0001\) yield 0.02--0.04\,\msun per star during the TP-AGB phase \citep{Karakas2010}. Fast-rotating WR stars (15--120\,\msun), with rotation velocity \(v=300~\kmps\) and \(Z=0.00013\) produce 0.006--0.08\,\msun of nitrogen per star \citep{LimongiChieffi2018}---of the
same order as the yield per an SAGB or IAGB star. According to the Kroupa IMF \citep{Kroupa2001}, for every 1\,\msun formed there are 0.012 IAGB stars, 0.003 SAGB stars and 0.004 and 0.001 of 15--40 and 40--120\,\msun stars. Multiplying
the yields per star by the number of stars of each type results in IAGB and SAGB nitrogen yields adding up to \(4\times10^{-4}\) (\msun of \(^{14}\text{N}\) per 1\,\msun of stars formed) and total WR yield of \(2 \times 10^{-4}\), meaning nitrogen enrichment from AGB and WR stars is comparable to within a factor of a few.

Even though both WR and AGB stars produce comparable IMF-averaged quantities of nitrogen, different delay time distributions mean that enrichment by these stellar types activates at different times. SAGB stars with 7\,\msun begin to reach the TP-AGB stage after at least 30--50\,Myr \citep{Doherty2014_iii,KarakasLugaro2016}, whereas the least massive WR stars start to deliver nitrogen almost immediately, with a lifetime of 5--10~Myr \citep{Crowther2007,LimongiChieffi2018}. Therefore, for starbursts with ages \(<\)\,30--50\,Myr their enrichment has to be dominated by low-metallicity WR stars, but for galaxies older than \(>\)\,30--50\,Myr, either both types or only AGB type stars can be responsible for nitrogen production.

Recent GCE modelling confirms these conjectures, showing that at low metallicity WR stars are the dominant source of nitrogen in the first 20\,Myr assuming standard IMF with stars below 120\,\msun \citep{Watanabe2026,KobayashiFerrara2024}. Their model predictions can explain the strongest abundances of \(-0.5 < \logNO < 0.0\) and metallicity \(6<\logOH<8\) found with JWST at \(6<z<12\), including some of the well-known objects with detailed observations, such as GN-z11 and RXCJ2248 \citep{Watanabe2026,Berg2026,KobayashiFerrara2024}. While WR stars eject nitrogen and then carbon via opacity-driven mass loss, only \NO is found to be enhanced in these systems and \CO matches the expected CNO yields. This suggests that the WR wind opacity is low and instead the mass loss is rotation-driven, ejecting only nitrogen without reaching into the deeper stellar layers containing triple-alpha-fused carbon---a signature of WN stars \citep{Crowther2007}. This is consistent with the low metallicities of these galaxies \citep{Meynet2005}. \cite{Berg2026} also argued for this scenario, strongly suggested by the presence of the blue WR bump with nitrogen emission and weakly-broadened \heiiwavelopt in the spectra of RXCJ2248\footnote{Galaxy RXCJ2248 is a nitrogen-enhanced galaxy lensed by a factor of \(\sim\)\,7 at \(z=6.1\) that has the deepest spectroscopic observations to date with around 30 hr of JWST/NIRSpec integration.} and the lack of the WR carbon bump at 5800\,\angstrom.

Although WR enrichment can successfully explain the observations, this mechanism is very short-lived \citep[e.g.,][]{KobayashiFerrara2024} and an additional mechanism is required to explain the relatively large number of galaxies with enhanced nitrogen, especially where stellar populations may be older than \(>\)\,30--50\,Myr \citep{Watanabe2026,ArellanoCordova2024b}, such as in our sample here (e.g., Figure~\ref{fig:NO_age_outflows_vs_models}). Indeed, high \NO from WR stars should be observable for only a few Myr before becoming diluted by oxygen produced in CCSN in \(<\)\,10~Myr. A proposed solution is a double starburst model in \cite{KobayashiFerrara2024}: pre-enrichment after the first burst places a galaxy in the region of \(Z\approx0.5 \,Z_\odot\) and \NO around local values, and the second burst (caused by an inflow of hydrogen gas) first enhances \NO via WR winds and then dilutes it with CCSN producing an \NO--\OH loop that can be matched to observations. This model allows a WR-enriched galaxy to spend around 10~Myr in the space of observed abundance ratios. Finally, for galaxies at \(>\)\,30--50\,Myr, the combined effect of an outflow decreasing the overall metallicity and AGB stars starting to produce primary nitrogen at these ages can explain the second episode of \NO enhancement \citep{Watanabe2026,Bhattacharya2026}---extending primary nitrogen enhancement to later cosmological epochs and explaining the fact that we find \NO-enhanced galaxies at intermediate redshifts and in galaxies with a Balmer break (Figure~\ref{fig:stacked_prisms},\,\ref{fig:NO_age_outflows_vs_models}) in this work.

\subsubsection{Model Comparisons (with standard assumptions)}
\label{sec:disc_no_cycle}

In this study, we have identified nitrogen emitters at \(4<z<8.5\)---extending the observational sample significantly to lower redshift than previously identified with JWST---and have shown that such systems span a wide range of chemical and physical properties, which suggests that multiple mechanisms may be at work depending on the type of galaxy or its star formation history. Below we discuss how their ages, outflows and presence of helium emission correspond to their chemical properties in the context of the different proposed GCE models.

In particular, we compare our sample at different redshifts with the double-burst model of \cite{KobayashiFerrara2024} without an outflow-induced gap between two bursts (summarised in \S\ref{sec:results_NO}) in Figure~\ref{fig:NO_age_outflows_vs_models}. The first burst is very extended in time with a 200\,Myr exponential timescale. After 200\,Myr, the model has an inflow of pristine gas at the time \(t_0\), from which we start the colour scheme in the figure. The inflow shifts the metallicity to lower \OH values and almost immediately leads to the second, very intense burst of star formation with the timescale of 0.2\,Myr. WR winds produced in this burst increase \NO in the first 1--3\,Myr, before the first massive stars start to explode driving up the metallicity and gradually decreasing \NO. Finally, the metallicity exceeds \(Z_{\odot}\) and secondary nitrogen production from metal-rich CCSN and primary nitrogen production in young AGB stars gradually increase \NO to the present day. We cut out the final part of the model track, as none of the known galaxies at high redshift populate that region of the diagram.

Our data are generally described well by this model in all redshift bins, with a key caveat. As the model describes only a brief period of \(\sim\)\,10~Myr, we expect it to apply most accurately to our galaxies at \(6<z<8.5\). Indeed, our galaxies falling on this track on average have the highest equivalent widths of \ha, and some have evident Balmer jumps (middle panel in Figure~\ref{fig:NO_age_outflows_vs_models}), consistent with the lowest mass-weighted ages (\(t_\text{age}\approx10\)~Myr) inferred from SED fitting (top panel in Figure~\ref{fig:NO_age_outflows_vs_models})
\footnote{It is worth noting that the second burst in KF24 is very intense and its stellar population in a real galaxy may outshine stars formed in the previous burst. Therefore, it is possible that the youngest mass-weighted ages we identify in this work do not reflect possible prior, lower-intensity star formation.
}. Besides, our stacked NOEGs appear to exhibit WR bump at 4687~\angstrom suggesting the presence of WR stars and very recent star formation (see \S\ref{sec:disc_WR_stars}).
At \(4<z<6\), on average \(>\)30\,Myr older galaxies should move to the region of secondary nitrogen enrichment according to the model, which is not observed. Therefore, to keep their metallicity low and \NO moderately-enhanced around the solar value, these galaxies must experience additional episodes of gas outflows or inflows. Indeed, we find evidence of outflows in some NOEGs in this region, as shown in the lower panel in Figure~\ref{fig:NO_age_outflows_vs_models}. This is similar to the recently considered variation of these GCE models in \cite{Bhattacharya2026} and the models in \cite{Watanabe2026}. Such a scenario was recently reproduced in \texttt{THESAN-ZOOM} simulations, where \cite{McClymont2026_MysteryNO} demonstrated that intermediate \NO excess can be observed after a feedback-driven outflow, when a galaxy becomes mini-quenched for a period of \(\lesssim\)\,100~Myr (see also \citealp{Rai2026} for evidence based on inferred star formation histories). In addition to diluting the metallicity, inter-burst outflows have been proposed as a viable mechanism for ejecting dust grains to explain the lack of dust attenuation in galaxies observed at \(z>10\) and their high luminosities \citep{Ferrara2026_RedMonsters,Ferrara2025_AFM}. Therefore, to explain our sample at \(4.0<z<6.0\), future models will require these additional outflow episodes in the \NO--\OH cycle, which we demonstrate in a schematic in Figure~\ref{fig:xkcd_no_cycle}. In the following sections, we will discuss signatures that indicate enrichment in our NOEGs on both of these timescales: \(<\)\,10~Myr and \(>\)\,10--40\,Myr.

\subsubsection{Wolf-Rayet Bump in \NO-enhanced Galaxies}
\label{sec:disc_WR_stars}

\begin{figure*}
\centering
\includegraphics[width=\linewidth]{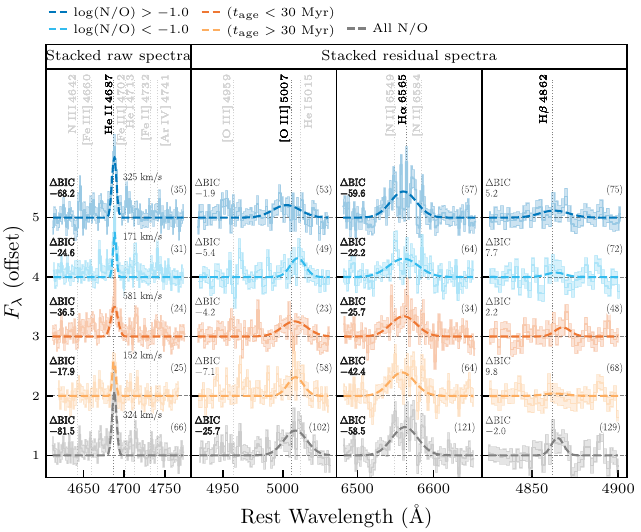}
\caption{Median spectroscopic stacks of: (a) \heiiwavelopt blue bump; (b) residuals of \oiiiwavelopt; (c) residuals of \(\ha\,\lambda 6565\); and (d) residuals of \(\hb\,\lambda 4862\). The residual emission in panels (b), (c), (d) is obtained by subtracting best-fit Gaussian line fits. The stacks of [\oiii], \ha and \hb lines exclude the galaxies with detected secondary broad components to investigate the presence of secondary components in the rest of galaxies. Each stack is shown in five bins from bottom to top: (1) total \NO sample; (2)-(3) mass-weighted ages \(t_\text{age}>30\)~Myr and \(t_\text{age}<30\)~Myr; and (4)-(5) \NO split around (\(\logNO=-1.0\)). For each stack, we show \(\Delta \text{BIC}=\text{BIC}_{\text{Lines+Continuum}}-\text{BIC}_{\text{Continuum}}\) of the best-fit models as evidence of the presence or absence of the lines. Note, in panel (b), the baseline model includes \(\mathrm{He\,\textsc{i}\,\lambda5015}\) and tests against the residual [\oiii] emission. The number of spectra combined in each stack is shown in brackets. Higher-\NO excess and younger-age stacks detect \heii with \(\text{FWHM}=300-600\)~\kmps (FWHM is annotated next to each line) and significantly detect residual \ha with \(\text{FWHM}=1800-2000\)~\kmps. Note, each spectrum is normalised by its median noise.
}
\label{fig:stacked_spectra}
\end{figure*}

WR stars are a key component of the chemical evolution models we discussed above (\S\ref{sec:disc_WR_AGB_stars}) to produce the observed \NO offset at the start of a starburst---for either a single or a double-burst model. We analyse our sample spectra for the presence of WR emission bumps around \heiiwavelopt (blue bump) and \civwavelopt (red bump), which are characteristic signatures of these WR stars. We detect the \heiiwavelopt line with \(\snr>3\) in four out of 66 NOEGs that have \heii spectra. However, Figure~\ref{fig:stacked_spectra} (left panel) reveals that \heii may be commonly present in the whole sample when we stack the spectra to achieve higher signal-to-noise. Our stacking, including all 66 spectra, splits galaxies into 5 bins, similarly to the analysis of PRISM spectra (Figure~\ref{fig:stacked_prisms}): (1) the total sample; (2)--(3) mass-weighted age younger or older than \(t_\text{age}=30\)~Myr; and (4)--(5) \NO ratio split around the median value \(\logNO=-0.96\).

Interestingly, the stacked spectra reveal a broadened \heii line in the high-\NO and young samples with \(\snr = 9.2\) and \(7.3\) and line widths \(\text{FWHM}=325^{+87}_{-79}\)~\kmps and \(581^{+234}_{-226}\)~\kmps, respectively. \niiiwavelopt and \(\mathrm{Fe\,\textsc{iii}]\,\lambda4702}\) can also be seen in the high-\NO spectrum, although higher signal-to-noise is required to significantly detect these faint lines. The intermediate width of the stacked \heii is consistent with \(\text{FWHM}(\heii)=530~\kmps\) in the WR bump in RXCJ2248 in \cite{Berg2026} at \(z=6.1\). Similarly to RXCJ2248, none of our stacks exhibit the red \civwavelopt bump, and our median high-\NO stack metallicity is equal to \(0.13\,Z_\odot\) (ranging between \(0.03-0.2\,Z_{\odot}\)), consistent with \(0.1\,Z_{\odot}\) for RXCJ2248. These features suggest the presence of low-metallicity WR stars in NOEGs whose rotation-driven winds can eject primary nitrogen without yielding carbon. This is in contrast, for example, to the Sunburst Arc at \(z=2.37\) \citep{RiveraThorsen2024}, which has \(\text{FWHM}(\heii)=1370~\kmps\), \(0.7\,Z_\odot\) gas metallicity and the carbon bump, indicating stronger line-driven winds that eject both nitrogen and carbon.

The stacked \heiiwavelopt in the lower-\NO and older galaxy groups is weaker and narrower than in their complementary samples, with \(\snr = 6.3\) and \(5.8\) and \(\text{FWHM}=150-170\)~\kmps, consistent with a typical nebular gas dispersion of these galaxies. Narrow \heii emission in this case may indicate ionisation by \(>\)\,54~eV photons from stripped binaries with \(M=10-20\,\msun\) stars \citep{Drout2023,Gotberg2019,Garnett1991} or X-ray binary stars \citep{Simmonds2021,Garnett1991}. As the stripped binaries can ionise the gas beyond \(>\)\,10~Myr after the starburst \citep{Gotberg2019}---longer than the WR stars can---this would be consistent with the longer mass-weighted ages we estimate for these galaxies. Therefore, contributions from WR winds in these galaxies are likely insignificant and stars with masses \(>\)\,20--25\msun are mostly gone. This enrichment stage would be consistent with the primary N enrichment by the ``AGB stars\,$+$\,outflows'' mechanism in NOEGs at \(4<z<6\).

Finally, our stacking demonstrates that detecting typically weak \heii emission originating from high-mass stars requires deep observations. By stacking 35 spectra for our high-\NO galaxies, we achieve the total integration time of around 23 hr of effective exposure time with NIRSpec/MSA. Future programs can use these findings to plan deep spectroscopic observations of WR features at high redshift.

\subsection{Are Nebular Outflows Associated with \NO excess?}
\label{sec:disc_nebular_outflows}

Outflows driven by stellar feedback can partly shape the local \NO--\OH relation. For example, local galaxies with higher \NO and larger stellar masses have outflows with the lowest loading factors, allowing them to retain nitrogen and high metallicity at \(\logOH \gtrsim 8.0\), whereas stronger outflows in lower-mass galaxies instead introduce scatter in \NO \citep{ArellanoCordova2025_density_feedback}. At high redshift, outflows are a necessary mechanism in GCE models using AGB stars to reproduce the strong nitrogen enhancement at \(\logOH \lesssim 8.0\): they lower \OH at \(>\)\,30~Myr allowing AGB stars to increase \NH \citep{Watanabe2026,McClymont2026_MysteryNO}. Do we find evidence of such outflows in our sample?

We investigate \oiiiwavelopt emission lines as the brightest optical tracer of ionised gas kinematics. To later compare outflow incidence in our galaxies with that in all star-forming JADES galaxies from \cite{Carniani2024}, we limit our sample to stellar masses of \(M_{\star}<10^{9.2}\,M_{\odot}\) and \(\text{SFR}<10~M_{\odot}\,\text{yr}^{-1}\). We detect a secondary component in the \([\oiii]\,\lambda 5007\) line in \(40 \pm 7\)\% (21 / 53) of galaxies in our \NO sample at \(\logOH < 8.2\) and at \(\logNO>-1.3\). This detection is based on a comparison of BIC values of single and double-Gaussian best-fit profiles. Our detection criteria are: \(\mathrm{BIC_{double}}-\mathrm{BIC_{single}}<-10\) and \(\snr>3\) (\S\ref{sec:data_spec_lines}). The median flux ratio of the secondary-to-primary components is \(0.43 \pm 0.11\), which is consistent with outflows in \(\log M_{\star}=4-7\) low-mass local galaxies \citep{Xu2022}. The individual ratios vary from \(0.08\pm0.04\) to \(1.13\pm0.51\) and \(1.35\pm0.50\)---if produced by an outflow, these ratios would indicate that more than half of the \opp gas in these galaxies is in the outflow. These lines have median \(\text{FWHM}(\oiii_\text{secondary})=196 \pm 66\)~\kmps and velocity offsets that vary from \(-110\) to \(100\)~\kmps for offsets detected at the 3\(\sigma\) level, with the median consistent with no offset. The symmetry of the total line profiles indicates either negligible dust extinction---so that all outflow velocities are unattenuated and symmetric around the systemic velocity---or a contribution from multiple nebular clumps in each galaxy. 

In a similar analysis, we find that secondary components in \ha lines are detected more frequently, in \(35\pm7\%\) (18 / 51) of galaxies with \NO that exclude our LRD and AGN selections (\S\ref{sec:data_agn_identification}). The median width of these components is \(\text{FWHM}=240 \pm 78~\kmps\), which is broader than the primary narrow component, and consistent with the broad [\oiii] lines within uncertainties. A median object does not show a significant offset, although in eight systems, where the offset is detected at 3\(\sigma\) level, we report a median redshift of \(\Delta v = 95 \pm 80~\kmps\). The median secondary-to-narrow flux ratio is \(0.32 \pm 0.06\), matching that of the [\oiii] broad components, although one galaxy has the ratio \(1.3\pm0.4\) with the line width \(\text{FWHM}=280\)~\kmps. Such a high ratio implies that more than half of the galaxy's nebular gas is in the outflow or dominated by stellar clump dispersions, or is possibly produced by a faint AGN. 

Interestingly, all broad [\oiii] and \ha emitters that we associate with outflows are in the moderate-to-strongly enhanced part of our sample with \(\logNO>-1.4\). Considering the union of these selections, we estimate that outflows take place in \(41 \pm 7 \%\) (22 / 54) of NOEGs, with 16\,/\,22 galaxies having both broad lines simultaneously. This fraction is consistent with the 25--40\% incidence of [\oiii] and \ha outflows in \cite{Carniani2024} in a similar redshift range, and matches their highest incidence estimates.

Our findings show that low stellar mass, metal-poor galaxies with \(\logNO>-1.4\) exhibit frequent outflows, but does selecting for \NO excess make an outflow incidence more likely in such a galaxy? We estimate that \(31 \pm 4\%\) (42 / 135) of our parent sample with nitrogen lines shows secondary [\oiii] and \ha components in the same mass and SFR range, compared to \(41 \pm 7 \%\) in NOEGs. The Fisher exact test for small samples shows that these ratios are the same with the probability \(p=0.059\). Therefore, there is tentative evidence suggesting a physical connection between outflow incidence and a moderate-to-high \NO excess in galaxies with low stellar masses and nitrogen lines, although a larger sample is required to reach higher confidence.

Depending on the mass-weighted ages of the galaxies with the outflow signatures, we interpret the tentative physical link between an outflow and \NO abundance differently. The median age of these galaxies is below 10~Myr. They likely experience an outflow as a result of CCSN feedback that starts to dilute their strong WR-driven \NO excess, moving these galaxies to the region of low \NO enhancement at approximately \(-1.2<\logNO<-0.6\). At the same time, five out of 22 galaxies are 20--40~Myr old---the time after the recent starburst when AGB enrichment of the primary N is expected to start. The outflow in this case enables the next cycle of low-metallicity \NO enrichment by removing some of the pre-enriched gas.

\subsection{Evidence of Partially Recombined Outflows?}
\label{sec:disc_nebular_outflows_fading}

The tentative connection between outflow incidence and \NO enhancement in \(M_{\star}<10^{9.2}\,M_{\odot}\) galaxies helps to explain the mechanism for producing the enhancement in around 40\% of our galaxies with \(\logNO>-1.4\), but do the rest of the high-\NO galaxies (many of which have mass-weighted ages greater than 10~Myr) experience pure nitrogen enrichment without outflows? Could some of the outflows be intrinsically weaker or predominantly neutral, preventing their detection?

We further search for possible faint [\oiii] emission from outflows by stacking their spectra. In particular, we stack the residuals after subtracting the [\oiii] doublet lines to be able to detect any unaccounted faint emission. To subtract the primary narrow lines accurately, we fit them using \texttt{emcee} Ensemble Sampler \citep{ForemanMackey2013_emcee}, with 64 walkers, 20,000 total steps and 10,000 warm-up steps, and verify each fit by eye. Then we combine the residual spectra and, to identify residual emission, we fit the stacked spectrum as a continuum (C) with and without a Gaussian line (L) and require \(\Delta \mathrm{BIC} = \mathrm{BIC_{L+C}}-\mathrm{BIC_{C}}<-10\) as strong evidence for the presence of the unaccounted components. We perform this test in separate subsamples, as described in \S\ref{sec:disc_WR_stars} when stacking \heii spectra, and demonstrate the results in Figure~\ref{fig:stacked_spectra}. 

The stacked spectra do not reveal any significant residual emission in \([\oiii]\,\lambda5007\). All stacked galaxies with \NO (lower panel in the second column, Figure~\ref{fig:stacked_spectra}) have residual flux likely produced by \(\hei\,\lambda5015\). Therefore, there is no evidence of ongoing [\oiii] outflows in the remaining 70\% of NOEGs.

However, when stacking the residual \ha spectra without a detected secondary component, we significantly detect broad residual emission. Similarly to stacking [\oiii] spectra described above, we subtract the primary \ha line and the \niiwavel doublet in individual spectra and stack the residuals. The third column in Figure~\ref{fig:stacked_spectra} demonstrates the median stacked spectrum with the secondary components detected in all groups with \(\snr=5-10\), \(\text{FWHM}=1800-2000\)~\kmps and all systematically offset by \(\approx-200\)~\kmps. The \ha broadening may arise from an AGN or an outflow not detectable in [\oiii].

As the nebular \oiiiwavelopt emission tends to be stronger than \ha in sufficiently metal-enriched starbursts, the reason for not detecting [\oiii] in the potential outflows is either dust attenuation or recombination \(\opp \rightarrow \ion{O}{0}\). A modest dust attenuation with \(A_V=1\) will remove around 1.3 times more [\oiii] flux than \ha \citep{Calzetti2000} in the already faint outflows. Partially recombined outflows in NOEGs is another plausible scenario: if outflows are required to reset the ISM metallicity for the subsequent nitrogen enrichment by AGB stars---as invoked in recent models \citep{Watanabe2026} and observed in simulations \citep{McClymont2026_MysteryNO}---then there must be a time delay between the outflow and when the excess nitrogen builds up and becomes observable. For a sufficiently long time delay, the ionised [\oiii] outflow in a NOEG may have already recombined and is no longer detectable. Several Myr after an outflow starts, any remaining massive stars die, the ionising spectrum softens and the 35--55 eV \(\op \rightarrow \opp\) ionisation begins to drop before the 13.6 eV \(\text{H}^0 \rightarrow \hp\) ionisation, leading to faster depopulation of \opp than \hp. On top of that, the existing \opp gas recombines 50 times faster than \hp as the ratio of recombination coefficients is \(\alpha_B^{\opp}/\alpha_B^H\approx50\) for \(\Te=10^4~\text{K}\), using \(\alpha_B^{\opp}=1.3\times10^{-11}~\text{cm}^3\,\text{s}^{-1}\) \citep{Nahar1999} and \(\alpha_B^H=2.6\times10^{-13}~\text{cm}^3\,\text{s}^{-1}\) \citep{OsterbrockFerland2006}, for the recombination timescale \(t_\text{rec}=1/(n_e \alpha_\text{rec})\). The combined effect of these factors will result in [\oiii] emission of the outflow fading sooner than that of \ha, explaining why broadening in \ha stacks is observed, while none is detected in [\oiii]. Tentative evidence of such faded post-outflow material at high redshift has been found in ALMA (Atacama Large Millimeter/submillimeter Array) observations of the galaxy GS-z14-0 \citep{Schouws2025,Carniani2025,Ferrara2024}.

By combining multiple observations, we find evidence that beyond the brightest nebular outflows detectable in individual galaxies, they may also be common in the rest of the high-\NO galaxies, albeit fainter and possibly partially recombined. This evidence suggests that the majority of NOEGs may be associated with outflows, either ongoing and fully ionised in 40\% of NOEGs, as discussed in the previous section, or launched recently and partially recombined, as would be consistent with the stacked nebular lines in the remaining 60\% of NOEGs. In the latter case, their detection requires deeper observations or stacked measurements.

Finally, we checked that the presence of significant residuals in our stacked residual spectra of \ha does not result from an inaccurate subtraction of the primary components. We exclude this possibility by inspecting the residual spectra by eye and verifying the quality of the best-fit models. Increasing the MCMC sampling density and length did not change the results, indicating that our fitting has converged. We also verify our procedure by showing that the stacked residual spectra of the \hb line---typically not exhibiting broadened components in spectra of high-redshift AGN (e.g., \citealp{Juodzbalis2026_AGN_census}) or nebular outflows--- do not exhibit significant residuals (last column in Figure~\ref{fig:stacked_spectra}).

\subsection{Synthesis: the Proposed Evolutionary Sequence}
\label{sec:disc_synthesis}

\begin{figure*}
\begin{center}
    \includegraphics[angle=0,width=0.7\textwidth]{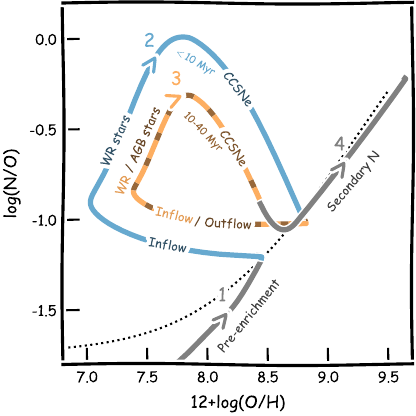}
\end{center}
\caption{Illustration of the \NO cycle of primary enrichment suggested to explain nitrogen enhancements observed in galaxies with strong ongoing and recent starbursts. The scenario involves WR stars and ``outflows\,\(+\)\,AGB stars'' driving nitrogen enrichment at different times, as discussed in \S\ref{sec:disc_no_cycle}. Briefly, after the initial enrichment (1), WR stars drive the \NO enhancement (2), quickly diluted by CCSNe, completing the first cycle within \(\sim10\)~Myr. This cycle may complete with a starburst-induced outflow, which resets the metallicity and enables the first AGB stars to produce \NO enhancement (3) at \(\sim30\)~Myr, diluted again by the remaining CCSNe, completing the second cycle. Any new inflows may produce another weaker cycle that eventually leads to secondary N production (4) by metal-rich AGB stars. The values used in this graphic only approximately correspond to JWST observations and discussed GCE models. The dotted line is the local relation from \protect\cite{Nicholls2017}.}
\label{fig:xkcd_no_cycle}
\end{figure*}

The evidence discussed above can be assembled into the following time-ordered sequence, illustrated schematically in Figure~\ref{fig:xkcd_no_cycle}. Compact, cluster-dominated star formation of the high-stellar-mass-density NOEGs (\S\ref{sec:disc_clusters}) and initial pre-enrichment to \(Z \lesssim 0.5\,Z_\odot\) set the stage; a pristine-gas inflow lowers \OH, and within \(\lesssim\)\,10~Myr of the resulting burst, rotation-driven WN winds raise \NO at fixed \OH, evidenced by the broad \heiiwavelopt bump and the absence of a carbon bump in our young, high-\NO stacks with several galaxies exhibiting Balmer jumps from nebular continuum (\S\ref{sec:disc_WR_stars}). CCSNe then dilute the excess within a few Myr, consistent with the CCSN-dominated \CO and \NeO ratios of our sample and with the delayed Type Ia enrichment implied by \ArO (\S\ref{sec:results_alpha_elements}, \S\ref{sec:disc_WR_AGB_stars}). Then, a feedback-driven outflow at \(\sim\)10--40~Myr ejects the metal-enriched gas lowering the \OH again---the tentatively elevated \(41 \pm 7\%\) outflow incidence and the broad residual \ha with no [\oiii] counterpart (\S\ref{sec:disc_nebular_outflows}, \S\ref{sec:disc_nebular_outflows_fading}) trace both its ongoing and faded phases. Beyond \(\sim\)30--50~Myr, IAGB and SAGB stars start to deliver the second episode of primary nitrogen into this diluted gas, accounting for the \NO-enhanced galaxies we find at \(4<z<6\) with \(t_\text{age}>30\)~Myr, Balmer breaks and narrow \heii. The increasing incidence of NOEGs with redshift (\S\ref{sec:number_density_noegs}) naturally signifies that the \NO enhancement cycles in the first \(\sim\)\,50~Myr comprise a larger fraction of the galaxies' age and a potentially growing contribution from cluster-dominated star formation. AGN are present at the rate expected for the parent population and are not required at any stage (\S\ref{sec:disc_agn}).

We acknowledge possible secondary effects and alternative scenarios behind the observed \NO--\OH distribution at high redshift. Wind-driven shocks can bias nitrogen abundances by altering UV line fluxes via photoionization \citep{Flury2025}. Although shocks do not rule out nitrogen enhancement in all galaxies \citep{Senchyna2024}, they can contribute to \(\niii]\), \niv line strengths in some galaxies, consistent with the higher incidence of winds that we tentatively find in NOEGs. Major mergers of metal-poor galaxies can dilute gas-phase metallicity (e.g., \citealp{Bustamante2018}), but since the effect is below 0.2--0.5 dex, it should be hidden in the random scatter in the \NO--\OH plane. Supermassive stars (\(10^3<M_\star<10^5~M_\odot\)) offer another alternative mechanism for the most extreme NOEGs, as they can reproduce some of the observed CNO abundance ratios \citep{Ebihara2026,Nandal2024a,NageleUmeda2023,Charbonnel2023}. However, because they distribute neon into the ISM less efficiently then CCSNe or WR stars \citep{Nandal2024a}, they cannot account for the full observed \NeC--\NO distribution \citep{Watanabe2026}, making them unlikely as a population-wide mechanism, but plausible as rare events producing the most extreme \([\NO]>1.5\) cases \citep{Watanabe2026}. Finally, \cite{Cameron2023_GNz11} discussed tidal disruption events as a mechanism producing the enhanced \NO in some galaxies with a supermassive black hole.

\section{Conclusion}
\label{sec:conclusion}

In this paper we investigate the origins of nitrogen abundance in low-metallicity galaxies at high redshift. We assemble the largest self-consistent sample to date at \(4<z<8.5\) using archival JWST data, comprising 134 \NO and \OH abundances (26 \Te-based and 108 using strong-line calibrations) and 76 NOEGs. Using C,\,N and alpha-element abundances of these galaxies in the context of their physical properties and morphologies, we find that they exhibit a diversity of star formation histories. At the same time, they share features that facilitate the observed \NO excess and point to the most likely sources of the primary nitrogen. We summarise our findings in the following:

\begin{itemize}
    
    \item \textbf{Incidence of \NO-excess galaxies}. NOEG cosmic number density increases with redshift in contrast to all star-forming galaxies. Although our sample draws from surveys with varying observational depths and complex selections, we distil it into a small homogeneous subsample with tractable selections from the JADES survey. The fraction of NOEGs rises from \(3^{+5}_{-1}\)\% at \(4<z<5\), to \(5^{+5}_{-2}\)\% at \(5<z<6\) and \(18^{+12}_{-6}\)\% at \(6<z<7\). This incidence is strongly elevated above the fraction of \(2.21 \pm 0.91 \%\) at \(z<0.5\) in DESI \citep{Bhattacharya2026}. We also note that most sources with nitrogen lines are NOEGs at higher redshift. We find that the nitrogen-line sample and low-metallicity NOEGs are biased to brighter (\(\Delta M_\text{UV}=0.6\)) and more massive (\(\Delta \log (M_\star/M_\odot)=0.6\)) systems, compared to total spectroscopic and photometric datasets, and therefore their incidence may be underestimated.
    
    \item \textbf{CCSN-dominated enrichment of C, O, Ar, Ne at \(z>5\)}. While nitrogen abundances exceed the expected yields of CCSN, \CO ratios of our NOEGs agree with them. \NeO abundances of NOEGs are consistent with solar values and typical CCSN yields at all times. \ArO tentatively decreases with redshift, from supersolar values at \(z=4-5\) to \((\ArO)=0.33 \pm 0.14\,(\ArO)_{\odot}\) at \(z=5-6\) and becomes undetected at \(z=6-8.5\). As around 30\% of argon is produced in Type Ia SNe in GCE models \citep{Kobayashi2020}, this suggests decreasing Type Ia SN rates in high-redshift NOEGs, CCSN-dominated yields and younger stellar populations.

    \item \textbf{Compactness and star formation density of NOEGs}. The effective radius and star formation rate of a galaxy are together responsible for approximately 20\% of the variation in \NO in low-metallicity systems. Due to a large scatter in the data (physical and statistical) these correlations are tentative at \(2-3\sigma\) significance. We find that \NO and the incidence of NOEGs increase with redshift; together with the on-average increase in gas density with redshift \citep{Topping2025_ElectronDensities}, this suggests that these galaxies experience star-cluster-dominated star formation. The SFR surface density of our NOEGs is similar to or higher than that of local compact dwarfs with star-cluster-dominated SFR \citep{Adamo2011}.

    \item \textbf{Wolf-Rayet helium bump in NOEGs}. We detect \heiiwavelopt with \(\text{FWHM}=325\)~\kmps and tentatively other WR bump lines by median-combining 35 spectra (23~hr of NIRSpec observations) in galaxies with \(\logNO>-1.0\). At typical metallicities of \(0.13\,Z_{\odot}\) and with no carbon bump, the blue bump indicates the presence of low-metallicity WR stars whose rotation-driven winds produce primary nitrogen without enhancing \CO. Stacked \(\logNO<-1.0\) galaxies have a weaker and narrower \heii (\(\text{FWHM}=171\)~\kmps) which implies that other nitrogen-production mechanisms dominate or that they have different star formation histories. UV-optical continua, nebular lines, and the \heii bump in higher-\NO galaxies match those with mass-weighted ages \(t_\text{age}<30\)~Myr, while lower-\NO systems resemble older (\(t_\text{age}>30\)~Myr) galaxies. Galaxies with UV nitrogen lines are younger than those with optical lines and appear to be most strongly enhanced in \NH---they are the most likely candidates for WR galaxies (RXCJ2248 is a confirmed WR galaxy in \citealp{Berg2026}). We conclude that WN stars drive primary nitrogen excess in NOEGs within the first 10~Myr of a strong starburst.
    
    \item \textbf{Elevated Incidence of Ionised Outflows in NOEGs}. Beyond \(\sim\)10~Myr, GCE models invoke outflows and AGB stars to explain \NO enhancement \citep{Watanabe2026,Bhattacharya2026}, as also seen in simulations \citep{McClymont2026_MysteryNO}. We find that \(41 \pm 7\%\) of NOEGs at \(M_{\star}<10^{9.2}\,\msun\) exhibit outflows with broadened [\oiii] and \ha lines. This exceeds the outflow incidence of \(31\pm4\%\) in the parent sample with nitrogen lines at \(4<z<8.5\). With \(p=0.059\), there is tentative evidence for a physical link between \NO enhancement and outflows. In NOEGs without individually detected outflows, subtracting [\oiii] and \ha and stacking residuals reveals broad \ha (\(\text{FWHM}=1800\)--\(2000\)~\kmps) with no [\oiii] counterpart---consistent with dust-attenuated or [\oiii]-recombined outflows. These outflows, present in many NOEGs, would enable AGB-driven \NO enhancement in galaxies older than \(\sim\)20--40~Myr, consistent with models and simulations.
    
\end{itemize}

This study finds evidence supporting multiple recently proposed channels for \NO enrichment---including WR stars and AGB stars\,\(+\)\,outflows, and argues that their combined evolutionary effects likely produce more than one \NO--\OH cycle at \(\logOH<8.2\) and \(\logNO>-1.4\), as presented in Figure~\ref{fig:xkcd_no_cycle} and summarised in \S\ref{sec:disc_synthesis}. Our large coherent sample expands on the existing observations. It will help to construct new chemical evolution models of nitrogen-enhanced galaxies. In future work, the C, N, O and Ne abundances presented here, combined with He, can be used to test for the distinct yields of supermassive stars or tidal disruption events in individual NOEGs (e.g., \citealp{Watanabe2026,Ebihara2026}).

\section*{Acknowledgements}

We would like to thank Chiaki Kobayashi for providing their chemical evolution models, Nathan Adams and Darach Watson for insightful discussions. We acknowledge support from the ERC (European Research Council) Advanced Investigator Grant EPOCHS (788113), as well as two studentships from the STFC (Science and Technology Facilities Council).  This work is based on observations made with the NASA (National Aeronautics and Space Administration) / ESA (European Space Agency) \textit{Hubble Space Telescope} (HST) and NASA/ESA/CSA (Canadian Space Agency) \textit{James Webb Space Telescope} (JWST) obtained from the \texttt{Mikulski Archive for Space Telescopes} (\texttt{MAST}) at the \textit{Space Telescope Science Institute} (STScI), which is operated by the Association of Universities for Research in Astronomy, Inc., under NASA contract NAS 5-03127 for JWST, and NAS 5–26555 for HST. The authors thank all involved with the construction and operation of JWST, without whom this work would not be possible. DJA is an initiative of the Cosmic Dawn Center (DAWN), which is funded by the Danish National Research Foundation under grant DNRF140.

\section*{Data Availability}

The data products presented herein were retrieved from the Dawn JWST Archive (DJA) at: \url{https://dawn-cph.github.io/dja/} \citep{Heintz2025_PRIMAL,BrammerValentino_DJAv4}. These products include the publicly available NIRSpec spectra observed in the following programs: 1180, 1181 (PI: D. Eisenstein, \citealp{Eisenstein2026_jades}), 1199, 2758 (PI: M. Stiavelli, \citealp{Stiavelli2023,Morishita2024}), 1207 (PI: G. Rieke), 1210 (PI: N. Luetzgendorf, \citealp{Eisenstein2026_jades}), 1211 (PI: K. Isaak, \citealp{Maseda2024}), 1213, 1214, 1215 (PI: N. Luetzgendorf, \citealp{Maseda2024}), 1286 (PI: N. Luetzgendorf, \citealp{Eisenstein2026_jades}), 1287 (PI: K. Isaak, \citealp{Eisenstein2026_jades}), 1324 (PI: T. Treu, \citealp{Treu2022}), 1345 (PI: S. Finkelstein, \citealp{Finkelstein2023}), 1671 (PI: M. Maseda, \citealp{Maseda2023}), 1810 (PI: S. Belli, \citealp{Belli2025}), 1869 (PI: D Schaerer,\citealp{Schaerer2026}), 1871 (PI: J Chisholm, \citealp{Chisholm2024}), 1914 (PI: A. Shapley, \citealp{Shapley2025}), 2028 (PI: F. Wang), 2110 (PI: M. Kriek, \citealp{Slob2024}), 2478 (PI: D. Stark, \citealp{Topping2025_CIV_N_emitters}), 2674 (PI: P Arrabal Haro), 2736 (PI: K. Pontoppidan, \citealp{Pontoppidan2022}), 3117, 4713 (PI: A. C. Eilers, \citealp{Eilers2025}), 3215 (PI: D. Eisenstein, \citealp{DEugenio2025,Eisenstein2025_origins}), 3325 (PI: F. Wang), 3543 (PI: A. Carnall, \citealp{Carnall2024}), 4106 (PI: E. Nelson), 4233 (PI: A. de Graaff, \citealp{deGraaff2025_rubies}), 4246 (PI: A. Abdurro'uf, \citealp{Abdurrouf2024}), 4265 (PI: J G Lopez, \citealp{Solimano2025}), 4287 (PI: C Mason, \citealp{Whitler2026}), 9214 (PI: C Mason, \citealp{Chen2026,Vasan2026}), 4318 (PI: J Antwi-Danso), 4446 (PI: B. Frye, \citealp{Frye2024}), 4527 (PI: C. Willott, \citealp{Willott2022}), 4750 (PI: K. Nakajima, \citealp{Nakajima2026}), 4762, 9223 (PI: S. Fujimoto, \citealp{Fujimoto2025}), 6053 (PI: I. Wold, \citealp{Wold2026}), 8204 (PI: J. Greene, \citealp{Greene2026}).

We also make use of DJA photometric catalogs and image mosaics of the following fields and programs: GOODS-S (v7.2) from programs 1210 (L. Luetzgendorf, \citealp{Eisenstein2026_jades}), 1895 (PI: P. Oesch), 1963 (C. Williams);  GOODS-N (v7.3) from 1181 (PI: D. Eisenstein, \citealp{Eisenstein2026_jades}), 2514 (PI: C. Williams), 3577 (PI: E. Egami);  CEERS (v7.4) from 1345 (PI: S. Finkelstein, \citealp{Finkelstein2023}), 2234 (PI: E. Ba\~{n}ados); PRIMER (v7.0 in COSMOS field; v7.2 in UDS) from 1837 (PI: J. Dunlop);  Abell S1063 (v7.5) from 1840 (PI: J. Alvarez-Marquez), 3293 (PI: H. Atek, \citealp{Atek2025}); Abell 2744 (v7.2) from 1324 (PI: T. Treu, \citealp{Treu2022}), 2561 (PI: I. Labbe), and 2756 (PI: W. Chen), 2883 (PI: F. Sun), 3516 (PI: J. Matthee), 3538 (PI: E. Iani), and 4111 (PI: K. Suess); MACS 0647 (v7.0) from 1433 (PI: D. Coe); SMACS 0723 (v7.0) from 2736 (PI: K. Pontoppidan, \citealp{Pontoppidan2022}).

This work made use of the following software: \texttt{msaexp} \citep{Brammer2023_msaexp}, \texttt{specutils} \citep{specutils2025}, \texttt{PyNeb} \citep{Luridiana2015}, \texttt{bagpipes} \citep{Carnall2018_bagpipes_spec,Carnall2019_bagpipes}, BPASS \citep{Stanway2018_bpass}, \texttt{CLOUDY} \citep{Ferland2017_CLOUDY}, \texttt{synthesizer} \citep{Lovell2025_synthesizer,Roper2026_synthesizer}, \texttt{nautilus} \citep{Lange2023_nautilus}, \texttt{pysersic} \citep{Pasha2023_pysersic}, \texttt{EXPANSE} \citep{Harvey2025_EXPANSE}, \texttt{aperpy} \citep{Weaver2024_aperpy}, \texttt{sep} \citep{sep_package}, \texttt{emcee} \citep{ForemanMackey2013_emcee}, \texttt{SciPy} \citep{Virtanen2020_scipy} and \texttt{JAX} \citep{Bradbury2018_jax}.

The catalogues with line fluxes, nebular properties and abundances are made public at this repository: \href{https://doi.org/10.5281/zenodo.22148302}{10.5281/zenodo.22148302}.



\bibliographystyle{mnras}
\bibliography{_main} 

@ARTICLE{Nakajima2026,
       author = {{Nakajima}, Kimihiko and {Ouchi}, Masami and {Harikane}, Yuichi and {Vanzella}, Eros and {Ono}, Yoshiaki and {Isobe}, Yuki and {Nishigaki}, Moka and {Tsujimoto}, Takuji and {Nakamura}, Fumitaka and {Xu}, Yi and {Umeda}, Hiroya and {Zhang}, Yechi},
        title = "{An ultra-faint, chemically primitive galaxy forming in the reionization era}",
      journal = {\nat},
         year = 2026,
        month = may,
       volume = {653},
       number = {8114},
        pages = {363-367},
          doi = {10.1038/s41586-026-10374-1},
archivePrefix = {arXiv},
       eprint = {2506.11846},
 primaryClass = {astro-ph.GA},
       adsurl = {https://ui.adsabs.harvard.edu/abs/2026Natur.653..363N}
}

@ARTICLE{Ji2026,
       author = {{Ji}, Xihan and {Belokurov}, Vasily and {Maiolino}, Roberto and {Monty}, Stephanie and {Isobe}, Yuki and {Kravtsov}, Andrey and {McClymont}, William and {{\"U}bler}, Hannah},
        title = "{Connecting JWST discovered N/O-enhanced galaxies to globular clusters: evidence from chemical imprints}",
      journal = {\mnras},
         year = 2026,
        month = jan,
       volume = {545},
       number = {3},
          eid = {staf2110},
        pages = {staf2110},
          doi = {10.1093/mnras/staf2110},
archivePrefix = {arXiv},
       eprint = {2505.12505},
 primaryClass = {astro-ph.GA},
       adsurl = {https://ui.adsabs.harvard.edu/abs/2026MNRAS.545f2110J}
}

@ARTICLE{Greene2026,
       author = {{Greene}, Jenny E. and {Liu}, Hanpu and {Hviding}, Raphael E. and {Naidu}, Rohan P. and {Setton}, David J. and {Wang}, Bingjie and {Labbe}, Ivo and {Kokorev}, Vasily and {Lin}, Xiaojing and {Akins}, Hollis B. and {de Graaff}, Anna and {Leja}, Joel and {Matthee}, Jorryt and {Torralba}, Alberto and {Glazebrook}, Karl and {Bezanson}, Rachel and {Boogaard}, Leindert A. and {Brammer}, Gabriel and {Casey}, Caitlin M. and {Chisholm}, John and {Fan}, Xiaohui and {Furtak}, Lukas and {Fujimoto}, Seiji and {Jiang}, Yan-Fei and {Ma}, Yilun and {Nanayakkara}, Themiya and {Nelson}, Erica J. and {Pan}, Richard and {Quataert}, Eliot and {Taylor}, Anthony J. and {Williams}, Christina and {Zitrin}, Adi},
        title = "{Calcium Triplet Absorption is Common around Little Red Dots}",
      journal = {arXiv e-prints},
         year = 2026,
        month = sep,
          eid = {arXiv:2609.25349},
        pages = {arXiv:2609.25349},
          doi = {10.48550/arXiv.2609.25349},
archivePrefix = {arXiv},
       eprint = {2609.25349},
 primaryClass = {astro-ph.GA},
       adsurl = {https://ui.adsabs.harvard.edu/abs/2026arXiv260925349G}
}

@ARTICLE{Fujimoto2025,
       author = {{Fujimoto}, Seiji and {Asada}, Yoshihisa and {Naidu}, Rohan P. and {Chisholm}, John and {Atek}, Hakim and {Brammer}, Gabriel and {Berg}, Danielle A. and {Schaerer}, Daniel and {Kokorev}, Vasily and {Furtak}, Lukas J. and {Richard}, Johan and {Venditti}, Alessandra and {Bromm}, Volker and {Adamo}, Angela and {Claeyssens}, Adelaide and {Dessauges-Zavadsky}, Miroslava and {Fei}, Qinyue and {Hsiao}, Tiger Yu-Yang and {Korber}, Damien and {Munoz}, Julian B. and {Pan}, Richard and {Saldana-Lopez}, Alberto},
        title = "{GLIMPSE-D: An Exotic Balmer-Jump Object at z=6.20? Revisiting Photometric Selection and the Cosmic Abundance of Pop III Galaxies}",
      journal = {arXiv e-prints},
         year = 2025,
        month = dec,
          eid = {arXiv:2512.11790},
        pages = {arXiv:2512.11790},
          doi = {10.48550/arXiv.2512.11790},
archivePrefix = {arXiv},
       eprint = {2512.11790},
 primaryClass = {astro-ph.GA},
       adsurl = {https://ui.adsabs.harvard.edu/abs/2025arXiv251211790F}
}

@ARTICLE{Eilers2025,
       author = {{Eilers}, Anna-Christina and {Yue}, Minghao and {Matthee}, Jorryt and {Hennawi}, Joseph F. and {Davies}, Frederick B. and {Simcoe}, Robert A. and {Teague}, Richard and {Bordoloi}, Rongmon and {Brammer}, Gabriel and {Kang}, Yi and {Kashino}, Daichi and {Mackenzie}, Ruari and {Naidu}, Rohan P. and {Navarrete}, Benjam{\'\i}n},
        title = "{The Light Echo of a High-redshift Quasar Mapped with Ly{\ensuremath{\alpha}} Tomography}",
      journal = {\apjl},
         year = 2025,
        month = oct,
       volume = {991},
       number = {2},
          eid = {L40},
        pages = {L40},
          doi = {10.3847/2041-8213/ae057a},
archivePrefix = {arXiv},
       eprint = {2509.05417},
 primaryClass = {astro-ph.GA},
       adsurl = {https://ui.adsabs.harvard.edu/abs/2025ApJ...991L..40E}
}

@ARTICLE{Whitler2026,
       author = {{Whitler}, Lily and {Stark}, Daniel P. and {Mason}, Charlotte A. and {Tang}, Mengtao and {Chen}, Zuyi and {Lu}, Ting-Yi and {Prieto-Lyon}, Gonzalo and {Hutter}, Anne},
        title = "{Deep JWST spectroscopy of galaxies in a candidate ionized bubble at z = 8.7: probing reionization at pMpc scales with Ly {\ensuremath{\alpha}} emission}",
      journal = {\mnras},
         year = 2026,
        month = may,
       volume = {548},
       number = {2},
          eid = {stag639},
        pages = {stag639},
          doi = {10.1093/mnras/stag639},
archivePrefix = {arXiv},
       eprint = {2510.12019},
 primaryClass = {astro-ph.GA},
       adsurl = {https://ui.adsabs.harvard.edu/abs/2026MNRAS.548ag639W}
}

@ARTICLE{Tang2025,
       author = {{Tang}, Mengtao and {Stark}, Daniel P. and {Plat}, Ad{\`e}le and {Feltre}, Anna and {Katz}, Harley and {Senchyna}, Peter and {Mason}, Charlotte A. and {Whitler}, Lily and {Chen}, Zuyi and {Topping}, Michael W.},
        title = "{JWST/NIRSpec Observations of High-ionization Emission Lines in Galaxies at High Redshift}",
      journal = {\apj},
         year = 2025,
        month = oct,
       volume = {991},
       number = {2},
          eid = {217},
        pages = {217},
          doi = {10.3847/1538-4357/adfd57},
archivePrefix = {arXiv},
       eprint = {2505.06359},
 primaryClass = {astro-ph.GA},
       adsurl = {https://ui.adsabs.harvard.edu/abs/2025ApJ...991..217T}
}

@ARTICLE{Vasan2026,
       author = {{Keerthi Vasan G.}, C. and {Senchyna}, Peter and {Mason}, Charlotte A. and {Chen}, Zuyi and {Stark}, Daniel P. and {Jones}, Tucker and {Whitler}, Lily and {Glazer}, Kelsey S. and {Aravena}, Manuel and {Gonzalez-Lopez}, Jorge and {Endsley}, Ryan and {Gelli}, Viola and {Tang}, Mengtao and {Topping}, Michael W.},
        title = "{JWST absorption line spectroscopy with SPURS: ISM covering fractions and kinematics in individual galaxies at $z=5-9$}",
      journal = {arXiv e-prints},
         year = 2026,
        month = jun,
          eid = {arXiv:2606.06609},
        pages = {arXiv:2606.06609},
          doi = {10.48550/arXiv.2606.06609},
archivePrefix = {arXiv},
       eprint = {2606.06609},
 primaryClass = {astro-ph.GA},
       adsurl = {https://ui.adsabs.harvard.edu/abs/2026arXiv260606609K}
}

@ARTICLE{Chen2026,
       author = {{Chen}, Zuyi and {Stark}, Daniel P. and {Mason}, Charlotte A. and {Plat}, Adele and {Gelli}, Viola and {Senchyna}, Peter and {Keerthi Vasan G.}, C. and {Endsley}, Ryan and {Tang}, Mengtao and {Topping}, Michael W. and {Whitler}, Lily},
        title = "{SPURS: Bursty Star Formation in an Extremely Luminous Weak Emission Line Galaxy at $z=9.3$}",
      journal = {arXiv e-prints},
         year = 2026,
        month = apr,
          eid = {arXiv:2604.21516},
        pages = {arXiv:2604.21516},
          doi = {10.48550/arXiv.2604.21516},
archivePrefix = {arXiv},
       eprint = {2604.21516},
 primaryClass = {astro-ph.GA},
       adsurl = {https://ui.adsabs.harvard.edu/abs/2026arXiv260421516C}
}

@ARTICLE{ArellanoCordova2025_density_feedback,
       author = {{Arellano-C{\'o}rdova}, K.~Z. and {Berg}, D.~A. and {Mingozzi}, M. and {James}, B.~L. and {Vincenzo}, F. and {Rogers}, N.~S.~J. and {Skillman}, E.~D. and {Amor{\'\i}n}, R.~O. and {Cullen}, F. and {Flury}, S.~R. and {Abril-Melgarejo}, V. and {Chisholm}, J. and {Heckman}, T. and {Hayes}, M.~J. and {Hernandez}, S. and {Kumari}, N. and {Kobayashi}, C. and {Leitherer}, C. and {Martin}, C.~L. and {Martinez}, Z. and {Nanayakkara}, T. and {Parker}, K.~S. and {Senchyna}, P. and {Scarlata}, C. and {Stephenson}, M.~G. and {Wofford}, A. and {Xu}, X. and {Zhu}, P.},
        title = "{CLASSY XII: nitrogen enrichment shaped by gas density and feedback}",
      journal = {\mnras},
         year = 2025,
        month = dec,
       volume = {544},
       number = {2},
        pages = {1588-1607},
          doi = {10.1093/mnras/staf1723},
archivePrefix = {arXiv},
       eprint = {2507.11658},
 primaryClass = {astro-ph.GA},
       adsurl = {https://ui.adsabs.harvard.edu/abs/2025MNRAS.544.1588A}
}

@ARTICLE{Flury2025,
       author = {{Flury}, Sophia R. and {Arellano-C{\'o}rdova}, Karla Z. and {Moran}, Edward C. and {Einsig}, Alaina},
        title = "{New ionization models and the shocking nitrogen excess at z > 5}",
      journal = {\mnras},
         year = 2025,
        month = nov,
       volume = {543},
       number = {4},
        pages = {3367-3381},
          doi = {10.1093/mnras/staf1615},
archivePrefix = {arXiv},
       eprint = {2412.06763},
 primaryClass = {astro-ph.GA},
       adsurl = {https://ui.adsabs.harvard.edu/abs/2025MNRAS.543.3367F}
}

@ARTICLE{Rai2026,
       author = {{Rai}, Raunaq Singh and {Roberts-Borsani}, Guido},
        title = "{Unveiling and Characterising Ubiquitous Nitrogen Enhancement in $6 \leq z \leq 10$ Galaxies with JWST Spectroscopy}",
      journal = {arXiv e-prints},
         year = 2026,
        month = aug,
          eid = {arXiv:2608.10063},
        pages = {arXiv:2608.10063},
          doi = {10.48550/arXiv.2608.10063},
archivePrefix = {arXiv},
       eprint = {2608.10063},
 primaryClass = {astro-ph.GA},
       adsurl = {https://ui.adsabs.harvard.edu/abs/2026arXiv260810063R}
}

@ARTICLE{Bustamante2018,
       author = {{Bustamante}, Sebasti{\'a}n and {Sparre}, Martin and {Springel}, Volker and {Grand}, Robert J.~J.},
        title = "{Merger-induced metallicity dilution in cosmological galaxy formation simulations}",
      journal = {\mnras},
         year = 2018,
        month = sep,
       volume = {479},
       number = {3},
        pages = {3381-3392},
          doi = {10.1093/mnras/sty1692},
archivePrefix = {arXiv},
       eprint = {1712.03250},
 primaryClass = {astro-ph.GA},
       adsurl = {https://ui.adsabs.harvard.edu/abs/2018MNRAS.479.3381B}
}

@ARTICLE{Carniani2025,
       author = {{Carniani}, Stefano and {D'Eugenio}, Francesco and {Ji}, Xihan and {Parlanti}, Eleonora and {Scholtz}, Jan and {Sun}, Fengwu and {Venturi}, Giacomo and {Bakx}, Tom J.~L.~C. and {Curti}, Mirko and {Maiolino}, Roberto and {Tacchella}, Sandro and {Zavala}, Jorge A. and {Hainline}, Kevin and {Witstok}, Joris and {Johnson}, Benjamin D. and {Alberts}, Stacey and {Bunker}, Andrew J. and {Charlot}, St{\'e}phane and {Eisenstein}, Daniel J. and {Helton}, Jakob M. and {Jakobsen}, Peter and {Kumari}, Nimisha and {Robertson}, Brant and {Saxena}, Aayush and {{\"U}bler}, Hannah and {Williams}, Christina C. and {Willmer}, Christopher N.~A. and {Willott}, Chris},
        title = "{The eventful life of a luminous galaxy at z = 14: metal enrichment, feedback, and low gas fraction?}",
      journal = {\aap},
         year = 2025,
        month = apr,
       volume = {696},
          eid = {A87},
        pages = {A87},
          doi = {10.1051/0004-6361/202452451},
archivePrefix = {arXiv},
       eprint = {2409.20533},
 primaryClass = {astro-ph.GA},
       adsurl = {https://ui.adsabs.harvard.edu/abs/2025A&A...696A..87C}
}

@ARTICLE{Schouws2025,
       author = {{Schouws}, Sander and {Bouwens}, Rychard J. and {Algera}, Hiddo and {Smit}, Renske and {Kumari}, Nimisha and {Rowland}, Lucie E. and {van Leeuwen}, Ivana and {Sommovigo}, Laura and {Ferrara}, Andrea and {Oesch}, Pascal A. and {Ormerod}, Katherine and {Stefanon}, Mauro and {Herard-Demanche}, Thomas and {Hodge}, Jacqueline and {Fudamoto}, Yoshinobu and {R{\"o}ttgering}, Huub and {van der Werf}, Paul},
        title = "{Deep Constraints on [CII]158$\mu$m in JADES-GS-z14-0: Further Evidence for a Galaxy with Low Gas Content at z=14.2}",
      journal = {arXiv e-prints},
         year = 2025,
        month = feb,
          eid = {arXiv:2502.01610},
        pages = {arXiv:2502.01610},
          doi = {10.48550/arXiv.2502.01610},
archivePrefix = {arXiv},
       eprint = {2502.01610},
 primaryClass = {astro-ph.GA},
       adsurl = {https://ui.adsabs.harvard.edu/abs/2025arXiv250201610S}
}

@ARTICLE{Ferrara2024,
       author = {{Ferrara}, A.},
        title = "{The eventful life of GS-z14-0, the most distant galaxy at redshift z = 14.32}",
      journal = {\aap},
         year = 2024,
        month = sep,
       volume = {689},
          eid = {A310},
        pages = {A310},
          doi = {10.1051/0004-6361/202450944},
archivePrefix = {arXiv},
       eprint = {2405.20370},
 primaryClass = {astro-ph.GA},
       adsurl = {https://ui.adsabs.harvard.edu/abs/2024A&A...689A.310F}
}

@ARTICLE{Ferrara2026_RedMonsters,
       author = {{Ferrara}, A. and {Rodighiero}, G. and {Carniani}, S. and {Zhang}, Z. and {Kohandel}, M. and {Das}, B.},
        title = "{No Blue without Red: Evolutionary Properties of Super-Early Galaxies}",
      journal = {arXiv e-prints},
         year = 2026,
        month = may,
          eid = {arXiv:2605.22914},
        pages = {arXiv:2605.22914},
          doi = {10.48550/arXiv.2605.22914},
archivePrefix = {arXiv},
       eprint = {2605.22914},
 primaryClass = {astro-ph.GA},
       adsurl = {https://ui.adsabs.harvard.edu/abs/2026arXiv260522914F}
}

@ARTICLE{Ferrara2025_AFM,
       author = {{Ferrara}, A. and {Carniani}, S. and {di Mascia}, F. and {Bouwens}, R.~J. and {Oesch}, P. and {Schouws}, S.},
        title = "{ALMA observations of super-early galaxies: Attenuation-free model predictions}",
      journal = {\aap},
         year = 2025,
        month = feb,
       volume = {694},
          eid = {A215},
        pages = {A215},
          doi = {10.1051/0004-6361/202452368},
archivePrefix = {arXiv},
       eprint = {2409.17223},
 primaryClass = {astro-ph.GA},
       adsurl = {https://ui.adsabs.harvard.edu/abs/2025A&A...694A.215F}
}

@article{Roper2026_synthesizer,
    author = {Roper, Will J. and Lovell, Christopher C. and Vijayan, Aswin and Wilkins, Stephen and Akins, Hollis and Berger, Sabrina and Sant Fournier, Connor and Harvey, Thomas and Iyer, Kartheik and Leonardi, Marco and Newman, Sophie and Pautasso, Borja and Perry, Ashley and Seeyave, Louise and Sommovigo, Laura and Punyasheel, Paurush and d'Hautefort, Adrien Aufan Stoffels and Rawlings, Alex},
    journal = {Journal of Open Source Software},
    doi = {10.21105/joss.09436},
    year = {2026},
    publisher = {Open Journals},
    title = {Synthesizer: Synthetic Observables for Modern Astronomy},
    volume = {11},
    number = {119},
    pages = {9436},
}

@ARTICLE{Atek2025,
       author = {{Atek}, Hakim and {Chisholm}, John and {Kokorev}, Vasily and {Endsley}, Ryan and {Pan}, Richard and {Furtak}, Lukas and {Chemerynska}, Iryna and {Richard}, Johan and {Claeyssens}, Ad{\'e}la{\"\i}de and {Oesch}, Pascal and {Fujimoto}, Seiji and {Naidu}, Rohan and {Korber}, Damien and {Schaerer}, Daniel and {Blaizot}, Jeremy and {Rosdahl}, Joki and {Adamo}, Angela and {Asada}, Yoshihisa and {Basu}, Arghyadeep and {Beauchesne}, Benjamin and {Berg}, Danielle and {Bezanson}, Rachel and {Bouwens}, Rychard and {Brammer}, Gabriel and {Dessauges-Zavadsky}, Miroslava and {Ellien}, Ama{\"e}l and {Ezziati}, Meriam and {Fei}, Qinyue and {Goovaerts}, Ilias and {Heurtier}, Sylvain and {Hsiao}, Tiger Yu-Yang and {Jecmen}, Michelle and {Khullar}, Gourav and {Kneib}, Jean-Paul and {Labb{\'e}}, Ivo and {Leclercq}, Floriane and {Marques-Chaves}, Rui and {Mason}, Charlotte and {McQuinn}, Kristen B.~W. and {Mu{\~n}oz}, Julian B. and {Natarajan}, Priyamvada and {Saldana-Lopez}, Alberto and {Stephenson}, Mabel G. and {Trebitsch}, Maxime and {Volonteri}, Marta and {Weibel}, Andrea and {Zitrin}, Adi},
        title = "{JWST's GLIMPSE: an overview of the deepest probe of early galaxy formation and cosmic reionization}",
      journal = {arXiv e-prints},
         year = 2025,
        month = nov,
          eid = {arXiv:2511.07542},
        pages = {arXiv:2511.07542},
          doi = {10.48550/arXiv.2511.07542},
archivePrefix = {arXiv},
       eprint = {2511.07542},
 primaryClass = {astro-ph.GA},
       adsurl = {https://ui.adsabs.harvard.edu/abs/2025arXiv251107542A}
}

@ARTICLE{Drout2023,
       author = {{Drout}, M.~R. and {G{\"o}tberg}, Y. and {Ludwig}, B.~A. and {Groh}, J.~H. and {de Mink}, S.~E. and {O'Grady}, A.~J.~G. and {Smith}, N.},
        title = "{An observed population of intermediate-mass helium stars that have been stripped in binaries}",
      journal = {Science},
         year = 2023,
        month = dec,
       volume = {382},
       number = {6676},
        pages = {1287-1291},
          doi = {10.1126/science.ade4970},
archivePrefix = {arXiv},
       eprint = {2307.00061},
 primaryClass = {astro-ph.SR},
       adsurl = {https://ui.adsabs.harvard.edu/abs/2023Sci...382.1287D}
}

@ARTICLE{Gotberg2019,
       author = {{G{\"o}tberg}, Y. and {de Mink}, S.~E. and {Groh}, J.~H. and {Leitherer}, C. and {Norman}, C.},
        title = "{The impact of stars stripped in binaries on the integrated spectra of stellar populations}",
      journal = {\aap},
         year = 2019,
        month = sep,
       volume = {629},
          eid = {A134},
        pages = {A134},
          doi = {10.1051/0004-6361/201834525},
archivePrefix = {arXiv},
       eprint = {1908.06102},
 primaryClass = {astro-ph.GA},
       adsurl = {https://ui.adsabs.harvard.edu/abs/2019A&A...629A.134G}
}

@ARTICLE{Simmonds2021,
       author = {{Simmonds}, Charlotte and {Schaerer}, Daniel and {Verhamme}, Anne},
        title = "{Can nebular He II emission be explained by ultra-luminous X-ray sources?}",
      journal = {\aap},
         year = 2021,
        month = dec,
       volume = {656},
          eid = {A127},
        pages = {A127},
          doi = {10.1051/0004-6361/202141856},
archivePrefix = {arXiv},
       eprint = {2108.12438},
 primaryClass = {astro-ph.GA},
       adsurl = {https://ui.adsabs.harvard.edu/abs/2021A&A...656A.127S}
}

@ARTICLE{Garnett1991,
       author = {{Garnett}, Donald R. and {Kennicutt}, Jr., Robert C. and {Chu}, You-Hua and {Skillman}, Evan D.},
        title = "{H II Regions With He II Emission}",
      journal = {\pasp},
         year = 1991,
        month = aug,
       volume = {103},
        pages = {850},
          doi = {10.1086/132892},
       adsurl = {https://ui.adsabs.harvard.edu/abs/1991PASP..103..850G}
}

@ARTICLE{Meynet2005,
       author = {{Meynet}, G. and {Maeder}, A.},
        title = "{Stellar evolution with rotation. XI. Wolf-Rayet star populations at different metallicities}",
      journal = {\aap},
         year = 2005,
        month = jan,
       volume = {429},
        pages = {581-598},
          doi = {10.1051/0004-6361:20047106},
archivePrefix = {arXiv},
       eprint = {astro-ph/0408319},
 primaryClass = {astro-ph},
       adsurl = {https://ui.adsabs.harvard.edu/abs/2005A&A...429..581M}
}

@ARTICLE{Crowther2007,
       author = {{Crowther}, Paul A.},
        title = "{Physical Properties of Wolf-Rayet Stars}",
      journal = {\araa},
         year = 2007,
        month = sep,
       volume = {45},
       number = {1},
        pages = {177-219},
          doi = {10.1146/annurev.astro.45.051806.110615},
archivePrefix = {arXiv},
       eprint = {astro-ph/0610356},
 primaryClass = {astro-ph},
       adsurl = {https://ui.adsabs.harvard.edu/abs/2007ARA&A..45..177C}
}

@ARTICLE{Karakas2010,
       author = {{Karakas}, A.~I.},
        title = "{Updated stellar yields from asymptotic giant branch models}",
      journal = {\mnras},
         year = 2010,
        month = apr,
       volume = {403},
       number = {3},
        pages = {1413-1425},
          doi = {10.1111/j.1365-2966.2009.16198.x},
archivePrefix = {arXiv},
       eprint = {0912.2142},
 primaryClass = {astro-ph.SR},
       adsurl = {https://ui.adsabs.harvard.edu/abs/2010MNRAS.403.1413K}
}

@ARTICLE{Claeyssens2025,
       author = {{Claeyssens}, Ad{\'e}la{\"\i}de and {Adamo}, Angela and {Messa}, Matteo and {Dessauges-Zavadsky}, Miroslava and {Richard}, Johan and {Kramarenko}, Ivan and {Matthee}, Jorryt and {Naidu}, Rohan P.},
        title = "{Tracing star formation across cosmic time at tens of parsec-scales in the lensing cluster field Abell 2744}",
      journal = {\mnras},
         year = 2025,
        month = mar,
       volume = {537},
       number = {3},
        pages = {2535-2558},
          doi = {10.1093/mnras/staf058},
archivePrefix = {arXiv},
       eprint = {2410.10974},
 primaryClass = {astro-ph.GA},
       adsurl = {https://ui.adsabs.harvard.edu/abs/2025MNRAS.537.2535C}
}

@ARTICLE{Maiolino2024,
       author = {{Maiolino}, Roberto and {Scholtz}, Jan and {Curtis-Lake}, Emma and {Carniani}, Stefano and {Baker}, William and {de Graaff}, Anna and {Tacchella}, Sandro and {{\"U}bler}, Hannah and {D'Eugenio}, Francesco and {Witstok}, Joris and {Curti}, Mirko and {Arribas}, Santiago and {Bunker}, Andrew J. and {Charlot}, St{\'e}phane and {Chevallard}, Jacopo and {Eisenstein}, Daniel J. and {Egami}, Eiichi and {Ji}, Zhiyuan and {Jones}, Gareth C. and {Lyu}, Jianwei and {Rawle}, Tim and {Robertson}, Brant and {Rujopakarn}, Wiphu and {Perna}, Michele and {Sun}, Fengwu and {Venturi}, Giacomo and {Williams}, Christina C. and {Willott}, Chris},
        title = "{JADES: The diverse population of infant black holes at 4 < z < 11: Merging, tiny, poor, but mighty}",
      journal = {\aap},
         year = 2024,
        month = nov,
       volume = {691},
          eid = {A145},
        pages = {A145},
          doi = {10.1051/0004-6361/202347640},
archivePrefix = {arXiv},
       eprint = {2308.01230},
 primaryClass = {astro-ph.GA},
       adsurl = {https://ui.adsabs.harvard.edu/abs/2024A&A...691A.145M}
}

@ARTICLE{ForbesBridges2010,
       author = {{Forbes}, Duncan A. and {Bridges}, Terry},
        title = "{Accreted versus in situ Milky Way globular clusters}",
      journal = {\mnras},
         year = 2010,
        month = may,
       volume = {404},
       number = {3},
        pages = {1203-1214},
          doi = {10.1111/j.1365-2966.2010.16373.x},
archivePrefix = {arXiv},
       eprint = {1001.4289},
 primaryClass = {astro-ph.GA},
       adsurl = {https://ui.adsabs.harvard.edu/abs/2010MNRAS.404.1203F}
}

@ARTICLE{Gehrels1986,
       author = {{Gehrels}, N.},
        title = "{Confidence Limits for Small Numbers of Events in Astrophysical Data}",
      journal = {\apj},
         year = 1986,
        month = apr,
       volume = {303},
        pages = {336},
          doi = {10.1086/164079},
       adsurl = {https://ui.adsabs.harvard.edu/abs/1986ApJ...303..336G}
}

@ARTICLE{Morishita2024_sizes,
       author = {{Morishita}, Takahiro and {Stiavelli}, Massimo and {Chary}, Ranga-Ram and {Trenti}, Michele and {Bergamini}, Pietro and {Chiaberge}, Marco and {Leethochawalit}, Nicha and {Roberts-Borsani}, Guido and {Shen}, Xuejian and {Treu}, Tommaso},
        title = "{Enhanced Subkiloparsec-scale Star Formation: Results from a JWST Size Analysis of 341 Galaxies at 5 < z < 14}",
      journal = {\apj},
         year = 2024,
        month = mar,
       volume = {963},
       number = {1},
          eid = {9},
        pages = {9},
          doi = {10.3847/1538-4357/ad1404},
archivePrefix = {arXiv},
       eprint = {2308.05018},
 primaryClass = {astro-ph.GA},
       adsurl = {https://ui.adsabs.harvard.edu/abs/2024ApJ...963....9M}
}

@ARTICLE{Harvey2025_epochs4,
       author = {{Harvey}, Thomas and {Conselice}, Christopher J. and {Adams}, Nathan J. and {Austin}, Duncan and {Juod{\v{z}}balis}, Ignas and {Trussler}, James and {Li}, Qiong and {Ormerod}, Katherine and {Ferreira}, Leonardo and {Lovell}, Christopher C. and {Duan}, Qiao and {Westcott}, Lewi and {Harris}, Honor and {Bhatawdekar}, Rachana and {Coe}, Dan and {Cohen}, Seth H. and {Caruana}, Joseph and {Cheng}, Cheng and {Driver}, Simon P. and {Frye}, Brenda and {Furtak}, Lukas J. and {Grogin}, Norman A. and {Hathi}, Nimish P. and {Holwerda}, Benne W. and {Jansen}, Rolf A. and {Koekemoer}, Anton M. and {Marshall}, Madeline A. and {Nonino}, Mario and {Vijayan}, Aswin P. and {Wilkins}, Stephen M. and {Windhorst}, Rogier and {Willmer}, Christopher N.~A. and {Yan}, Haojing and {Zitrin}, Adi},
        title = "{EPOCHS. IV. SED Modeling Assumptions and Their Impact on the Stellar Mass Function at 6.5 {\ensuremath{\leq}} z {\ensuremath{\leq}} 13.5 Using PEARLS and Public JWST Observations}",
      journal = {\apj},
         year = 2025,
        month = jan,
       volume = {978},
       number = {1},
          eid = {89},
        pages = {89},
          doi = {10.3847/1538-4357/ad8c29},
archivePrefix = {arXiv},
       eprint = {2403.03908},
 primaryClass = {astro-ph.GA},
       adsurl = {https://ui.adsabs.harvard.edu/abs/2025ApJ...978...89H}
}

@ARTICLE{Conselice2025,
       author = {{Conselice}, Christopher J. and {Adams}, Nathan and {Harvey}, Thomas and {Austin}, Duncan and {Ferreira}, Leonardo and {Ormerod}, Katherine and {Duan}, Qiao and {Trussler}, James and {Li}, Qiong and {Juod{\v{z}}balis}, Ignas and {Westcott}, Lewi and {Harris}, Honor and {Seeyave}, Louise T.~C. and {Bluck}, Asa F.~L. and {Windhorst}, Rogier A. and {Bhatawdekar}, Rachana and {Coe}, Dan and {Cohen}, Seth H. and {Cheng}, Cheng and {Driver}, Simon P. and {Frye}, Brenda and {Furtak}, Lukas J. and {Grogin}, Norman A. and {Hathi}, Nimish P. and {Holwerda}, Benne W. and {Jansen}, Rolf A. and {Koekemoer}, Anton M. and {Marshall}, Madeline A. and {Nonino}, Mario and {Robotham}, Aaron and {Summers}, Jake and {Wilkins}, Stephen M. and {Willmer}, Christopher N.~A. and {Yan}, Haojing and {Zitrin}, Adi},
        title = "{EPOCHS. I. The Discovery and Star-forming Properties of Galaxies in the Epoch of Reionization at 6.5 < z < 18 with PEARLS and Public JWST Data}",
      journal = {\apj},
         year = 2025,
        month = apr,
       volume = {983},
       number = {1},
          eid = {30},
        pages = {30},
          doi = {10.3847/1538-4357/ada608},
archivePrefix = {arXiv},
       eprint = {2407.14973},
 primaryClass = {astro-ph.GA},
       adsurl = {https://ui.adsabs.harvard.edu/abs/2025ApJ...983...30C}
}

@ARTICLE{ArellanoCordova2024_classy,
       author = {{Arellano-C{\'o}rdova}, Karla Z. and {Berg}, Danielle A. and {Mingozzi}, Matilde and {James}, Bethan L. and {Rogers}, Noah S.~J. and {Skillman}, Evan D. and {Cullen}, Fergus and {Alexander}, Ryan K. and {Amor{\'\i}n}, Ricardo O. and {Chisholm}, John and {Hayes}, Matthew and {Heckman}, Timothy and {Hernandez}, Svea and {Kumari}, Nimisha and {Leitherer}, Claus and {Martin}, Crystal L. and {Maseda}, Michael and {Nanayakkara}, Themiya and {Parker}, Kaelee and {Ravindranath}, Swara and {Strom}, Allison L. and {Vincenzo}, Fiorenzo and {Wofford}, Aida},
        title = "{CLASSY. IX. The Chemical Evolution of the Ne, S, Cl, and Ar Elements}",
      journal = {\apj},
         year = 2024,
        month = jun,
       volume = {968},
       number = {2},
          eid = {98},
        pages = {98},
          doi = {10.3847/1538-4357/ad34cf},
archivePrefix = {arXiv},
       eprint = {2403.08401},
 primaryClass = {astro-ph.GA},
       adsurl = {https://ui.adsabs.harvard.edu/abs/2024ApJ...968...98A}
}

@ARTICLE{Bhattacharya2025_argon,
       author = {{Bhattacharya}, Souradeep and {Arnaboldi}, Magda and {Gerhard}, Ortwin and {Kobayashi}, Chiaki and {Saha}, Kanak},
        title = "{Unveiling Galaxy Chemical Enrichment Mechanisms Out to z {\ensuremath{\sim}} 8 from Direct Determination of O and Ar Abundances from JWST/NIRSPEC Spectroscopy}",
      journal = {\apjl},
         year = 2025,
        month = apr,
       volume = {983},
       number = {2},
          eid = {L30},
        pages = {L30},
          doi = {10.3847/2041-8213/adc735},
archivePrefix = {arXiv},
       eprint = {2408.13396},
 primaryClass = {astro-ph.GA},
       adsurl = {https://ui.adsabs.harvard.edu/abs/2025ApJ...983L..30B}
}

@ARTICLE{Isobe2023_densities,
       author = {{Isobe}, Yuki and {Ouchi}, Masami and {Nakajima}, Kimihiko and {Harikane}, Yuichi and {Ono}, Yoshiaki and {Xu}, Yi and {Zhang}, Yechi and {Umeda}, Hiroya},
        title = "{Redshift Evolution of Electron Density in the Interstellar Medium at z   0-9 Uncovered with JWST/NIRSpec Spectra and Line-spread Function Determinations}",
      journal = {\apj},
         year = 2023,
        month = oct,
       volume = {956},
       number = {2},
          eid = {139},
        pages = {139},
          doi = {10.3847/1538-4357/acf376},
archivePrefix = {arXiv},
       eprint = {2301.06811},
 primaryClass = {astro-ph.GA},
       adsurl = {https://ui.adsabs.harvard.edu/abs/2023ApJ...956..139I}
}

@ARTICLE{Finkelstein2023,
       author = {{Finkelstein}, Steven L. and {Bagley}, Micaela B. and {Ferguson}, Henry C. and {Wilkins}, Stephen M. and {Kartaltepe}, Jeyhan S. and {Papovich}, Casey and {Yung}, L.~Y. Aaron and {Arrabal Haro}, Pablo and {Behroozi}, Peter and {Dickinson}, Mark and {Kocevski}, Dale D. and {Koekemoer}, Anton M. and {Larson}, Rebecca L. and {Le Bail}, Aur{\'e}lien and {Morales}, Alexa M. and {P{\'e}rez-Gonz{\'a}lez}, Pablo G. and {Burgarella}, Denis and {Dav{\'e}}, Romeel and {Hirschmann}, Michaela and {Somerville}, Rachel S. and {Wuyts}, Stijn and {Bromm}, Volker and {Casey}, Caitlin M. and {Fontana}, Adriano and {Fujimoto}, Seiji and {Gardner}, Jonathan P. and {Giavalisco}, Mauro and {Grazian}, Andrea and {Grogin}, Norman A. and {Hathi}, Nimish P. and {Hutchison}, Taylor A. and {Jha}, Saurabh W. and {Jogee}, Shardha and {Kewley}, Lisa J. and {Kirkpatrick}, Allison and {Long}, Arianna S. and {Lotz}, Jennifer M. and {Pentericci}, Laura and {Pierel}, Justin D.~R. and {Pirzkal}, Nor and {Ravindranath}, Swara and {Ryan}, Russell E. and {Trump}, Jonathan R. and {Yang}, Guang and {Bhatawdekar}, Rachana and {Bisigello}, Laura and {Buat}, V{\'e}ronique and {Calabr{\`o}}, Antonello and {Castellano}, Marco and {Cleri}, Nikko J. and {Cooper}, M.~C. and {Croton}, Darren and {Daddi}, Emanuele and {Dekel}, Avishai and {Elbaz}, David and {Franco}, Maximilien and {Gawiser}, Eric and {Holwerda}, Benne W. and {Huertas-Company}, Marc and {Jaskot}, Anne E. and {Leung}, Gene C.~K. and {Lucas}, Ray A. and {Mobasher}, Bahram and {Pandya}, Viraj and {Tacchella}, Sandro and {Weiner}, Benjamin J. and {Zavala}, Jorge A.},
        title = "{CEERS Key Paper. I. An Early Look into the First 500 Myr of Galaxy Formation with JWST}",
      journal = {\apjl},
         year = 2023,
        month = mar,
       volume = {946},
       number = {1},
          eid = {L13},
        pages = {L13},
          doi = {10.3847/2041-8213/acade4},
archivePrefix = {arXiv},
       eprint = {2211.05792},
 primaryClass = {astro-ph.GA},
       adsurl = {https://ui.adsabs.harvard.edu/abs/2023ApJ...946L..13F}
}

@ARTICLE{Maseda2024,
       author = {{Maseda}, Michael V. and {de Graaff}, Anna and {Franx}, Marijn and
         {Rix}, Hans-Walter and {Carniani}, Stefano and {Laseter}, Isaac and
         {Dudzevi{\v{c}}i{\={u}}t{\.{e}}}, Ugn{\.{e}} and {Rawle}, Tim and
         {Parlanti}, Eleonora and {Arribas}, Santiago and {Bunker}, Andrew J. and
         {Cameron}, Alex J. and {Charlot}, Stephane and {Curti}, Mirko and
         {D'Eugenio}, Francesco and {Jones}, Gareth C. and {Kumari}, Nimisha and
         {Maiolino}, Roberto and {{\"U}bler}, Hannah and {Saxena}, Aayush and
         {Smit}, Renske and {Willott}, Chris and {Witstok}, Joris},
        title = "{The NIRSpec Wide GTO Survey}",
      journal = {\aap},
         year = 2024,
        month = sep,
       volume = {689},
          eid = {A73},
        pages = {A73},
          doi = {10.1051/0004-6361/202449914},
archivePrefix = {arXiv},
       eprint = {2403.05506},
 primaryClass = {astro-ph.GA},
       adsurl = {https://ui.adsabs.harvard.edu/abs/2024A%26A...689A..73M}
}

@ARTICLE{Morishita2024,
       author = {{Morishita}, Takahiro and {Stiavelli}, Massimo and {Grillo}, Claudio and
         {Rosati}, Piero and {Schuldt}, Stefan and {Trenti}, Michele and
         {Bergamini}, Pietro and {Boyett}, Kit and {Chary}, Ranga-Ram and
         {Leethochawalit}, Nicha and {Roberts-Borsani}, Guido and {Treu}, Tommaso and
         {Vanzella}, Eros},
        title = "{Diverse Oxygen Abundance in Early Galaxies Unveiled by Auroral Line
                  Analysis with JWST}",
      journal = {\apj},
         year = 2024,
        month = aug,
       volume = {971},
          eid = {43},
        pages = {43},
          doi = {10.3847/1538-4357/ad5290},
archivePrefix = {arXiv},
       eprint = {2402.14084},
 primaryClass = {astro-ph.GA},
       adsurl = {https://ui.adsabs.harvard.edu/abs/2024ApJ...971...43M}
}

@ARTICLE{Stiavelli2023,
       author = {{Stiavelli}, Massimo and {Morishita}, Takahiro and {Chiaberge}, Marco and {Grillo}, Claudio and {Leethochawalit}, Nicha and {Rosati}, Piero and {Schuldt}, Stefan and {Trenti}, Michele and {Treu}, Tommaso},
        title = "{The Puzzling Properties of the MACS1149-JD1 Galaxy at z = 9.11}",
      journal = {\apjl},
         year = 2023,
        month = nov,
       volume = {957},
       number = {2},
          eid = {L18},
        pages = {L18},
          doi = {10.3847/2041-8213/ad0159},
archivePrefix = {arXiv},
       eprint = {2308.14696},
 primaryClass = {astro-ph.GA},
       adsurl = {https://ui.adsabs.harvard.edu/abs/2023ApJ...957L..18S}
}

@ARTICLE{Treu2022,
       author = {{Treu}, T. and {Roberts-Borsani}, G. and {Bradac}, M. and {Brammer}, G. and
         {Fontana}, A. and {Henry}, A. and {Mason}, C. and {Morishita}, T. and
         {Pentericci}, L. and {Wang}, X. and {Acebron}, A. and {Bagley}, M. and
         {Bergamini}, P. and {Belfiori}, D. and {Bonchi}, A. and {Boyett}, K. and
         {Boutsia}, K. and {Calabr{\'o}}, A. and {Caminha}, G.~B. and
         {Castellano}, M. and {Dressler}, A. and {Glazebrook}, K. and {Grillo}, C. and
         {Jacobs}, C. and {Jones}, T. and {Kelly}, P.~L. and {Leethochawalit}, N. and
         {Malkan}, M.~A. and {Marchesini}, D. and {Mascia}, S. and {Mercurio}, A. and
         {Merlin}, E. and {Nanayakkara}, T. and {Nonino}, M. and {Paris}, D. and
         {Poggianti}, B. and {Rosati}, P. and {Santini}, P. and {Scarlata}, C. and
         {Shipley}, H.~V. and {Strait}, V. and {Trenti}, M. and {Tubthong}, C. and
         {Vanzella}, E. and {Vulcani}, B. and {Yang}, L.},
        title = "{The GLASS-JWST Early Release Science Program. I. Survey Design and
                  Release Plans}",
      journal = {\apj},
         year = 2022,
        month = aug,
       volume = {935},
       number = {2},
          eid = {110},
        pages = {110},
          doi = {10.3847/1538-4357/ac8158},
archivePrefix = {arXiv},
       eprint = {2206.07978},
 primaryClass = {astro-ph.GA},
       adsurl = {https://ui.adsabs.harvard.edu/abs/2022ApJ...935..110T}
}

@ARTICLE{Maseda2023,
       author = {{Maseda}, Michael V. and {Lewis}, Zach and {Matthee}, Jorryt and
         {Hennawi}, Joseph F. and {Boogaard}, Leindert and {Feltre}, Anna and
         {Nanayakkara}, Themiya and {Bacon}, Roland and {Barger}, Amy and
         {Brinchmann}, Jarle and {Franx}, Marijn and {Hashimoto}, Takuya and
         {Inami}, Hanae and {Kusakabe}, Haruka and {Leclercq}, Floriane and
         {Rowland}, Lucie and {Taylor}, Anthony J. and {Tremonti}, Christy and
         {Urrutia}, Tanya and {Schaye}, Joop and {Simmonds}, Charlotte and
         {Vitte}, Elo{\"i}se},
        title = "{JWST/NIRSpec Measurements of Extremely Low Metallicities in High
                  Equivalent Width Ly$\alpha$ Emitters}",
      journal = {\apj},
         year = 2023,
        month = oct,
       volume = {956},
       number = {1},
          eid = {11},
        pages = {11},
          doi = {10.3847/1538-4357/acf12b},
archivePrefix = {arXiv},
       eprint = {2304.08511},
 primaryClass = {astro-ph.GA},
       adsurl = {https://ui.adsabs.harvard.edu/abs/2023ApJ...956...11M}
}

@ARTICLE{Pontoppidan2022,
       author = {{Pontoppidan}, Klaus M. and {Barrientes}, Jaclyn and {Blome}, Claire and
         {Braun}, Hannah and {Brown}, Matthew and {Carruthers}, Margaret and
         {Coe}, Dan and {DePasquale}, Joseph and {Espinoza}, N{\'e}stor and
         {Marin}, Macarena Garcia and {Gordon}, Karl D. and {Henry}, Alaina and
         {Hustak}, Leah and {James}, Andi and {Jenkins}, Ann and
         {Koekemoer}, Anton M. and {LaMassa}, Stephanie and {Law}, David and
         {Lockwood}, Alexandra and {Moro-Martin}, Amaya and {Mullally}, Susan E. and
         {Pagan}, Alyssa and {Player}, Dani and {Proffitt}, Charles and
         {Pulliam}, Christine and {Ramsay}, Leah and {Ravindranath}, Swara and
         {Reid}, Neill and {Robberto}, Massimo and {Sabbi}, Elena and
         {Ubeda}, Leonardo and {Balogh}, Michael and {Flanagan}, Kathryn and
         {Gardner}, Jonathan and {Hasan}, Hashima and {Meinke}, Bonnie and
         {Nota}, Antonella},
        title = "{The JWST Early Release Observations}",
      journal = {\apjl},
         year = 2022,
        month = sep,
       volume = {936},
          eid = {L14},
        pages = {L14},
          doi = {10.3847/2041-8213/ac8a4e},
archivePrefix = {arXiv},
       eprint = {2207.13067},
 primaryClass = {astro-ph.GA},
       adsurl = {https://ui.adsabs.harvard.edu/abs/2022ApJ...936L..14P}
}

@ARTICLE{Carnall2024,
       author = {{Carnall}, A.~C. and {Cullen}, F. and {McLure}, R.~J. and
         {McLeod}, D.~J. and {Begley}, R. and {Donnan}, C.~T. and {Dunlop}, J.~S. and
         {Shapley}, A.~E. and {Rowlands}, K. and {Almaini}, O. and
         {Arellano-C{\'o}rdova}, K.~Z. and {Barrufet}, L. and {Cimatti}, A. and
         {Ellis}, R.~S. and {Grogin}, N.~A. and {Hamadouche}, M.~L. and
         {Illingworth}, G.~D. and {Koekemoer}, A.~M. and {Leung}, H.-H. and
         {Lovell}, C.~C. and {P{\'e}rez-Gonz{\'a}lez}, P.~G. and {Santini}, P. and
         {Stanton}, T.~M. and {Wild}, V.},
        title = "{The JWST EXCELS Survey: Too Much, Too Young, Too Fast? Ultra-massive
                  Quiescent Galaxies at $3 < z < 5$}",
      journal = {\mnras},
         year = 2024,
        month = oct,
       volume = {534},
       number = {1},
        pages = {325},
          doi = {10.1093/mnras/stae2092},
archivePrefix = {arXiv},
       eprint = {2405.02242},
 primaryClass = {astro-ph.GA},
       adsurl = {https://ui.adsabs.harvard.edu/abs/2024MNRAS.534..325C}
}

@ARTICLE{Belli2025,
       author = {{Belli}, Sirio and {Bugiani}, Letizia and {Park}, Minjung and
         {Mendel}, J.~Trevor and {Davies}, Rebecca L. and {Khoram}, Amir H. and
         {Johnson}, Benjamin D. and {Leja}, Joel and {Tacchella}, Sandro and
         {Brown}, Vanessa and {Conroy}, Charlie and {Emami}, Razieh and {Li}, Yijia and
         {Liboni}, Caterina and {Maheson}, Gabriel and {Mathews}, Elijah P. and
         {Naidu}, Rohan P. and {Nelson}, Erica J. and {Terrazas}, Bryan A. and
         {Weinberger}, Rainer},
        title = "{The Blue Jay Survey: Deep JWST Spectroscopy for a Representative
                  Sample of Galaxies at Cosmic Noon}",
      journal = {arXiv e-prints},
         year = 2025,
        month = oct,
          eid = {arXiv:2510.11775},
        pages = {arXiv:2510.11775},
          doi = {10.48550/arXiv.2510.11775},
archivePrefix = {arXiv},
       eprint = {2510.11775},
 primaryClass = {astro-ph.GA},
       adsurl = {https://ui.adsabs.harvard.edu/abs/2025arXiv251011775B}
}

@ARTICLE{Shapley2025,
       author = {{Shapley}, Alice E. and {Sanders}, Ryan L. and {Topping}, Michael W. and
         {Reddy}, Naveen A. and {Berg}, Danielle A. and {Bouwens}, Rychard J. and
         {Brammer}, Gabriel and {Carnall}, Adam C. and {Cullen}, Fergus and
         {Dav{\'e}}, Romeel and {Dunlop}, James S. and {Ellis}, Richard S. and
         {F{\"o}rster Schreiber}, N.~M. and {Furlanetto}, Steven R. and
         {Glazebrook}, Karl and {Illingworth}, Garth D. and {Jones}, Tucker and
         {Kriek}, Mariska and {McLeod}, Derek J. and {McLure}, Ross J. and
         {Narayanan}, Desika and {Oesch}, Pascal and {Pahl}, Anthony J. and
         {Pettini}, Max and {Schaerer}, Daniel and {Stark}, Daniel P. and
         {Steidel}, Charles C. and {Tang}, Mengtao and {Clarke}, Leonardo and
         {Donnan}, Callum T. and {Kehoe}, Emily},
        title = "{The AURORA Survey: A New Era of Emission-line Diagrams with JWST/NIRSpec}",
      journal = {\apj},
         year = 2025,
        month = feb,
       volume = {980},
       number = {2},
          eid = {242},
        pages = {242},
          doi = {10.3847/1538-4357/adad68},
archivePrefix = {arXiv},
       eprint = {2407.00157},
 primaryClass = {astro-ph.GA},
       adsurl = {https://ui.adsabs.harvard.edu/abs/2025ApJ...980..242S}
}

@ARTICLE{Slob2024,
       author = {{Slob}, Martje and {Kriek}, Mariska and {Beverage}, Aliza G. and
         {Suess}, Katherine A. and {Barro}, Guillermo and {Bezanson}, Rachel and
         {Brammer}, Gabriel and {Cheng}, Chloe M. and {Conroy}, Charlie and
         {de Graaff}, Anna and {F{\"o}rster Schreiber}, Natascha M. and {Franx}, Marijn and
         {Lorenz}, Brian and {Mancera Pi{\~n}a}, Pavel E. and {Marchesini}, Danilo and
         {Muzzin}, Adam and {Newman}, Andrew B. and {Price}, Sedona H. and
         {Shapley}, Alice E. and {Stefanon}, Mauro and {van Dokkum}, Pieter and
         {Weisz}, Daniel R.},
        title = "{The JWST-SUSPENSE Ultradeep Spectroscopic Program: Survey Overview and
                  Star Formation Histories of Quiescent Galaxies at $1 < z < 3$}",
      journal = {\apj},
         year = 2024,
        month = oct,
       volume = {973},
       number = {2},
          eid = {131},
        pages = {131},
          doi = {10.3847/1538-4357/ad65ff},
archivePrefix = {arXiv},
       eprint = {2404.12432},
 primaryClass = {astro-ph.GA},
       adsurl = {https://ui.adsabs.harvard.edu/abs/2024ApJ...973..131S}
}

@ARTICLE{Topping2025_CIV_N_emitters,
       author = {{Topping}, Michael W. and {Stark}, Daniel P. and {Senchyna}, Peter and {Chen}, Zuyi and {Zitrin}, Adi and {Endsley}, Ryan and {Charlot}, St{\'e}phane and {Furtak}, Lukas J. and {Maseda}, Michael V. and {Plat}, Adele and {Smit}, Renske and {Mainali}, Ramesh and {Chevallard}, Jacopo and {Molyneux}, Stephen and {Rigby}, Jane R.},
        title = "{Deep Rest-UV JWST/NIRSpec Spectroscopy of Early Galaxies: The Demographics of C IV and N-emitters in the Reionization Era}",
      journal = {\apj},
         year = 2025,
        month = feb,
       volume = {980},
       number = {2},
          eid = {225},
        pages = {225},
          doi = {10.3847/1538-4357/ada95c},
archivePrefix = {arXiv},
       eprint = {2407.19009},
 primaryClass = {astro-ph.GA},
       adsurl = {https://ui.adsabs.harvard.edu/abs/2025ApJ...980..225T}
}

@ARTICLE{DEugenio2025,
       author = {{D'Eugenio}, Francesco and {Cameron}, Alex J. and {Scholtz}, Jan and {Carniani}, Stefano and {Willott}, Chris J. and {Curtis-Lake}, Emma and {Bunker}, Andrew J. and {Parlanti}, Eleonora and {Maiolino}, Roberto and {Willmer}, Christopher N.~A. and {Jakobsen}, Peter and {Robertson}, Brant E. and {Johnson}, Benjamin D. and {Tacchella}, Sandro and {Cargile}, Phillip A. and {Rawle}, Tim and {Arribas}, Santiago and {Chevallard}, Jacopo and {Curti}, Mirko and {Egami}, Eiichi and {Eisenstein}, Daniel J. and {Kumari}, Nimisha and {Looser}, Tobias J. and {Rieke}, Marcia J. and {Rodr{\'\i}guez Del Pino}, Bruno and {Saxena}, Aayush and {{\"U}bler}, Hannah and {Venturi}, Giacomo and {Witstok}, Joris and {Baker}, William M. and {Bhatawdekar}, Rachana and {Bonaventura}, Nina and {Boyett}, Kristan and {Charlot}, Stephane and {Danhaive}, A. Lola and {Hainline}, Kevin N. and {Hausen}, Ryan and {Helton}, Jakob M. and {Ji}, Xihan and {Ji}, Zhiyuan and {Jones}, Gareth C. and {Juod{\v{z}}balis}, Ignas and {Maseda}, Michael V. and {P{\'e}rez-Gonz{\'a}lez}, Pablo G. and {Perna}, Michele and {Pusk{\'a}s}, D{\'a}vid and {Shivaei}, Irene and {Silcock}, Maddie S. and {Simmonds}, Charlotte and {Smit}, Renske and {Sun}, Fengwu and {Villanueva}, Natalia C. and {Williams}, Christina C. and {Zhu}, Yongda},
        title = "{JADES Data Release 3: NIRSpec/Microshutter Assembly Spectroscopy for 4000 Galaxies in the GOODS Fields}",
      journal = {\apjs},
         year = 2025,
        month = mar,
       volume = {277},
       number = {1},
          eid = {4},
        pages = {4},
          doi = {10.3847/1538-4365/ada148},
archivePrefix = {arXiv},
       eprint = {2404.06531},
 primaryClass = {astro-ph.GA},
       adsurl = {https://ui.adsabs.harvard.edu/abs/2025ApJS..277....4D}
}

@ARTICLE{Eisenstein2025_origins,
       author = {{Eisenstein}, Daniel J. and {Johnson}, Benjamin D. and {Robertson}, Brant and {Tacchella}, Sandro and {Hainline}, Kevin and {Jakobsen}, Peter and {Maiolino}, Roberto and {Bonaventura}, Nina and {Bunker}, Andrew J. and {Cameron}, Alex J. and {Cargile}, Phillip A. and {Curtis-Lake}, Emma and {Hausen}, Ryan and {Pusk{\'a}s}, D{\'a}vid and {Rieke}, Marcia and {Sun}, Fengwu and {Willmer}, Christopher N.~A. and {Willott}, Chris and {Alberts}, Stacey and {Arribas}, Santiago and {Baker}, William M. and {Baum}, Stefi and {Bhatawdekar}, Rachana and {Carniani}, Stefano and {Charlot}, Stephane and {Chen}, Zuyi and {Chevallard}, Jacopo and {Curti}, Mirko and {DeCoursey}, Christa and {D'Eugenio}, Francesco and {de Graaff}, Anna and {Egami}, Eiichi and {Helton}, Jakob M. and {Ji}, Zhiyuan and {Jones}, Gareth C. and {Kumari}, Nimisha and {L{\"u}tzgendorf}, Nora and {Laseter}, Isaac and {Looser}, Tobias J. and {Lyu}, Jianwei and {Maseda}, Michael V. and {Nelson}, Erica and {Parlanti}, Eleonora and {Rauscher}, Bernard J. and {Rawle}, Tim and {Rieke}, George and {Rix}, Hans-Walter and {Rujopakarn}, Wiphu and {Sandles}, Lester and {Saxena}, Aayush and {Scholtz}, Jan and {Sharpe}, Katherine and {Shivaei}, Irene and {Simmonds}, Charlotte and {Smit}, Renske and {Topping}, Michael W. and {{\"U}bler}, Hannah and {Venturi}, Giacomo and {Williams}, Christina C. and {Witstok}, Joris and {Woodrum}, Charity},
        title = "{The JADES Origins Field: A New JWST Deep Field in the JADES Second NIRCam Data Release}",
      journal = {\apjs},
         year = 2025,
        month = dec,
       volume = {281},
       number = {2},
          eid = {50},
        pages = {50},
          doi = {10.3847/1538-4365/ae1137},
archivePrefix = {arXiv},
       eprint = {2310.12340},
 primaryClass = {astro-ph.GA},
       adsurl = {https://ui.adsabs.harvard.edu/abs/2025ApJS..281...50E}
}

@ARTICLE{Frye2024,
       author = {{Frye}, Brenda L. and {Pascale}, Massimo and {Pierel}, Justin and {Chen}, Wenlei and {Foo}, Nicholas and {Leimbach}, Reagen and {Garuda}, Nikhil and {Cohen}, Seth H. and {Kamieneski}, Patrick S. and {Windhorst}, Rogier A. and {Koekemoer}, Anton M. and {Kelly}, Pat and {Summers}, Jake and {Engesser}, Michael and {Liu}, Daizhong and {Furtak}, Lukas J. and {Polletta}, Maria del Carmen and {Harrington}, Kevin C. and {Willner}, S.~P. and {Diego}, Jose M. and {Jansen}, Rolf A. and {Coe}, Dan and {Conselice}, Christopher J. and {Dai}, Liang and {Dole}, Herv{\'e} and {D'Silva}, Jordan C.~J. and {Driver}, Simon P. and {Grogin}, Norman A. and {Marshall}, Madeline A. and {Meena}, Ashish K. and {Nonino}, Mario and {Ortiz}, Rafael and {Pirzkal}, Nor and {Robotham}, Aaron and {Ryan}, Russell E. and {Strolger}, Lou and {Tompkins}, Scott and {Willmer}, Christopher N.~A. and {Yan}, Haojing and {Yun}, Min S. and {Zitrin}, Adi},
        title = "{The JWST Discovery of the Triply Imaged Type Ia ``Supernova H0pe'' and Observations of the Galaxy Cluster PLCK G165.7+67.0}",
      journal = {\apj},
         year = 2024,
        month = feb,
       volume = {961},
       number = {2},
          eid = {171},
        pages = {171},
          doi = {10.3847/1538-4357/ad1034},
archivePrefix = {arXiv},
       eprint = {2309.07326},
 primaryClass = {astro-ph.GA},
       adsurl = {https://ui.adsabs.harvard.edu/abs/2024ApJ...961..171F}
}

@ARTICLE{Valentino2023,
       author = {{Valentino}, Francesco and {Brammer}, Gabriel and {Gould}, Katriona M.~L. and {Kokorev}, Vasily and {Fujimoto}, Seiji and {Jespersen}, Christian Kragh and {Vijayan}, Aswin P. and {Weaver}, John R. and {Ito}, Kei and {Tanaka}, Masayuki and {Ilbert}, Olivier and {Magdis}, Georgios E. and {Whitaker}, Katherine E. and {Faisst}, Andreas L. and {Gallazzi}, Anna and {Gillman}, Steven and {Gim{\'e}nez-Arteaga}, Clara and {G{\'o}mez-Guijarro}, Carlos and {Kubo}, Mariko and {Heintz}, Kasper E. and {Hirschmann}, Michaela and {Oesch}, Pascal and {Onodera}, Masato and {Rizzo}, Francesca and {Lee}, Minju and {Strait}, Victoria and {Toft}, Sune},
        title = "{An Atlas of Color-selected Quiescent Galaxies at z > 3 in Public JWST Fields}",
      journal = {\apj},
         year = 2023,
        month = apr,
       volume = {947},
       number = {1},
          eid = {20},
        pages = {20},
          doi = {10.3847/1538-4357/acbefa},
archivePrefix = {arXiv},
       eprint = {2302.10936},
 primaryClass = {astro-ph.GA},
       adsurl = {https://ui.adsabs.harvard.edu/abs/2023ApJ...947...20V}
}

@ARTICLE{Pollock2026,
       author = {{Pollock}, Clara L. and {Gottumukkala}, Rashmi and {Heintz}, Kasper E. and {Brammer}, Gabriel B. and {Roberts-Borsani}, Guido and {Oesch}, Pascal A. and {Witstok}, Joris and {Arellano-C{\'o}rdova}, Karla Z. and {Cullen}, Fergus and {Scholte}, Dirk and {Terp}, Chamilla and {Rowland}, Lucie and {Sneppen}, Albert and {Ito}, Kei and {Valentino}, Francesco and {Matthee}, Jorryt and {Watson}, Darach and {Toft}, Sune},
        title = "{Novel z {\ensuremath{\sim}} 10 auroral line measurements extend the gradual offset of the fundamental metallicity relation deep into the first gigayear of cosmic time}",
      journal = {\aap},
         year = 2026,
        month = apr,
       volume = {708},
          eid = {A203},
        pages = {A203},
          doi = {10.1051/0004-6361/202556032},
archivePrefix = {arXiv},
       eprint = {2506.15779},
 primaryClass = {astro-ph.GA},
       adsurl = {https://ui.adsabs.harvard.edu/abs/2026A&A...708A.203P}
}

@ARTICLE{Valentino2025,
       author = {{Valentino}, F. and {Heintz}, K.~E. and {Brammer}, G. and {Ito}, K. and {Kokorev}, V. and {Whitaker}, K.~E. and {Gallazzi}, A. and {de Graaff}, A. and {Weibel}, A. and {Frye}, B.~L. and {Kamieneski}, P.~S. and {Jin}, S. and {Ceverino}, D. and {Faisst}, A. and {Farcy}, M. and {Fujimoto}, S. and {Gillman}, S. and {Gottumukkala}, R. and {Hamadouche}, M. and {Harrington}, K.~C. and {Hirschmann}, M. and {Jespersen}, C.~K. and {Kakimoto}, T. and {Kubo}, M. and {Lagos}, C. d. P. and {Lee}, M. and {Magdis}, G.~E. and {Man}, A.~W.~S. and {Onodera}, M. and {Rizzo}, F. and {Shimakawa}, R. and {Setton}, D.~J. and {Tanaka}, M. and {Toft}, S. and {Wu}, P.-F. and {Zhu}, P.},
        title = "{Gas outflows in two recently quenched galaxies at z = 4 and 7}",
      journal = {\aap},
         year = 2025,
        month = jul,
       volume = {699},
          eid = {A358},
        pages = {A358},
          doi = {10.1051/0004-6361/202553908},
archivePrefix = {arXiv},
       eprint = {2503.01990},
 primaryClass = {astro-ph.GA},
       adsurl = {https://ui.adsabs.harvard.edu/abs/2025A&A...699A.358V}
}

@ARTICLE{Rynkun2019,
       author = {{Rynkun}, P. and {Gaigalas}, G. and {J{\"o}nsson}, P.},
        title = "{Energies and E1, M1, E2, M2 transition rates for states of the 2s$^{2}$2p$^{3}$, 2s2p$^{4}$, and 2p$^{5}$ configurations in nitrogen-like ions between F III and Zn XXIV}",
      journal = {\aap},
         year = 2019,
       volume = {623},
          eid = {A155},
          doi = {10.1051/0004-6361/201834931},
       adsurl = {https://ui.adsabs.harvard.edu/abs/2019A&A...623A.155R}
}

@ARTICLE{RamsbottomBell1997,
       author = {{Ramsbottom}, C.~A. and {Bell}, K.~L.},
        title = "{Effective Collision Strengths for Electron Impact Excitation of Ar~{\sc iv}}",
      journal = {Atomic Data and Nuclear Data Tables},
         year = 1997,
       volume = {66},
        pages = {1-48},
          doi = {10.1006/adnd.1997.0739},
       adsurl = {https://ui.adsabs.harvard.edu/abs/1997ADNDT..66....1R}
}

@ARTICLE{Zeippen1982,
       author = {{Zeippen}, C.~J.},
        title = "{Transition probabilities for forbidden lines in the 2p$^{3}$ configuration}",
      journal = {\mnras},
         year = 1982,
        month = jan,
       volume = {198},
        pages = {111-125},
          doi = {10.1093/mnras/198.1.111},
       adsurl = {https://ui.adsabs.harvard.edu/abs/1982MNRAS.198..111Z}
}

@BOOK{Wiese1996,
       author = {{Wiese}, W.~L. and {Fuhr}, J.~R. and {Deters}, T.~M.},
        title = "{Atomic Transition Probabilities of Carbon, Nitrogen, and Oxygen}",
       series = {Journal of Physical and Chemical Reference Data Monograph, No.~7},
         year = 1996,
    publisher = {American Institute of Physics},
      address = {New York},
       adsurl = {https://ui.adsabs.harvard.edu/abs/1996atpc.book.....W}
}

@ARTICLE{Storey2000,
       author = {{Storey}, P.~J. and {Zeippen}, C.~J.},
        title = "{Theoretical values for the [O~{\sc iii}] 5007/4959 line-intensity ratio and homologous cases}",
      journal = {\mnras},
         year = 2000,
        month = mar,
       volume = {312},
        pages = {813-816},
          doi = {10.1046/j.1365-8711.2000.03184.x},
       adsurl = {https://ui.adsabs.harvard.edu/abs/2000MNRAS.312..813S}
}

@ARTICLE{FroeseFischer2004,
       author = {{Froese Fischer}, C. and {Tachiev}, G.},
        title = "{Breit-Pauli energy levels, lifetimes, and transition probabilities for the beryllium-like to neon-like sequences}",
      journal = {Atomic Data and Nuclear Data Tables},
         year = 2004,
        month = may,
       volume = {87},
        pages = {1-184},
          doi = {10.1016/j.adt.2003.11.002},
       adsurl = {https://ui.adsabs.harvard.edu/abs/2004ADNDT..87....1F}
}

@ARTICLE{Galavis1998,
       author = {{Galavis}, M.~E. and {Mendoza}, C. and {Zeippen}, C.~J.},
        title = "{Atomic data from the IRON Project. {XXIX}. Radiative rates for transitions within the n = 2 complex in ions of the boron isoelectronic sequence}",
      journal = {\aaps},
         year = 1998,
        month = sep,
       volume = {131},
        pages = {499-504},
          doi = {10.1051/aas:1998286},
       adsurl = {https://ui.adsabs.harvard.edu/abs/1998A&AS..131..499G}
}

@ARTICLE{Glass1983,
       author = {{Glass}, R.},
        title = "{Electric quadrupole transitions in the beryllium isoelectronic sequence}",
      journal = {\apss},
         year = 1983,
       volume = {92},
        pages = {307-319},
          doi = {10.1007/BF00651295},
       adsurl = {https://ui.adsabs.harvard.edu/abs/1983Ap%26SS..92..307G}
}

@ARTICLE{Nussbaumer1978,
       author = {{Nussbaumer}, H. and {Storey}, P.~J.},
        title = "{[VERIFY TITLE IN ADS]}",
      journal = {\aap},
         year = 1978,
       volume = {64},
        pages = {139},
       adsurl = {https://ui.adsabs.harvard.edu/abs/1978A%26A....64..139N}
}

@ARTICLE{Galavis1997,
       author = {{Galavis}, M.~E. and {Mendoza}, C. and {Zeippen}, C.~J.},
        title = "{Atomic data from the IRON Project. {XXII}. Radiative rates for forbidden transitions within the ground configuration of ions in the carbon and oxygen isoelectronic sequences}",
      journal = {\aaps},
         year = 1997,
       volume = {123},
        pages = {159-171},
          doi = {10.1051/aas:1997156},
       adsurl = {https://ui.adsabs.harvard.edu/abs/1997A%26AS..123..159G}
}

@ARTICLE{Kisielius2009,
       author = {{Kisielius}, R. and {Storey}, P.~J. and {Ferland}, G.~J. and {Keenan}, F.~P.},
        title = "{Electron-impact excitation of O~{\sc ii} fine-structure levels}",
      journal = {\mnras},
         year = 2009,
        month = aug,
       volume = {397},
        pages = {903-912},
          doi = {10.1111/j.1365-2966.2009.14989.x},
       adsurl = {https://ui.adsabs.harvard.edu/abs/2009MNRAS.397..903K}
}

@ARTICLE{Palay2012,
       author = {{Palay}, E. and {Nahar}, S.~N. and {Pradhan}, A.~K. and {Eissner}, W.},
        title = "{Improved collision strengths and line ratios for forbidden [O~{\sc iii}] far-infrared and optical lines}",
      journal = {\mnras},
         year = 2012,
        month = jun,
       volume = {423},
        pages = {L35-L39},
          doi = {10.1111/j.1745-3933.2012.01253.x},
       adsurl = {https://ui.adsabs.harvard.edu/abs/2012MNRAS.423L..35P}
}

@ARTICLE{Aggarwal1999,
       author = {{Aggarwal}, K.~M. and {Keenan}, F.~P.},
        title = "{Excitation Rate Coefficients for Fine-Structure Transitions in O~{\sc iii}}",
      journal = {\apjs},
         year = 1999,
        month = sep,
       volume = {123},
        pages = {311-341},
          doi = {10.1086/313232},
       adsurl = {https://ui.adsabs.harvard.edu/abs/1999ApJS..123..311A}
}

@ARTICLE{Tayal2011,
       author = {{Tayal}, S.~S.},
        title = "{Electron Excitation Collision Strengths for Singly Ionized Nitrogen}",
      journal = {\apjs},
         year = 2011,
        month = aug,
       volume = {195},
          eid = {12},
          doi = {10.1088/0067-0049/195/2/12},
       adsurl = {https://ui.adsabs.harvard.edu/abs/2011ApJS..195...12T}
}

@ARTICLE{Blum1992,
       author = {{Blum}, R.~D. and {Pradhan}, A.~K.},
        title = "{Rate Coefficients for the Excitation of Infrared and Ultraviolet Lines in C~{\sc ii}, N~{\sc iii}, and O~{\sc iv}}",
      journal = {\apjs},
         year = 1992,
        month = feb,
       volume = {80},
        pages = {425-440},
          doi = {10.1086/191674},
       adsurl = {https://ui.adsabs.harvard.edu/abs/1992ApJS...80..425B}
}

@ARTICLE{Ramsbottom1994,
       author = {{Ramsbottom}, C.~A. and {Berrington}, K.~A. and {Hibbert}, A. and {Bell}, K.~L.},
        title = "{Electron impact excitation rates for transitions involving the n = 2 and n = 3 levels of beryllium-like N~{\sc iv}}",
      journal = {Physica Scripta},
         year = 1994,
        month = sep,
       volume = {50},
        pages = {246-253},
          doi = {10.1088/0031-8949/50/3/005},
       adsurl = {https://ui.adsabs.harvard.edu/abs/1994PhyS...50..246R}
}

@ARTICLE{Berrington1985,
       author = {{Berrington}, K.~A. and {Burke}, P.~G. and {Dufton}, P.~L. and {Kingston}, A.~E.},
        title = "{Electron-Impact-Excitation Collision Strengths for Be-like Ions. {II}. Intermediate-Energy Region and Collision Rates}",
      journal = {Atomic Data and Nuclear Data Tables},
         year = 1985,
       volume = {33},
        pages = {195-210},
          doi = {10.1016/0092-640X(85)90017-8},
       adsurl = {https://ui.adsabs.harvard.edu/abs/1985ADNDT..33..195B}
}

@ARTICLE{MunozBurgos2009,
       author = {{Munoz Burgos}, J.~M. and {Loch}, S.~D. and {Ballance}, C.~P. and {Boivin}, R.~F.},
        title = "{Electron-impact excitation of Ar$^{2+}$}",
      journal = {\aap},
         year = 2009,
       volume = {500},
        pages = {1253-1261},
          doi = {10.1051/0004-6361/200911743},
       adsurl = {https://ui.adsabs.harvard.edu/abs/2009A%26A...500.1253M}
}

@ARTICLE{McLaughlin2000,
       author = {{McLaughlin}, B.~M. and {Bell}, K.~L.},
        title = "{Electron collisional excitation of Ne~{\sc iii}: $(1s^22s^22p^4\ ^3P_{2,1,0},\ ^1D_2,\ ^1S_0)$ fine-structure transitions}",
      journal = {Journal of Physics B: Atomic, Molecular and Optical Physics},
         year = 2000,
        month = feb,
       volume = {33},
        pages = {597-607},
          doi = {10.1088/0953-4075/33/4/301},
       adsurl = {https://ui.adsabs.harvard.edu/abs/2000JPhB...33..597M}
}

@ARTICLE{Pilyugin2009,
       author = {{Pilyugin}, L.~S. and {Mattsson}, L. and {V{\'\i}lchez}, J.~M. and {Cedr{\'e}s}, B.},
        title = "{On the electron temperatures in high-metallicity HII regions}",
      journal = {\mnras},
         year = 2009,
        month = sep,
       volume = {398},
       number = {1},
        pages = {485-496},
          doi = {10.1111/j.1365-2966.2009.15182.x},
archivePrefix = {arXiv},
       eprint = {0907.0084},
 primaryClass = {astro-ph.CO},
       adsurl = {https://ui.adsabs.harvard.edu/abs/2009MNRAS.398..485P}
}

@ARTICLE{PeimbertCostero1969,
       author = {{Peimbert}, M. and {Costero}, R.},
        title = "{Chemical Abundances in Galactic HII Regions}",
      journal = {Boletin de los Observatorios Tonantzintla y Tacubaya},
         year = 1969,
        month = may,
       volume = {5},
        pages = {3-22},
       adsurl = {https://ui.adsabs.harvard.edu/abs/1969BOTT....5....3P}
}

@ARTICLE{Abdurrouf2024,
       author = {{Abdurrouf} and {Larson}, Rebecca L. and {Coe}, Dan and {Hsiao}, Tiger Yu-Yang and {{\'A}lvarez-M{\'a}rquez}, Javier and {G{\'o}mez}, Alejandro Crespo and {Adamo}, Angela and {Bhatawdekar}, Rachana and {Bik}, Arjan and {Bradley}, Larry D. and {Conselice}, Christopher J. and {Dayal}, Pratika and {Diego}, Jose M. and {Fujimoto}, Seiji and {Furtak}, Lukas J. and {Hutchison}, Taylor A. and {Jung}, Intae and {Killi}, Meghana and {Kokorev}, Vasily and {Mingozzi}, Matilde and {Norman}, Colin and {Resseguier}, Tom and {Ricotti}, Massimo and {Rigby}, Jane R. and {Vanzella}, Eros and {Welch}, Brian and {Windhorst}, Rogier A. and {Xu}, Xinfeng and {Zitrin}, Adi},
        title = "{JWST NIRSpec High-resolution Spectroscopy of MACS0647─JD at z = 10.167: Resolved [O II] Doublet and Electron Density in an Early Galaxy}",
      journal = {\apj},
         year = 2024,
        month = sep,
       volume = {973},
       number = {1},
          eid = {47},
        pages = {47},
          doi = {10.3847/1538-4357/ad6001},
archivePrefix = {arXiv},
       eprint = {2404.16201},
 primaryClass = {astro-ph.GA},
       adsurl = {https://ui.adsabs.harvard.edu/abs/2024ApJ...973...47A}
}

@misc{specutils2025,
  author       = {Nicholas Earl and
                  Erik Tollerud and
                  Ricky O'Steen and
                  brechmos and
                  Wolfgang Kerzendorf and
                  Ivo Busko and
                  shaileshahuja and
                  P. L. Lim and
                  Dan D'Avella and
                  Thomas Robitaille and
                  Adam Ginsburg and
                  Derek Homeier and
                  Brigitta Sipőcz and
                  Jesse Averbukh and
                  Brian Cherinka and
                  James Tocknell and
                  Sara Ogaz and
                  Robel Geda and
                  James Davies and
                  Kyle Conroy and
                  Hans Moritz Günther and
                  Kyle Barbary and
                  Kelle Cruz and
                  Jonathan Foster and
                  Michael Droettboom and
                  Duy Nguyen and
                  E. M. Bray and
                  Andy Casey and
                  Henry Ferguson},
  title        = {astropy/specutils: v2.1.0},
  month        = jul,
  year         = 2025,
  publisher    = {Zenodo},
  version      = {v2.1.0},
  doi          = {10.5281/zenodo.16615456},
  url          = {https://doi.org/10.5281/zenodo.16615456},
}

@ARTICLE{Rusakov2026,
       author = {{Rusakov}, V. and {Watson}, D. and {Nikopoulos}, G.~P. and {Brammer}, G. and {Gottumukkala}, R. and {Harvey}, T. and {Heintz}, K.~E. and {Damgaard}, R. and {Sim}, S.~A. and {Sneppen}, A. and {Vijayan}, A.~P. and {Adams}, N. and {Austin}, D. and {Conselice}, C.~J. and {Goolsby}, C.~M. and {Toft}, S. and {Witstok}, J.},
        title = "{Little red dots as young supermassive black holes in dense ionized cocoons}",
      journal = {\nat},
         year = 2026,
        month = jan,
       volume = {649},
       number = {8097},
        pages = {574-579},
          doi = {10.1038/s41586-025-09900-4},
archivePrefix = {arXiv},
       eprint = {2503.16595},
 primaryClass = {astro-ph.GA},
       adsurl = {https://ui.adsabs.harvard.edu/abs/2026Natur.649..574R}
}

@ARTICLE{Ito2026,
       author = {{Ito}, K. and {Valentino}, F. and {Brammer}, G. and {Hamadouche}, M.~L. and {Whitaker}, K.~E. and {Kokorev}, V. and {Zhu}, P. and {Kakimoto}, T. and {Wu}, P.-F. and {Antwi-Danso}, J. and {Baker}, W.~M. and {Ceverino}, D. and {Faisst}, A.~L. and {Farcy}, M. and {Fujimoto}, S. and {Gallazzi}, A. and {Gillman}, S. and {Gottumukkala}, R. and {Heintz}, K.~E. and {Hirschmann}, M. and {Jespersen}, C.~K. and {Kubo}, M. and {Lee}, M. and {Magdis}, G. and {Onodera}, M. and {Shimakawa}, R. and {Tanaka}, M. and {Toft}, S. and {Weaver}, J.~R.},
        title = "{DeepDive: A deep dive into the physics of the first massive quiescent galaxies in the Universe}",
      journal = {\aap},
         year = 2026,
        month = jun,
       volume = {710},
          eid = {A364},
        pages = {A364},
          doi = {10.1051/0004-6361/202556137},
archivePrefix = {arXiv},
       eprint = {2506.22642},
 primaryClass = {astro-ph.GA},
       adsurl = {https://ui.adsabs.harvard.edu/abs/2026A&A...710A.364I}
}

@ARTICLE{PerezMontero2009,
       author = {{P{\'e}rez-Montero}, Enrique and {Contini}, Thierry},
        title = "{The impact of the nitrogen-to-oxygen ratio on ionized nebula diagnostics based on [NII] emission lines}",
      journal = {\mnras},
         year = 2009,
        month = sep,
       volume = {398},
       number = {2},
        pages = {949-960},
          doi = {10.1111/j.1365-2966.2009.15145.x},
archivePrefix = {arXiv},
       eprint = {0905.4621},
 primaryClass = {astro-ph.CO},
       adsurl = {https://ui.adsabs.harvard.edu/abs/2009MNRAS.398..949P}
}

@ARTICLE{Sarkar2025,
       author = {{Sarkar}, Arnab and {Chakraborty}, Priyanka and {Vogelsberger}, Mark and {McDonald}, Michael and {Torrey}, Paul and {Garcia}, Alex M. and {Khullar}, Gourav and {Ferland}, Gary J. and {Forman}, William and {Wolk}, Scott and {Schneider}, Benjamin and {Bautz}, Mark and {Miller}, Eric and {Grant}, Catherine and {ZuHone}, John},
        title = "{Unveiling the Cosmic Chemistry: Revisiting the Mass─Metallicity Relation with JWST/NIRSpec at 4 < z < 10}",
      journal = {\apj},
         year = 2025,
        month = jan,
       volume = {978},
       number = {2},
          eid = {136},
        pages = {136},
          doi = {10.3847/1538-4357/ad8f32},
archivePrefix = {arXiv},
       eprint = {2408.07974},
 primaryClass = {astro-ph.GA},
       adsurl = {https://ui.adsabs.harvard.edu/abs/2025ApJ...978..136S}
}

@ARTICLE{Laseter2024,
       author = {{Laseter}, Isaac H. and {Maseda}, Michael V. and {Curti}, Mirko and {Maiolino}, Roberto and {D'Eugenio}, Francesco and {Cameron}, Alex J. and {Looser}, Tobias J. and {Arribas}, Santiago and {Baker}, William M. and {Bhatawdekar}, Rachana and {Boyett}, Kristan and {Bunker}, Andrew J. and {Carniani}, Stefano and {Charlot}, Stephane and {Chevallard}, Jacopo and {Curtis-lake}, Emma and {Egami}, Eiichi and {Eisenstein}, Daniel J. and {Hainline}, Kevin and {Hausen}, Ryan and {Ji}, Zhiyuan and {Kumari}, Nimisha and {Perna}, Michele and {Rawle}, Tim and {Rix}, Hans-Walter and {Robertson}, Brant and {Rodr{\'\i}guez Del Pino}, Bruno and {Sandles}, Lester and {Scholtz}, Jan and {Smit}, Renske and {Tacchella}, Sandro and {{\"U}bler}, Hannah and {Williams}, Christina C. and {Willott}, Chris and {Witstok}, Joris},
        title = "{JADES: Detecting [OIII]{\ensuremath{\lambda}}4363 emitters and testing strong line calibrations in the high-z Universe with ultra-deep JWST/NIRSpec spectroscopy up to z {\ensuremath{\sim}} 9.5}",
      journal = {\aap},
         year = 2024,
        month = jan,
       volume = {681},
          eid = {A70},
        pages = {A70},
          doi = {10.1051/0004-6361/202347133},
archivePrefix = {arXiv},
       eprint = {2306.03120},
 primaryClass = {astro-ph.GA},
       adsurl = {https://ui.adsabs.harvard.edu/abs/2024A&A...681A..70L}
}

@ARTICLE{Setton2025,
       author = {{Setton}, David J. and {Greene}, Jenny E. and {de Graaff}, Anna and {Ma}, Yilun and {Leja}, Joel and {Matthee}, Jorryt and {Bezanson}, Rachel and {Boogaard}, Leindert A. and {Cleri}, Nikko J. and {Katz}, Harley and {Labbe}, Ivo and {Maseda}, Michael V. and {McConachie}, Ian and {Miller}, Tim B. and {Price}, Sedona H. and {Suess}, Katherine A. and {van Dokkum}, Pieter and {Wang}, Bingjie and {Weibel}, Andrea and {Whitaker}, Katherine E. and {Williams}, Christina C.},
        title = "{Little Red Dots at an Inflection Point: Ubiquitous V-shaped Turnover Consistently Occurs at the Balmer Limit}",
      journal = {\apj},
         year = 2025,
        month = dec,
       volume = {995},
       number = {1},
          eid = {118},
        pages = {118},
          doi = {10.3847/1538-4357/ae1500},
archivePrefix = {arXiv},
       eprint = {2411.03424},
 primaryClass = {astro-ph.GA},
       adsurl = {https://ui.adsabs.harvard.edu/abs/2025ApJ...995..118S}
}

@ARTICLE{Matthee2024,
       author = {{Matthee}, Jorryt and {Naidu}, Rohan P. and {Brammer}, Gabriel and {Chisholm}, John and {Eilers}, Anna-Christina and {Goulding}, Andy and {Greene}, Jenny and {Kashino}, Daichi and {Labbe}, Ivo and {Lilly}, Simon J. and {Mackenzie}, Ruari and {Oesch}, Pascal A. and {Weibel}, Andrea and {Wuyts}, Stijn and {Xiao}, Mengyuan and {Bordoloi}, Rongmon and {Bouwens}, Rychard and {van Dokkum}, Pieter and {Illingworth}, Garth and {Kramarenko}, Ivan and {Maseda}, Michael V. and {Mason}, Charlotte and {Meyer}, Romain A. and {Nelson}, Erica J. and {Reddy}, Naveen A. and {Shivaei}, Irene and {Simcoe}, Robert A. and {Yue}, Minghao},
        title = "{Little Red Dots: An Abundant Population of Faint Active Galactic Nuclei at z {\ensuremath{\sim}} 5 Revealed by the EIGER and FRESCO JWST Surveys}",
      journal = {\apj},
         year = 2024,
        month = mar,
       volume = {963},
       number = {2},
          eid = {129},
        pages = {129},
          doi = {10.3847/1538-4357/ad2345},
archivePrefix = {arXiv},
       eprint = {2306.05448},
 primaryClass = {astro-ph.GA},
       adsurl = {https://ui.adsabs.harvard.edu/abs/2024ApJ...963..129M}
}

@ARTICLE{Kokorev2024,
       author = {{Kokorev}, Vasily and {Caputi}, Karina I. and {Greene}, Jenny E. and {Dayal}, Pratika and {Trebitsch}, Maxime and {Cutler}, Sam E. and {Fujimoto}, Seiji and {Labb{\'e}}, Ivo and {Miller}, Tim B. and {Iani}, Edoardo and {Navarro-Carrera}, Rafael and {Rinaldi}, Pierluigi},
        title = "{A Census of Photometrically Selected Little Red Dots at 4 < z < 9 in JWST Blank Fields}",
      journal = {\apj},
         year = 2024,
        month = jun,
       volume = {968},
       number = {1},
          eid = {38},
        pages = {38},
          doi = {10.3847/1538-4357/ad4265},
archivePrefix = {arXiv},
       eprint = {2401.09981},
 primaryClass = {astro-ph.GA},
       adsurl = {https://ui.adsabs.harvard.edu/abs/2024ApJ...968...38K}
}

@ARTICLE{deGraaff2026,
       author = {{de Graaff}, Anna and {Hviding}, Raphael E. and {Naidu}, Rohan P. and {Greene}, Jenny E. and {Miller}, Tim B. and {Leja}, Joel and {Matthee}, Jorryt and {Brammer}, Gabriel and {Katz}, Harley and {Bezanson}, Rachel and {Boogaard}, Leindert A. and {Bose}, Sownak and {Chisholm}, John and {Cleri}, Nikko J. and {Dayal}, Pratika and {Feldmann}, Robert and {Fudamoto}, Yoshinobu and {Fujimoto}, Seiji and {Furtak}, Lukas J. and {Glazebrook}, Karl and {Gottumukkala}, Rashmi and {Heintz}, Kasper E. and {Kokorev}, Vasily and {Labbe}, Ivo and {Maseda}, Michael V. and {McConachie}, Ian and {Nanayakkara}, Themiya and {Nelson}, Erica and {Nowaczyk}, Przemys{\l}aw and {Oesch}, Pascal A. and {Rix}, Hans-Walter and {Setton}, David J. and {Torralba}, Alberto and {Walter}, Fabian and {Wang}, Bingjie and {Weibel}, Andrea and {van der Wel}, Arjen},
        title = "{Little red dots host black hole stars: a unified family of gas-reddened AGN revealed by JWST/NIRSpec spectroscopy}",
      journal = {\mnras},
         year = 2026,
        month = sep,
       volume = {551},
       number = {3},
          eid = {stag1567},
        pages = {stag1567},
          doi = {10.1093/mnras/stag1567},
archivePrefix = {arXiv},
       eprint = {2511.21820},
 primaryClass = {astro-ph.GA},
       adsurl = {https://ui.adsabs.harvard.edu/abs/2026MNRAS.551g1567D}
}

@ARTICLE{Ferland2017_CLOUDY,
       author = {{Ferland}, G.~J. and {Chatzikos}, M. and {Guzm{\'a}n}, F. and {Lykins}, M.~L. and {van Hoof}, P.~A.~M. and {Williams}, R.~J.~R. and {Abel}, N.~P. and {Badnell}, N.~R. and {Keenan}, F.~P. and {Porter}, R.~L. and {Stancil}, P.~C.},
        title = "{The 2017 Release Cloudy}",
      journal = {\rmxaa},
         year = 2017,
        month = oct,
       volume = {53},
        pages = {385-438},
          doi = {10.48550/arXiv.1705.10877},
archivePrefix = {arXiv},
       eprint = {1705.10877},
 primaryClass = {astro-ph.GA},
       adsurl = {https://ui.adsabs.harvard.edu/abs/2017RMxAA..53..385F}
}

@ARTICLE{Xu2022,
       author = {{Xu}, Yi and {Ouchi}, Masami and {Rauch}, Michael and {Nakajima}, Kimihiko and {Harikane}, Yuichi and {Sugahara}, Yuma and {Komiyama}, Yutaka and {Kusakabe}, Haruka and {Fujimoto}, Seiji and {Isobe}, Yuki and {Kim}, Ji Hoon and {Ono}, Yoshiaki and {Zahedy}, Fakhri S.},
        title = "{EMPRESS. VI. Outflows Investigated in Low-mass Galaxies with M $_{{\ensuremath{*}}}$ = {}10$^{4}$-{}10$^{7}$ M $_{☉}$: Weak Feedback in Low-mass Galaxies?}",
      journal = {\apj},
         year = 2022,
        month = apr,
       volume = {929},
       number = {2},
          eid = {134},
        pages = {134},
          doi = {10.3847/1538-4357/ac5e32},
archivePrefix = {arXiv},
       eprint = {2112.08045},
 primaryClass = {astro-ph.GA},
       adsurl = {https://ui.adsabs.harvard.edu/abs/2022ApJ...929..134X}
}

@ARTICLE{RiveraThorsen2024,
       author = {{Rivera-Thorsen}, T. Emil and {Chisholm}, J. and {Welch}, B. and {Rigby}, J.~R. and {Hutchison}, T. and {Florian}, M. and {Sharon}, K. and {Choe}, S. and {Dahle}, H. and {Bayliss}, M.~B. and {Khullar}, G. and {Gladders}, M. and {Hayes}, M. and {Adamo}, A. and {Owens}, M.~R. and {Kim}, K.},
        title = "{The Sunburst Arc with JWST: I. Detection of Wolf-Rayet stars injecting nitrogen into a low-metallicity, z = 2.37 proto-globular cluster leaking ionizing photons}",
      journal = {\aap},
         year = 2024,
        month = oct,
       volume = {690},
          eid = {A269},
        pages = {A269},
          doi = {10.1051/0004-6361/202450359},
archivePrefix = {arXiv},
       eprint = {2404.08884},
 primaryClass = {astro-ph.GA},
       adsurl = {https://ui.adsabs.harvard.edu/abs/2024A&A...690A.269R}
}

@ARTICLE{Hamann1993,
       author = {{Hamann}, Fred and {Ferland}, Gary},
        title = "{The Chemical Evolution of QSOs and the Implications for Cosmology and Galaxy Formation}",
      journal = {\apj},
         year = 1993,
        month = nov,
       volume = {418},
        pages = {11},
          doi = {10.1086/173366},
       adsurl = {https://ui.adsabs.harvard.edu/abs/1993ApJ...418...11H}
}

@ARTICLE{Uebler2023_GS3073,
       author = {{{\"U}bler}, Hannah and {Maiolino}, Roberto and {Curtis-Lake}, Emma and {P{\'e}rez-Gonz{\'a}lez}, Pablo G. and {Curti}, Mirko and {Perna}, Michele and {Arribas}, Santiago and {Charlot}, St{\'e}phane and {Marshall}, Madeline A. and {D'Eugenio}, Francesco and {Scholtz}, Jan and {Bunker}, Andrew and {Carniani}, Stefano and {Ferruit}, Pierre and {Jakobsen}, Peter and {Rix}, Hans-Walter and {Rodr{\'\i}guez Del Pino}, Bruno and {Willott}, Chris J. and {Boeker}, Torsten and {Cresci}, Giovanni and {Jones}, Gareth C. and {Kumari}, Nimisha and {Rawle}, Tim},
        title = "{GA-NIFS: A massive black hole in a low-metallicity AGN at z {\ensuremath{\sim}} 5.55 revealed by JWST/NIRSpec IFS}",
      journal = {\aap},
         year = 2023,
        month = sep,
       volume = {677},
          eid = {A145},
        pages = {A145},
          doi = {10.1051/0004-6361/202346137},
archivePrefix = {arXiv},
       eprint = {2302.06647},
 primaryClass = {astro-ph.GA},
       adsurl = {https://ui.adsabs.harvard.edu/abs/2023A&A...677A.145U}
}

@ARTICLE{Ji2024_GS3073,
       author = {{Ji}, Xihan and {{\"U}bler}, Hannah and {Maiolino}, Roberto and {D'Eugenio}, Francesco and {Arribas}, Santiago and {Bunker}, Andrew J. and {Charlot}, St{\'e}phane and {Perna}, Michele and {Rodr{\'\i}guez Del Pino}, Bruno and {B{\"o}ker}, Torsten and {Cresci}, Giovanni and {Curti}, Mirko and {Kumari}, Nimisha and {Lamperti}, Isabella},
        title = "{GA-NIFS: an extremely nitrogen-loud and chemically stratified galaxy at z   5.55}",
      journal = {\mnras},
         year = 2024,
        month = nov,
       volume = {535},
       number = {1},
        pages = {881-908},
          doi = {10.1093/mnras/stae2375},
archivePrefix = {arXiv},
       eprint = {2404.04148},
 primaryClass = {astro-ph.GA},
       adsurl = {https://ui.adsabs.harvard.edu/abs/2024MNRAS.535..881J}
}

@ARTICLE{Curti2025,
       author = {{Curti}, Mirko and {Witstok}, Joris and {Jakobsen}, Peter and {Kobayashi}, Chiaki and {Curtis-Lake}, Emma and {Hainline}, Kevin and {Ji}, Xihan and {D'Eugenio}, Francesco and {Chevallard}, Jacopo and {Maiolino}, Roberto and {Scholtz}, Jan and {Carniani}, Stefano and {Arribas}, Santiago and {Baker}, William M. and {Bhatawdekar}, Rachana and {Boyett}, Kristan and {Bunker}, Andrew J. and {Cameron}, Alex and {Cargile}, Phillip A. and {Charlot}, St{\'e}phane and {Eisenstein}, Daniel J. and {Ji}, Zhiyuan and {Johnson}, Benjamin D. and {Kumari}, Nimisha and {Maseda}, Michael V. and {Robertson}, Brant and {Silcock}, Maddie S. and {Tacchella}, Sandro and {{\"U}bler}, Hannah and {Venturi}, Giacomo and {Williams}, Christina C. and {Willmer}, Christopher N.~A. and {Willott}, Chris},
        title = "{JADES: The star formation and chemical enrichment history of a luminous galaxy at z {\ensuremath{\sim}} 9.43 probed by ultra-deep JWST/NIRSpec spectroscopy}",
      journal = {\aap},
         year = 2025,
        month = may,
       volume = {697},
          eid = {A89},
        pages = {A89},
          doi = {10.1051/0004-6361/202451410},
archivePrefix = {arXiv},
       eprint = {2407.02575},
 primaryClass = {astro-ph.GA},
       adsurl = {https://ui.adsabs.harvard.edu/abs/2025A&A...697A..89C}
}

@ARTICLE{Adamo2015,
       author = {{Adamo}, A. and {Kruijssen}, J.~M.~D. and {Bastian}, N. and {Silva-Villa}, E. and {Ryon}, J.},
        title = "{Probing the role of the galactic environment in the formation of stellar clusters, using M83 as a test bench}",
      journal = {\mnras},
         year = 2015,
        month = sep,
       volume = {452},
       number = {1},
        pages = {246-260},
          doi = {10.1093/mnras/stv1203},
archivePrefix = {arXiv},
       eprint = {1505.07475},
 primaryClass = {astro-ph.GA},
       adsurl = {https://ui.adsabs.harvard.edu/abs/2015MNRAS.452..246A}
}

@ARTICLE{Pilyugin2012,
       author = {{Pilyugin}, L.~S. and {Grebel}, E.~K. and {Mattsson}, L.},
        title = "{'Counterpart' method for abundance determinations in H II regions}",
      journal = {\mnras},
         year = 2012,
        month = aug,
       volume = {424},
       number = {3},
        pages = {2316-2329},
          doi = {10.1111/j.1365-2966.2012.21398.x},
archivePrefix = {arXiv},
       eprint = {1205.5716},
 primaryClass = {astro-ph.CO},
       adsurl = {https://ui.adsabs.harvard.edu/abs/2012MNRAS.424.2316P}
}

@ARTICLE{Nahar1999,
       author = {{Nahar}, Sultana N.},
        title = "{Electron-Ion Recombination Rate Coefficients, Photoionization Cross Sections, and Ionization Fractions for Astrophysically Abundant Elements. II. Oxygen Ions}",
      journal = {\apjs},
         year = 1999,
        month = jan,
       volume = {120},
       number = {1},
        pages = {131-145},
          doi = {10.1086/313173},
       adsurl = {https://ui.adsabs.harvard.edu/abs/1999ApJS..120..131N}
}

@ARTICLE{Watanabe2026,
       author = {{Watanabe}, Kuria and {Ouchi}, Masami and {Nakajima}, Kimihiko and {Tominaga}, Nozomu and {Harikane}, Yuichi and {Ishigaki}, Miho N. and {Isobe}, Yuki and {Nakane}, Minami and {Nishigaki}, Moka and {Nomoto}, Ken'ichi and {Ono}, Yoshiaki and {Onodera}, Masato and {Suzuki}, Akihiro and {Takahashi}, Koh and {Takeda}, Yui and {Yanagisawa}, Hiroto},
        title = "{Chemical Abundance Ratios of Nitrogen-rich Galaxies Identified at z \raisebox{-0.5ex}\textasciitilde 6-12: Observational Demographics and Models}",
      journal = {\apj},
         year = 2026,
        month = sep,
       volume = {1008},
       number = {2},
          eid = {188},
        pages = {188},
          doi = {10.3847/1538-4357/ae91d9},
archivePrefix = {arXiv},
       eprint = {2603.21570},
 primaryClass = {astro-ph.GA},
       adsurl = {https://ui.adsabs.harvard.edu/abs/2026ApJ..1008..188W}
}

@ARTICLE{Kroupa2001,
       author = {{Kroupa}, Pavel},
        title = "{On the variation of the initial mass function}",
      journal = {\mnras},
         year = 2001,
        month = apr,
       volume = {322},
       number = {2},
        pages = {231-246},
          doi = {10.1046/j.1365-8711.2001.04022.x},
archivePrefix = {arXiv},
       eprint = {astro-ph/0009005},
 primaryClass = {astro-ph},
       adsurl = {https://ui.adsabs.harvard.edu/abs/2001MNRAS.322..231K}
}

@ARTICLE{LimongiChieffi2018,
       author = {{Limongi}, Marco and {Chieffi}, Alessandro},
        title = "{Presupernova Evolution and Explosive Nucleosynthesis of Rotating Massive Stars in the Metallicity Range -3 {\ensuremath{\leq}} [Fe/H] {\ensuremath{\leq}} 0}",
      journal = {\apjs},
         year = 2018,
        month = jul,
       volume = {237},
       number = {1},
          eid = {13},
        pages = {13},
          doi = {10.3847/1538-4365/aacb24},
archivePrefix = {arXiv},
       eprint = {1805.09640},
 primaryClass = {astro-ph.SR},
       adsurl = {https://ui.adsabs.harvard.edu/abs/2018ApJS..237...13L}
}

@ARTICLE{KarakasLugaro2016,
       author = {{Karakas}, Amanda I. and {Lugaro}, Maria},
        title = "{Stellar Yields from Metal-rich Asymptotic Giant Branch Models}",
      journal = {\apj},
         year = 2016,
        month = jul,
       volume = {825},
       number = {1},
          eid = {26},
        pages = {26},
          doi = {10.3847/0004-637X/825/1/26},
archivePrefix = {arXiv},
       eprint = {1604.02178},
 primaryClass = {astro-ph.SR},
       adsurl = {https://ui.adsabs.harvard.edu/abs/2016ApJ...825...26K}
}

@ARTICLE{Doherty2014_iii,
       author = {{Doherty}, Carolyn L. and {Gil-Pons}, Pilar and {Lau}, Herbert H.~B. and {Lattanzio}, John C. and {Siess}, Lionel and {Campbell}, Simon W.},
        title = "{Super and massive AGB stars - III. Nucleosynthesis in metal-poor and very metal-poor stars - Z = 0.001 and 0.0001}",
      journal = {\mnras},
         year = 2014,
        month = jun,
       volume = {441},
       number = {1},
        pages = {582-598},
          doi = {10.1093/mnras/stu571},
archivePrefix = {arXiv},
       eprint = {1403.5054},
 primaryClass = {astro-ph.SR},
       adsurl = {https://ui.adsabs.harvard.edu/abs/2014MNRAS.441..582D}
}

@ARTICLE{Croxall2016,
       author = {{Croxall}, Kevin V. and {Pogge}, Richard W. and {Berg}, Danielle A. and {Skillman}, Evan D. and {Moustakas}, John},
        title = "{CHAOS III: Gas-phase Abundances in NGC 5457}",
      journal = {\apj},
         year = 2016,
        month = oct,
       volume = {830},
       number = {1},
          eid = {4},
        pages = {4},
          doi = {10.3847/0004-637X/830/1/4},
archivePrefix = {arXiv},
       eprint = {1605.01612},
 primaryClass = {astro-ph.GA},
       adsurl = {https://ui.adsabs.harvard.edu/abs/2016ApJ...830....4C}
}

@ARTICLE{Berg2020,
       author = {{Berg}, Danielle A. and {Pogge}, Richard W. and {Skillman}, Evan D. and {Croxall}, Kevin V. and {Moustakas}, John and {Rogers}, Noah S.~J. and {Sun}, Jiayi},
        title = "{CHAOS IV: Gas-phase Abundance Trends from the First Four CHAOS Galaxies}",
      journal = {\apj},
         year = 2020,
        month = apr,
       volume = {893},
       number = {2},
          eid = {96},
        pages = {96},
          doi = {10.3847/1538-4357/ab7eab},
archivePrefix = {arXiv},
       eprint = {2001.05002},
 primaryClass = {astro-ph.GA},
       adsurl = {https://ui.adsabs.harvard.edu/abs/2020ApJ...893...96B}
}

@ARTICLE{Izotov2006,
       author = {{Izotov}, Y.~I. and {Stasi{\'n}ska}, G. and {Meynet}, G. and {Guseva}, N.~G. and {Thuan}, T.~X.},
        title = "{The chemical composition of metal-poor emission-line galaxies in the Data Release 3 of the Sloan Digital Sky Survey}",
      journal = {\aap},
         year = 2006,
        month = mar,
       volume = {448},
       number = {3},
        pages = {955-970},
          doi = {10.1051/0004-6361:20053763},
archivePrefix = {arXiv},
       eprint = {astro-ph/0511644},
 primaryClass = {astro-ph},
       adsurl = {https://ui.adsabs.harvard.edu/abs/2006A&A...448..955I}
}

@ARTICLE{Maoz2014,
       author = {{Maoz}, Dan and {Mannucci}, Filippo and {Nelemans}, Gijs},
        title = "{Observational Clues to the Progenitors of Type Ia Supernovae}",
      journal = {\araa},
         year = 2014,
        month = aug,
       volume = {52},
        pages = {107-170},
          doi = {10.1146/annurev-astro-082812-141031},
archivePrefix = {arXiv},
       eprint = {1312.0628},
 primaryClass = {astro-ph.CO},
       adsurl = {https://ui.adsabs.harvard.edu/abs/2014ARA&A..52..107M}
}

@ARTICLE{Martinez2025,
       author = {{Martinez}, Zorayda and {Berg}, Danielle A. and {James}, Bethan L. and {Arellano-C{\'o}rdova}, Karla Z. and {Stark}, Daniel P. and {Senchyna}, Peter and {Skillman}, Evan D. and {Rogers}, Noah S.~J. and {Chisholm}, John},
        title = "{Under Pressure: Decoding the Effect of High Densities on Derived Nebular Properties}",
      journal = {\apj},
         year = 2025,
        month = dec,
       volume = {995},
       number = {2},
          eid = {204},
        pages = {204},
          doi = {10.3847/1538-4357/ae17c6},
archivePrefix = {arXiv},
       eprint = {2510.21960},
 primaryClass = {astro-ph.GA},
       adsurl = {https://ui.adsabs.harvard.edu/abs/2025ApJ...995..204M}
}

@ARTICLE{Labbe2024,
       author = {{Labbe}, Ivo and {Greene}, Jenny E. and {Matthee}, Jorryt and {Treiber}, Helena and {Kokorev}, Vasily and {Miller}, Tim B. and {Kramarenko}, Ivan and {Setton}, David J. and {Ma}, Yilun and {Goulding}, Andy D. and {Bezanson}, Rachel and {Naidu}, Rohan P. and {Williams}, Christina C. and {Atek}, Hakim and {Brammer}, Gabriel and {Cutler}, Sam E. and {Chemerynska}, Iryna and {Cloonan}, Aidan P. and {Dayal}, Pratika and {de Graaff}, Anna and {Fudamoto}, Yoshinobu and {Fujimoto}, Seiji and {Furtak}, Lukas J. and {Glazebrook}, Karl and {Heintz}, Kasper E. and {Leja}, Joel and {Marchesini}, Danilo and {Nanayakkara}, Themiya and {Nelson}, Erica J. and {Oesch}, Pascal A. and {Pan}, Richard and {Price}, Sedona H. and {Shivaei}, Irene and {Sobral}, David and {Suess}, Katherine A. and {van Dokkum}, Pieter and {Wang}, Bingjie and {Weaver}, John R. and {Whitaker}, Katherine E. and {Zitrin}, Adi},
        title = "{An unambiguous AGN and a Balmer break in an Ultraluminous Little Red Dot at z=4.47 from Ultradeep UNCOVER and All the Little Things Spectroscopy}",
      journal = {arXiv e-prints},
         year = 2024,
        month = dec,
          eid = {arXiv:2412.04557},
        pages = {arXiv:2412.04557},
          doi = {10.48550/arXiv.2412.04557},
archivePrefix = {arXiv},
       eprint = {2412.04557},
 primaryClass = {astro-ph.GA},
       adsurl = {https://ui.adsabs.harvard.edu/abs/2024arXiv241204557L}
}

@ARTICLE{Ebihara2026,
       author = {{Ebihara}, S. and {Fujii}, M. and {Saitoh}, T. and {Hirai}, Y. and {Umeda}, H. and {Isobe}, Y. and {Nagele}, C.},
        title = "{Nitrogen enhancement of GN-z11 by metal pollution from supermassive stars}",
      journal = {\aap},
         year = 2026,
        month = jun,
       volume = {710},
          eid = {A239},
        pages = {A239},
          doi = {10.1051/0004-6361/202558704},
       adsurl = {https://ui.adsabs.harvard.edu/abs/2026A&A...710A.239E}
}

@ARTICLE{RobertsBorsani2024,
       author = {{Roberts-Borsani}, Guido and {Treu}, Tommaso and {Shapley}, Alice and {Fontana}, Adriano and {Pentericci}, Laura and {Castellano}, Marco and {Morishita}, Takahiro and {Bergamini}, Pietro and {Rosati}, Piero},
        title = "{Between the Extremes: A JWST Spectroscopic Benchmark for High-redshift Galaxies Using {\ensuremath{\sim}}500 Confirmed Sources at z {\ensuremath{\geq}} 5}",
      journal = {\apj},
         year = 2024,
        month = dec,
       volume = {976},
       number = {2},
          eid = {193},
        pages = {193},
          doi = {10.3847/1538-4357/ad85d3},
archivePrefix = {arXiv},
       eprint = {2403.07103},
 primaryClass = {astro-ph.GA},
       adsurl = {https://ui.adsabs.harvard.edu/abs/2024ApJ...976..193R}
}

@ARTICLE{McClymont2025_dust,
       author = {{McClymont}, William and {Tacchella}, Sandro and {D'Eugenio}, Francesco and {Witten}, Callum and {Ji}, Xihan and {Smith}, Aaron and {Maiolino}, Roberto and {Arribas}, Santiago and {Scholtz}, Jan and {Simmonds}, Charlotte and {Witstok}, Joris},
        title = "{The density-bounded twilight of starbursts in the early Universe}",
      journal = {\mnras},
         year = 2025,
        month = jun,
       volume = {540},
       number = {1},
        pages = {190-203},
          doi = {10.1093/mnras/staf745},
archivePrefix = {arXiv},
       eprint = {2405.15859},
 primaryClass = {astro-ph.GA},
       adsurl = {https://ui.adsabs.harvard.edu/abs/2025MNRAS.540..190M}
}

@ARTICLE{Solimano2025,
       author = {{Solimano}, M. and {Gonz{\'a}lez-L{\'o}pez}, J. and {Aravena}, M. and {Alcalde Pampliega}, B. and {Assef}, R.~J. and {B{\'e}thermin}, M. and {Boquien}, M. and {Bovino}, S. and {Casey}, C.~M. and {Cassata}, P. and {da Cunha}, E. and {Davies}, R.~L. and {De Looze}, I. and {Ding}, X. and {D{\'\i}az-Santos}, T. and {Faisst}, A.~L. and {Ferrara}, A. and {Fisher}, D.~B. and {F{\"o}rster-Schreiber}, N.~M. and {Fujimoto}, S. and {Ginolfi}, M. and {Gruppioni}, C. and {Guaita}, L. and {Hathi}, N. and {Herrera-Camus}, R. and {Ibar}, E. and {Inami}, H. and {Jones}, G.~C. and {Koekemoer}, A.~M. and {Lee}, L.~L. and {Li}, J. and {Liu}, D. and {Liu}, Z. and {Molina}, J. and {Ogle}, P. and {Posses}, A.~C. and {Pozzi}, F. and {Rela{\~n}o}, M. and {Riechers}, D.~A. and {Romano}, M. and {Spilker}, J. and {Sulzenauer}, N. and {Telikova}, K. and {Vallini}, L. and {Vasan}, K.~G.~C. and {Veilleux}, S. and {Vergani}, D. and {Villanueva}, V. and {Wang}, W. and {Yan}, L. and {Zamorani}, G.},
        title = "{A hidden active galactic nucleus powering bright [O III] nebulae in a protocluster at z = 4.5 revealed by JWST}",
      journal = {\aap},
         year = 2025,
        month = jan,
       volume = {693},
          eid = {A70},
        pages = {A70},
          doi = {10.1051/0004-6361/202451551},
archivePrefix = {arXiv},
       eprint = {2407.13020},
 primaryClass = {astro-ph.GA},
       adsurl = {https://ui.adsabs.harvard.edu/abs/2025A&A...693A..70S}
}

@ARTICLE{Shivaei2026,
       author = {{Shivaei}, Irene and {Naidu}, Rohan P. and {Rodr{\'\i}guez Montero}, Francisco and {Matsumoto}, Kosei and {Leja}, Joel and {Matthee}, Jorryt and {Johnson}, Benjamin D. and {Oesch}, Pascal A. and {Chevallard}, Jacopo and {Adamo}, Angela and {Bodansky}, Sarah and {Bunker}, Andrew J. and {Covelo Paz}, Alba and {Di Cesare}, Claudia and {Egami}, Eiichi and {Furtak}, Lukas J. and {Heintz}, Kasper E. and {Kramarenko}, Ivan and {Meyer}, Romain A. and {Reddy}, Naveen A. and {Rinaldi}, Pierluigi and {Tacchella}, Sandro and {Torralba}, Alberto and {Witstok}, Joris and {Wozniak}, Michael A. and {Xiao}, Mengyuan},
        title = "{Diversity and evolution of dust-attenuation curves from redshift z {\ensuremath{\sim}} 1 to 9}",
      journal = {\aap},
         year = 2026,
        month = aug,
       volume = {712},
          eid = {A132},
        pages = {A132},
          doi = {10.1051/0004-6361/202557064},
archivePrefix = {arXiv},
       eprint = {2509.01795},
 primaryClass = {astro-ph.GA},
       adsurl = {https://ui.adsabs.harvard.edu/abs/2026A&A...712A.132S}
}

@ARTICLE{Calzetti2000,
       author = {{Calzetti}, Daniela and {Armus}, Lee and {Bohlin}, Ralph C. and {Kinney}, Anne L. and {Koornneef}, Jan and {Storchi-Bergmann}, Thaisa},
        title = "{The Dust Content and Opacity of Actively Star-forming Galaxies}",
      journal = {\apj},
         year = 2000,
        month = apr,
       volume = {533},
       number = {2},
        pages = {682-695},
          doi = {10.1086/308692},
archivePrefix = {arXiv},
       eprint = {astro-ph/9911459},
 primaryClass = {astro-ph},
       adsurl = {https://ui.adsabs.harvard.edu/abs/2000ApJ...533..682C}
}

@ARTICLE{Chiappini1997,
       author = {{Chiappini}, C. and {Matteucci}, F. and {Gratton}, R.},
        title = "{The Chemical Evolution of the Galaxy: The Two-Infall Model}",
      journal = {\apj},
         year = 1997,
        month = mar,
       volume = {477},
       number = {2},
        pages = {765-780},
          doi = {10.1086/303726},
archivePrefix = {arXiv},
       eprint = {astro-ph/9609199},
 primaryClass = {astro-ph},
       adsurl = {https://ui.adsabs.harvard.edu/abs/1997ApJ...477..765C}
}

@ARTICLE{Adamo2011,
       author = {{Adamo}, A. and {{\"O}stlin}, G. and {Zackrisson}, E.},
        title = "{Probing cluster formation under extreme conditions: massive star clusters in blue compact galaxies}",
      journal = {\mnras},
         year = 2011,
        month = nov,
       volume = {417},
       number = {3},
        pages = {1904-1912},
          doi = {10.1111/j.1365-2966.2011.19377.x},
archivePrefix = {arXiv},
       eprint = {1107.0725},
 primaryClass = {astro-ph.CO},
       adsurl = {https://ui.adsabs.harvard.edu/abs/2011MNRAS.417.1904A}
}

@ARTICLE{deGraaff2025_rubies,
       author = {{de Graaff}, Anna and {Brammer}, Gabriel and {Weibel}, Andrea and {Lewis}, Zach and {Maseda}, Michael V. and {Oesch}, Pascal A. and {Bezanson}, Rachel and {Boogaard}, Leindert A. and {Cleri}, Nikko J. and {Cooper}, Olivia R. and {Gottumukkala}, Rashmi and {Greene}, Jenny E. and {Hirschmann}, Michaela and {Hviding}, Raphael E. and {Katz}, Harley and {Labb{\'e}}, Ivo and {Leja}, Joel and {Matthee}, Jorryt and {McConachie}, Ian and {Miller}, Tim B. and {Naidu}, Rohan P. and {Price}, Sedona H. and {Rix}, Hans-Walter and {Setton}, David J. and {Suess}, Katherine A. and {Wang}, Bingjie and {Whitaker}, Katherine E. and {Williams}, Christina C.},
        title = "{RUBIES: A complete census of the bright and red distant Universe with JWST/NIRSpec}",
      journal = {\aap},
         year = 2025,
        month = may,
       volume = {697},
          eid = {A189},
        pages = {A189},
          doi = {10.1051/0004-6361/202452186},
archivePrefix = {arXiv},
       eprint = {2409.05948},
 primaryClass = {astro-ph.GA},
       adsurl = {https://ui.adsabs.harvard.edu/abs/2025A&A...697A.189D}
}

@ARTICLE{Eisenstein2026_jades,
       author = {{Eisenstein}, Daniel J. and {Willott}, Chris and {Alberts}, Stacey and {Arribas}, Santiago and {Bonaventura}, Nina and {Bunker}, Andrew J. and {Cameron}, Alex J. and {Carniani}, Stefano and {Charlot}, Stephane and {Curtis-Lake}, Emma and {D'Eugenio}, Francesco and {Ferruit}, Pierre and {Giardino}, Giovanna and {Hainline}, Kevin and {Hausen}, Ryan and {Jakobsen}, Peter and {Johnson}, Benjamin D. and {Maiolino}, Roberto and {Rauscher}, Bernard J. and {Rieke}, Marcia and {Rieke}, George and {Rix}, Hans-Walter and {Robertson}, Brant and {Stark}, Daniel P. and {Tacchella}, Sandro and {Williams}, Christina C. and {Willmer}, Christopher N.~A. and {Baker}, William M. and {Baum}, Stefi and {Bhatawdekar}, Rachana and {Boyett}, Kristan and {Chen}, Zuyi and {Chevallard}, Jacopo and {Circosta}, Chiara and {Curti}, Mirko and {Danhaive}, A. Lola and {DeCoursey}, Christa and {Endsley}, Ryan and {de Graaff}, Anna and {Dressler}, Alan and {Egami}, Eiichi and {Helton}, Jakob M. and {Hviding}, Raphael E. and {Ji}, Zhiyuan and {Jones}, Gareth C. and {Kumari}, Nimisha and {L{\"u}tzgendorf}, Nora and {Laseter}, Isaac and {Looser}, Tobias J. and {Lyu}, Jianwei and {Maseda}, Michael V. and {Nelson}, Erica and {Parlanti}, Eleonora and {Perna}, Michele and {Pusk{\'a}s}, D{\'a}vid and {Rawle}, Tim and {Rodr{\'\i}guez Del Pino}, Bruno and {Rujopakarn}, Wiphu and {Sandles}, Lester and {Saxena}, Aayush and {Scholtz}, Jan and {Sharpe}, Katherine and {Shivaei}, Irene and {Silcock}, Maddie S. and {Simmonds}, Charlotte and {Skarbinski}, Maya and {Smit}, Renske and {Stone}, Meredith and {Suess}, Katherine A. and {Sun}, Fengwu and {Tang}, Mengtao and {Topping}, Michael W. and {{\"U}bler}, Hannah and {Villanueva}, Natalia C. and {Wallace}, Imaan E.~B. and {Whitler}, Lily and {Witstok}, Joris and {Woodrum}, Charity},
        title = "{Overview of the JWST Advanced Deep Extragalactic Survey (JADES)}",
      journal = {\apjs},
         year = 2026,
        month = mar,
       volume = {283},
       number = {1},
          eid = {6},
        pages = {6},
          doi = {10.3847/1538-4365/ae3163},
archivePrefix = {arXiv},
       eprint = {2306.02465},
 primaryClass = {astro-ph.GA},
       adsurl = {https://ui.adsabs.harvard.edu/abs/2026ApJS..283....6E}
}

@ARTICLE{Scholtz2026_jades,
       author = {{Scholtz}, J. and {Carniani}, S. and {Parlanti}, E. and {D'Eugenio}, F. and {Curtis-Lake}, E. and {Jakobsen}, P. and {Bunker}, A.~J. and {Cameron}, A.~J. and {Arribas}, S. and {Baker}, W.~M. and {Charlot}, S. and {Chevellard}, J. and {Circosta}, C. and {Curti}, M. and {Duan}, Q. and {Eisenstein}, D.~J. and {Hainline}, K. and {Ji}, Z. and {Johnson}, B.~D. and {Jones}, G.~C. and {Kumari}, N. and {Maiolino}, R. and {Maseda}, M.~V. and {Perna}, M. and {P{\'e}rez-Gonz{\'a}lez}, P.~G. and {Rawle}, T. and {Rieke}, M. and {Rinaldi}, P. and {Robertson}, B. and {Saxena}, A. and {Shivaei}, I. and {Silcock}, M.~S. and {Sun}, Y. and {Rodr{\'\i}guez Del Pino}, B. and {Tacchella}, S. and {{\"U}bler}, H. and {Venturi}, G. and {Williams}, C.~C. and {Willmer}, C.~N.~A. and {Willott}, C. and {Witstok}, J.},
        title = "{JADES Data Release 4 ─ Paper II. Data reduction, analysis, and emission-line fluxes of the complete spectroscopic sample}",
      journal = {\mnras},
         year = 2026,
        month = jul,
       volume = {549},
       number = {4},
          eid = {stag939},
        pages = {stag939},
          doi = {10.1093/mnras/stag939},
archivePrefix = {arXiv},
       eprint = {2510.01034},
 primaryClass = {astro-ph.GA},
       adsurl = {https://ui.adsabs.harvard.edu/abs/2026MNRAS.549ag939S}
}

@ARTICLE{ForemanMackey2013_emcee,
       author = {{Foreman-Mackey}, Daniel and {Hogg}, David W. and {Lang}, Dustin and {Goodman}, Jonathan},
        title = "{emcee: The MCMC Hammer}",
      journal = {\pasp},
         year = 2013,
        month = mar,
       volume = {125},
       number = {925},
        pages = {306},
          doi = {10.1086/670067},
archivePrefix = {arXiv},
       eprint = {1202.3665},
 primaryClass = {astro-ph.IM},
       adsurl = {https://ui.adsabs.harvard.edu/abs/2013PASP..125..306F}
}

@ARTICLE{Isobe2025,
       author = {{Isobe}, Yuki and {Maiolino}, Roberto and {D'Eugenio}, Francesco and {Curti}, Mirko and {Ji}, Xihan and {Juod{\v{z}}balis}, Ignas and {Scholtz}, Jan and {Feltre}, Anne and {Charlot}, St{\'e}phane and {{\"U}bler}, Hannah and {J. Bunker}, Andrew and {Carniani}, Stefano and {Curtis-Lake}, Emma and {Ji}, Zhiyuan and {Kumari}, Nimisha and {Rinaldi}, Pierluigi and {Robertson}, Brant and {Willott}, Chris and {Witstok}, Joris},
        title = "{JADES: nitrogen enhancement in high-redshift broad-line active galactic nuclei}",
      journal = {\mnras},
         year = 2025,
        month = jul,
       volume = {541},
       number = {1},
        pages = {L71-L79},
          doi = {10.1093/mnrasl/slaf056},
archivePrefix = {arXiv},
       eprint = {2502.12091},
 primaryClass = {astro-ph.GA},
       adsurl = {https://ui.adsabs.harvard.edu/abs/2025MNRAS.541L..71I}
}

@ARTICLE{Hayes2025,
       author = {{Hayes}, Matthew J. and {Saldana-Lopez}, Alberto and {Citro}, Annalisa and {James}, Bethan L. and {Mingozzi}, Matilde and {Scarlata}, Claudia and {Martinez}, Zorayda and {Berg}, Danielle A.},
        title = "{On the Average Ultraviolet Emission-line Spectra of High-redshift Galaxies: Hot and Cold, Carbon-poor, Nitrogen Modest, and Oozing Ionizing Photons}",
      journal = {\apj},
         year = 2025,
        month = mar,
       volume = {982},
       number = {1},
          eid = {14},
        pages = {14},
          doi = {10.3847/1538-4357/adaea1},
archivePrefix = {arXiv},
       eprint = {2411.09262},
 primaryClass = {astro-ph.GA},
       adsurl = {https://ui.adsabs.harvard.edu/abs/2025ApJ...982...14H}
}

@ARTICLE{Bhattacharya2026,
       author = {{Bhattacharya}, Souradeep and {Kobayashi}, Chiaki},
        title = "{The origin of extreme N-emitters in star-forming galaxies at z < 0.5 with DESI DR1}",
      journal = {\mnras},
         year = 2026,
        month = may,
       volume = {548},
       number = {1},
          eid = {stag545},
        pages = {stag545},
          doi = {10.1093/mnras/stag545},
archivePrefix = {arXiv},
       eprint = {2508.11998},
 primaryClass = {astro-ph.GA},
       adsurl = {https://ui.adsabs.harvard.edu/abs/2026MNRAS.548ag545B}
}

@ARTICLE{Carniani2024,
       author = {{Carniani}, Stefano and {Venturi}, Giacomo and {Parlanti}, Eleonora and {de Graaff}, Anna and {Maiolino}, Roberto and {Arribas}, Santiago and {Bonaventura}, Nina and {Boyett}, Kristan and {Bunker}, Andrew J. and {Cameron}, Alex J. and {Charlot}, Stephane and {Chevallard}, Jacopo and {Curti}, Mirko and {Curtis-Lake}, Emma and {Eisenstein}, Daniel J. and {Giardino}, Giovanna and {Hausen}, Ryan and {Kumari}, Nimisha and {Maseda}, Michael V. and {Nelson}, Erica and {Perna}, Michele and {Rix}, Hans-Walter and {Robertson}, Brant and {Del Pino}, Bruno Rodr{\'\i}guez and {Sandles}, Lester and {Scholtz}, Jan and {Simmonds}, Charlotte and {Smit}, Renske and {Tacchella}, Sandro and {{\"U}bler}, Hannah and {Williams}, Christina C. and {Willott}, Chris and {Witstok}, Joris},
        title = "{JADES: The incidence rate and properties of galactic outflows in low-mass galaxies across 3 < z < 9}",
      journal = {\aap},
         year = 2024,
        month = may,
       volume = {685},
          eid = {A99},
        pages = {A99},
          doi = {10.1051/0004-6361/202347230},
archivePrefix = {arXiv},
       eprint = {2306.11801},
 primaryClass = {astro-ph.GA},
       adsurl = {https://ui.adsabs.harvard.edu/abs/2024A&A...685A..99C}
}

@ARTICLE{KobayashiFerrara2024,
       author = {{Kobayashi}, Chiaki and {Ferrara}, Andrea},
        title = "{Rapid Chemical Enrichment by Intermittent Star Formation in GN-z11}",
      journal = {\apjl},
         year = 2024,
        month = feb,
       volume = {962},
       number = {1},
          eid = {L6},
        pages = {L6},
          doi = {10.3847/2041-8213/ad1de1},
archivePrefix = {arXiv},
       eprint = {2308.15583},
 primaryClass = {astro-ph.GA},
       adsurl = {https://ui.adsabs.harvard.edu/abs/2024ApJ...962L...6K}
}

@ARTICLE{Berg2026,
       author = {{Berg}, Danielle A. and {Naidu}, Rohan P. and {Chisholm}, John and {Atek}, Hakim and {Fujimoto}, Seiji and {Kokorev}, Vasily and {Furtak}, Lukas J. and {Kobayashi}, Chiaki and {Schaerer}, Daniel and {Adamo}, Angela and {Fei}, Qinyue and {Korber}, Damien and {Matthee}, Jorryt and {Marques-Chaves}, Rui and {Martinez}, Zorayda and {McQuinn}, Kristen. B.~W. and {Mu{\~n}oz}, Julian B. and {Oesch}, Pascal A. and {Saldana-Lopez}, Alberto and {Stark}, Daniel P. and {Stephenson}, Mabel G. and {Hsiao}, Tiger Yu-Yang},
        title = "{A Fleeting GLIMPSE of N/O Enrichment at Cosmic Dawn: Evidence for Wolf Rayet N Stars in a z = 6.1 Galaxy}",
      journal = {\apj},
         year = 2026,
        month = jun,
       volume = {1003},
       number = {2},
          eid = {112},
        pages = {112},
          doi = {10.3847/1538-4357/ae5e4c},
archivePrefix = {arXiv},
       eprint = {2511.13591},
 primaryClass = {astro-ph.GA},
       adsurl = {https://ui.adsabs.harvard.edu/abs/2026ApJ..1003..112B}
}

@ARTICLE{Nikopoulos2026,
       author = {{Nikopoulos}, G.~P. and {Watson}, D. and {Pollock}, C.~L. and {Sneppen}, A. and {Heintz}, K.~E. and {Witstok}, J. and {Brammer}, G.},
        title = "{The metallicities of little red dot host galaxies: LRDs are metal poor, but not pristine}",
      journal = {arXiv e-prints},
         year = 2026,
        month = jun,
          eid = {arXiv:2606.31515},
        pages = {arXiv:2606.31515},
          doi = {10.48550/arXiv.2606.31515},
archivePrefix = {arXiv},
       eprint = {2606.31515},
 primaryClass = {astro-ph.GA},
       adsurl = {https://ui.adsabs.harvard.edu/abs/2026arXiv260631515N}
}

@ARTICLE{Schaerer2026,
       author = {{Schaerer}, D. and {Izotov}, Y.~I. and {Marques-Chaves}, R. and {Steidel}, C.~C. and {Reddy}, N. and {Shapley}, A.~E. and {Mascia}, S. and {Chisholm}, J. and {Flury}, S.~R. and {Guseva}, N. and {Heckman}, T. and {Henry}, A. and {Inoue}, A.~K. and {Jung}, I. and {Kusakabe}, H. and {Mawatari}, K. and {Oesch}, P. and {{\"O}stlin}, G. and {Pentericci}, L. and {Roy}, N. and {Saldana-Lopez}, A. and {Sato}, R. and {Vanzella}, E. and {Verhamme}, A. and {Wang}, B.},
        title = "{Nitrogen abundances in star-forming galaxies 2.2 Gyr after the Big Bang are not elevated}",
      journal = {\aap},
         year = 2026,
        month = apr,
       volume = {708},
          eid = {A242},
        pages = {A242},
          doi = {10.1051/0004-6361/202556832},
archivePrefix = {arXiv},
       eprint = {2601.06968},
 primaryClass = {astro-ph.GA},
       adsurl = {https://ui.adsabs.harvard.edu/abs/2026A&A...708A.242S}
}

@ARTICLE{Chisholm2024,
       author = {{Chisholm}, J. and {Berg}, D.~A. and {Endsley}, R. and {Gazagnes}, S. and {Richardson}, C.~T. and {Lambrides}, E. and {Greene}, J. and {Finkelstein}, S. and {Flury}, S. and {Guseva}, N.~G. and {Henry}, A. and {Hutchison}, T.~A. and {Izotov}, Y.~I. and {Marques-Chaves}, R. and {Oesch}, P. and {Papovich}, C. and {Saldana-Lopez}, A. and {Schaerer}, D. and {Stephenson}, M.~G.},
        title = "{[Ne v] emission from a faint epoch of reionization-era galaxy: evidence for a narrow-line intermediate-mass black hole}",
      journal = {\mnras},
         year = 2024,
        month = nov,
       volume = {534},
       number = {3},
        pages = {2633-2652},
          doi = {10.1093/mnras/stae2199},
archivePrefix = {arXiv},
       eprint = {2402.18643},
 primaryClass = {astro-ph.GA},
       adsurl = {https://ui.adsabs.harvard.edu/abs/2024MNRAS.534.2633C}
}

@ARTICLE{Cataldi2026,
       author = {{Cataldi}, E. and {Belfiore}, F. and {Curti}, M. and {Moreschini}, B. and {Marconi}, A. and {Maiolino}, R. and {Feltre}, A. and {Ginolfi}, M. and {Mannucci}, F. and {Cresci}, G. and {Ji}, X. and {Amiri}, A. and {Arnaboldi}, M. and {Bertola}, E. and {Bracci}, C. and {Ceci}, M. and {Chakraborty}, A. and {Cullen}, F. and {D'Amato}, Q. and {Kobayashi}, C. and {Lamperti}, I. and {Marconcini}, C. and {Scialpi}, M. and {Ulivi}, L. and {Zanchettin}, M.~V.},
        title = "{Tracing nitrogen enrichment across cosmic time with JWST}",
      journal = {\aap},
         year = 2026,
        month = may,
       volume = {709},
          eid = {A199},
        pages = {A199},
          doi = {10.1051/0004-6361/202558415},
archivePrefix = {arXiv},
       eprint = {2512.07955},
 primaryClass = {astro-ph.GA},
       adsurl = {https://ui.adsabs.harvard.edu/abs/2026A&A...709A.199C}
}

@ARTICLE{Cameron2026,
       author = {{Cameron}, Alex J. and {Carreira}, Courtney and {Simmonds}, Charlotte and {Bunker}, Andrew J. and {Saxena}, Aayush and {Carniani}, Stefano and {Charlot}, St{\'e}phane and {Chevallard}, Jacopo and {Curtis-Lake}, Emma and {Hainline}, Kevin and {Hausen}, Ryan and {Ji}, Xihan and {Ji}, Zhiyuan and {Johnson}, Benjamin D. and {Rinaldi}, Pierluigi and {Robertson}, Brant and {Scholtz}, Jan and {Silcock}, Maddie S. and {Tacchella}, Sandro and {Trussler}, James A.~A. and {{\"U}bler}, Hannah and {Williams}, Christina C. and {Willmer}, Christopher N.~A. and {Willott}, Chris and {Witstok}, Joris},
        title = "{JADES: Evolution of nitrogen abundances in star-forming galaxies from z \raisebox{-0.5ex}\textasciitilde 1.5 - 7}",
      journal = {\mnras},
         year = 2026,
        month = aug,
          doi = {10.1093/mnras/stag1575},
archivePrefix = {arXiv},
       eprint = {2601.15964},
 primaryClass = {astro-ph.GA},
       adsurl = {https://ui.adsabs.harvard.edu/abs/2026MNRAS.tmp.1475C}
}

@ARTICLE{Inoue2014,
       author = {{Inoue}, Akio K. and {Shimizu}, Ikkoh and {Iwata}, Ikuru and {Tanaka}, Masayuki},
        title = "{An updated analytic model for attenuation by the intergalactic medium}",
      journal = {\mnras},
         year = 2014,
        month = aug,
       volume = {442},
       number = {2},
        pages = {1805-1820},
          doi = {10.1093/mnras/stu936},
archivePrefix = {arXiv},
       eprint = {1402.0677},
 primaryClass = {astro-ph.CO},
       adsurl = {https://ui.adsabs.harvard.edu/abs/2014MNRAS.442.1805I}
}

@ARTICLE{Leja2019,
       author = {{Leja}, Joel and {Carnall}, Adam C. and {Johnson}, Benjamin D. and {Conroy}, Charlie and {Speagle}, Joshua S.},
        title = "{How to Measure Galaxy Star Formation Histories. II. Nonparametric Models}",
      journal = {\apj},
         year = 2019,
        month = may,
       volume = {876},
       number = {1},
          eid = {3},
        pages = {3},
          doi = {10.3847/1538-4357/ab133c},
archivePrefix = {arXiv},
       eprint = {1811.03637},
 primaryClass = {astro-ph.GA},
       adsurl = {https://ui.adsabs.harvard.edu/abs/2019ApJ...876....3L}
}

@ARTICLE{Charlot2000,
       author = {{Charlot}, St{\'e}phane and {Fall}, S. Michael},
        title = "{A Simple Model for the Absorption of Starlight by Dust in Galaxies}",
      journal = {\apj},
         year = 2000,
        month = aug,
       volume = {539},
       number = {2},
        pages = {718-731},
          doi = {10.1086/309250},
archivePrefix = {arXiv},
       eprint = {astro-ph/0003128},
 primaryClass = {astro-ph},
       adsurl = {https://ui.adsabs.harvard.edu/abs/2000ApJ...539..718C}
}

@ARTICLE{Stanway2018_bpass,
       author = {{Stanway}, E.~R. and {Eldridge}, J.~J.},
        title = "{Re-evaluating old stellar populations}",
      journal = {\mnras},
         year = 2018,
        month = sep,
       volume = {479},
       number = {1},
        pages = {75-93},
          doi = {10.1093/mnras/sty1353},
archivePrefix = {arXiv},
       eprint = {1805.08784},
 primaryClass = {astro-ph.GA},
       adsurl = {https://ui.adsabs.harvard.edu/abs/2018MNRAS.479...75S}
}

@ARTICLE{Wold2026,
       author = {{Wold}, Isak and {Malhotra}, Sangeeta and {Rhoads}, James},
        title = "{Strong [OIII]$+$H$β$ emitters dominated the ionizing budget at $z\sim7$}",
      journal = {arXiv e-prints},
         year = 2026,
        month = jun,
          eid = {arXiv:2606.05323},
        pages = {arXiv:2606.05323},
          doi = {10.48550/arXiv.2606.05323},
archivePrefix = {arXiv},
       eprint = {2606.05323},
 primaryClass = {astro-ph.GA},
       adsurl = {https://ui.adsabs.harvard.edu/abs/2026arXiv260605323W}
}

@ARTICLE{Willott2022,
       author = {{Willott}, Chris J. and {Doyon}, Ren{\'e} and {Albert}, Loic and {Brammer}, Gabriel B. and {Dixon}, William V. and {Muzic}, Koraljka and {Ravindranath}, Swara and {Scholz}, Aleks and {Abraham}, Roberto and {Artigau}, {\'E}tienne and {Brada{\v{c}}}, Maru{\v{s}}a and {Goudfrooij}, Paul and {Hutchings}, John B. and {Iyer}, Kartheik G. and {Jayawardhana}, Ray and {LaMassa}, Stephanie and {Martis}, Nicholas and {Meyer}, Michael R. and {Morishita}, Takahiro and {Mowla}, Lamiya and {Muzzin}, Adam and {Noirot}, Ga{\"e}l and {Pacifici}, Camilla and {Rowlands}, Neil and {Sarrouh}, Ghassan and {Sawicki}, Marcin and {Taylor}, Joanna M. and {Volk}, Kevin and {Zabl}, Johannes},
        title = "{The Near-infrared Imager and Slitless Spectrograph for the James Webb Space Telescope. II. Wide Field Slitless Spectroscopy}",
      journal = {\pasp},
         year = 2022,
        month = feb,
       volume = {134},
       number = {1032},
          eid = {025002},
        pages = {025002},
          doi = {10.1088/1538-3873/ac5158},
archivePrefix = {arXiv},
       eprint = {2202.01714},
 primaryClass = {astro-ph.IM},
       adsurl = {https://ui.adsabs.harvard.edu/abs/2022PASP..134b5002W}
}

@ARTICLE{Weaver2024_aperpy,
       author = {{Weaver}, John R. and {Cutler}, Sam E. and {Pan}, Richard and {Whitaker}, Katherine E. and {Labb{\'e}}, Ivo and {Price}, Sedona H. and {Bezanson}, Rachel and {Brammer}, Gabriel and {Marchesini}, Danilo and {Leja}, Joel and {Wang}, Bingjie and {Furtak}, Lukas J. and {Zitrin}, Adi and {Atek}, Hakim and {Chemerynska}, Iryna and {Coe}, Dan and {Dayal}, Pratika and {van Dokkum}, Pieter and {Feldmann}, Robert and {F{\"o}rster Schreiber}, Natascha M. and {Franx}, Marijn and {Fujimoto}, Seiji and {Fudamoto}, Yoshinobu and {Glazebrook}, Karl and {de Graaff}, Anna and {Greene}, Jenny E. and {Juneau}, St{\'e}phanie and {Kassin}, Susan and {Kriek}, Mariska and {Khullar}, Gourav and {Maseda}, Michael V. and {Mowla}, Lamiya A. and {Muzzin}, Adam and {Nanayakkara}, Themiya and {Nelson}, Erica J. and {Oesch}, Pascal A. and {Pacifici}, Camilla and {Papovich}, Casey and {Setton}, David J. and {Shapley}, Alice E. and {Shipley}, Heath V. and {Smit}, Renske and {Stefanon}, Mauro and {Taylor}, Edward N. and {Weibel}, Andrea and {Williams}, Christina C.},
        title = "{The UNCOVER Survey: A First-look HST + JWST Catalog of 60,000 Galaxies near A2744 and beyond}",
      journal = {\apjs},
         year = 2024,
        month = jan,
       volume = {270},
       number = {1},
          eid = {7},
        pages = {7},
          doi = {10.3847/1538-4365/ad07e0},
archivePrefix = {arXiv},
       eprint = {2301.02671},
 primaryClass = {astro-ph.GA},
       adsurl = {https://ui.adsabs.harvard.edu/abs/2024ApJS..270....7W}
}

@ARTICLE{Harvey2025_EXPANSE,
       author = {{Harvey}, Thomas and {Conselice}, Christopher J. and {Adams}, Nathan J. and {Austin}, Duncan and {Li}, Qiong and {Rusakov}, Vadim and {Westcott}, Lewi and {Goolsby}, Caio M. and {Lovell}, Christopher C. and {Cochrane}, Rachel K. and {Vijayan}, Aswin P. and {Trussler}, James},
        title = "{Behind the spotlight: a systematic assessment of outshining using NIRCam medium bands in the JADES Origins Field}",
      journal = {\mnras},
         year = 2025,
        month = oct,
       volume = {542},
       number = {4},
        pages = {2998-3027},
          doi = {10.1093/mnras/staf1396},
archivePrefix = {arXiv},
       eprint = {2504.05244},
 primaryClass = {astro-ph.GA},
       adsurl = {https://ui.adsabs.harvard.edu/abs/2025MNRAS.542.2998H}
}

@ARTICLE{Berg2021,
       author = {{Berg}, Danielle A. and {Chisholm}, John and {Erb}, Dawn K. and {Skillman}, Evan D. and {Pogge}, Richard W. and {Olivier}, Grace M.},
        title = "{Characterizing Extreme Emission-line Galaxies. I. A Four-zone Ionization Model for Very High-ionization Emission}",
      journal = {\apj},
         year = 2021,
        month = dec,
       volume = {922},
       number = {2},
          eid = {170},
        pages = {170},
          doi = {10.3847/1538-4357/ac141b},
archivePrefix = {arXiv},
       eprint = {2105.12765},
 primaryClass = {astro-ph.GA},
       adsurl = {https://ui.adsabs.harvard.edu/abs/2021ApJ...922..170B}
}

@ARTICLE{Juodzbalis2026_AGN_census,
       author = {{Juod{\v{z}}balis}, Ignas and {Maiolino}, Roberto and {Baker}, William M. and {Lake}, Emma Curtis and {Scholtz}, Jan and {D'Eugenio}, Francesco and {Trefoloni}, Bartolomeo and {Isobe}, Yuki and {Tacchella}, Sandro and {Bunker}, Andrew J. and {Carniani}, Stefano and {Charlot}, St{\'e}phane and {Jones}, Gareth C. and {Parlanti}, Eleonora and {Perna}, Michele and {Rinaldi}, Pierluigi and {Robertson}, Brant and {{\"U}bler}, Hannah and {Venturi}, Giacomo and {Willott}, Chris},
        title = "{JADES: comprehensive census of broad-line AGN from reionization to cosmic noon revealed by JWST}",
      journal = {\mnras},
         year = 2026,
        month = mar,
       volume = {546},
       number = {3},
          eid = {stag086},
        pages = {stag086},
          doi = {10.1093/mnras/stag086},
archivePrefix = {arXiv},
       eprint = {2504.03551},
 primaryClass = {astro-ph.GA},
       adsurl = {https://ui.adsabs.harvard.edu/abs/2026MNRAS.546ag086J}
}

@misc{BrammerValentino_DJAv4,
  author       = {Brammer, Gabriel and
                  Valentino, Francesco},
  title        = {The DAWN JWST Archive: Compilation of Public
                   NIRSpec Spectra},
  month        = may,
  year         = 2025,
  publisher    = {Zenodo},
  version      = {4.4},
  doi          = {10.5281/zenodo.15472354},
  url          = {https://doi.org/10.5281/zenodo.15472354},
}

@ARTICLE{Bentz2004,
       author = {{Bentz}, Misty C. and {Osmer}, Patrick S.},
        title = "{A Search for Nitrogen-Enriched Quasars in the Sloan Digital Sky Survey Early Data Release}",
      journal = {\aj},
         year = 2004,
        month = feb,
       volume = {127},
       number = {2},
        pages = {576-586},
          doi = {10.1086/381299},
archivePrefix = {arXiv},
       eprint = {astro-ph/0308486},
 primaryClass = {astro-ph},
       adsurl = {https://ui.adsabs.harvard.edu/abs/2004AJ....127..576B}
}

@ARTICLE{Kobayashi2023,
       author = {{Kobayashi}, Chiaki and {Bhattacharya}, Souradeep and {Arnaboldi}, Magda and {Gerhard}, Ortwin},
        title = "{On the {\ensuremath{\alpha}}/Fe Bimodality of the M31 Disks}",
      journal = {\apjl},
         year = 2023,
        month = oct,
       volume = {956},
       number = {1},
          eid = {L14},
        pages = {L14},
          doi = {10.3847/2041-8213/acf7c7},
archivePrefix = {arXiv},
       eprint = {2309.01707},
 primaryClass = {astro-ph.GA},
       adsurl = {https://ui.adsabs.harvard.edu/abs/2023ApJ...956L..14K}
}

@ARTICLE{Arnaboldi2022,
       author = {{Arnaboldi}, Magda and {Bhattacharya}, Souradeep and {Gerhard}, Ortwin and {Kobayashi}, Chiaki and {Freeman}, Kenneth C. and {Caldwell}, Nelson and {Hartke}, Johanna and {McConnachie}, Alan and {Guhathakurta}, Puragra},
        title = "{The survey of planetary nebulae in Andromeda (M31). V. Chemical enrichment of the thin and thicker discs of Andromeda: Oxygen to argon abundance ratios for planetary nebulae and HII regions}",
      journal = {\aap},
         year = 2022,
        month = oct,
       volume = {666},
          eid = {A109},
        pages = {A109},
          doi = {10.1051/0004-6361/202244258},
archivePrefix = {arXiv},
       eprint = {2208.02328},
 primaryClass = {astro-ph.GA},
       adsurl = {https://ui.adsabs.harvard.edu/abs/2022A&A...666A.109A}
}

@ARTICLE{Morel2026,
       author = {{Morel}, I. and {Schaerer}, D. and {Marques-Chaves}, R. and {Prantzos}, N. and {Charbonnel}, C. and {Brammer}, G. and {Xiao}, M. and {Dessauges-Zavadsky}, M.},
        title = "{Discovery of new N-emitters over a wide redshift range}",
      journal = {\aap},
         year = 2026,
        month = jun,
       volume = {710},
          eid = {A321},
        pages = {A321},
          doi = {10.1051/0004-6361/202558256},
       adsurl = {https://ui.adsabs.harvard.edu/abs/2026A&A...710A.321M}
}

@ARTICLE{Carnall2018_bagpipes_spec,
       author = {{Carnall}, A.~C. and {McLure}, R.~J. and {Dunlop}, J.~S. and {Dav{\'e}}, R.},
        title = "{Inferring the star formation histories of massive quiescent galaxies with BAGPIPES: evidence for multiple quenching mechanisms}",
      journal = {\mnras},
         year = 2018,
        month = nov,
       volume = {480},
       number = {4},
        pages = {4379-4401},
          doi = {10.1093/mnras/sty2169},
archivePrefix = {arXiv},
       eprint = {1712.04452},
 primaryClass = {astro-ph.GA},
       adsurl = {https://ui.adsabs.harvard.edu/abs/2018MNRAS.480.4379C}
}

@ARTICLE{Carnall2019_bagpipes,
       author = {{Carnall}, A.~C. and {McLure}, R.~J. and {Dunlop}, J.~S. and {Cullen}, F. and {McLeod}, D.~J. and {Wild}, V. and {Johnson}, B.~D. and {Appleby}, S. and {Dav{\'e}}, R. and {Amorin}, R. and {Bolzonella}, M. and {Castellano}, M. and {Cimatti}, A. and {Cucciati}, O. and {Gargiulo}, A. and {Garilli}, B. and {Marchi}, F. and {Pentericci}, L. and {Pozzetti}, L. and {Schreiber}, C. and {Talia}, M. and {Zamorani}, G.},
        title = "{The VANDELS survey: the star-formation histories of massive quiescent galaxies at 1.0 < z < 1.3}",
      journal = {\mnras},
         year = 2019,
        month = nov,
       volume = {490},
       number = {1},
        pages = {417-439},
          doi = {10.1093/mnras/stz2544},
archivePrefix = {arXiv},
       eprint = {1903.11082},
 primaryClass = {astro-ph.GA},
       adsurl = {https://ui.adsabs.harvard.edu/abs/2019MNRAS.490..417C}
}

@ARTICLE{Berg2019,
       author = {{Berg}, Danielle A. and {Erb}, Dawn K. and {Henry}, Richard B.~C. and {Skillman}, Evan D. and {McQuinn}, Kristen B.~W.},
        title = "{The Chemical Evolution of Carbon, Nitrogen, and Oxygen in Metal-poor Dwarf Galaxies}",
      journal = {\apj},
         year = 2019,
        month = mar,
       volume = {874},
       number = {1},
          eid = {93},
        pages = {93},
          doi = {10.3847/1538-4357/ab020a},
archivePrefix = {arXiv},
       eprint = {1901.08160},
 primaryClass = {astro-ph.GA},
       adsurl = {https://ui.adsabs.harvard.edu/abs/2019ApJ...874...93B}
}

@ARTICLE{Garnett1992,
       author = {{Garnett}, Donald R.},
        title = "{Electron Temperature Variations and the Measurement of Nebular Abundances}",
      journal = {\aj},
         year = 1992,
        month = apr,
       volume = {103},
        pages = {1330},
          doi = {10.1086/116146},
       adsurl = {https://ui.adsabs.harvard.edu/abs/1992AJ....103.1330G}
}

@ARTICLE{Scholte2025,
       author = {{Scholte}, D. and {Cullen}, F. and {Carnall}, A.~C. and {Arellano-C{\'o}rdova}, K.~Z. and {Stanton}, T.~M. and {Barrufet}, L. and {Begley}, R. and {Bondestam}, C. and {Donnan}, C.~T. and {Dunlop}, J.~S. and {Leung}, H.-H. and {McLeod}, D.~J. and {McLure}, R.~J. and {Moustakas}, J.~M. and {Pollock}, C.~L. and {Shapley}, A.~E. and {Stevenson}, S. and {Zou}, H.},
        title = "{The JWST EXCELS survey: probing strong-line diagnostics and the chemical evolution of galaxies over cosmic time using T$_{e}$-metallicities}",
      journal = {\mnras},
         year = 2025,
        month = jun,
       volume = {540},
       number = {2},
        pages = {1800-1826},
          doi = {10.1093/mnras/staf834},
archivePrefix = {arXiv},
       eprint = {2502.10499},
 primaryClass = {astro-ph.GA},
       adsurl = {https://ui.adsabs.harvard.edu/abs/2025MNRAS.540.1800S}
}

@ARTICLE{Luridiana2015,
       author = {{Luridiana}, V. and {Morisset}, C. and {Shaw}, R.~A.},
        title = "{PyNeb: a new tool for analyzing emission lines. I. Code description and validation of results}",
      journal = {\aap},
         year = 2015,
        month = jan,
       volume = {573},
          eid = {A42},
        pages = {A42},
          doi = {10.1051/0004-6361/201323152},
archivePrefix = {arXiv},
       eprint = {1410.6662},
 primaryClass = {astro-ph.IM},
       adsurl = {https://ui.adsabs.harvard.edu/abs/2015A&A...573A..42L}
}

@ARTICLE{Pettini1995,
       author = {{Pettini}, Max and {Lipman}, Keith and {Hunstead}, Richard W.},
        title = "{Element Abundances at High Redshifts: The N/O Ratio in a Primeval Galaxy}",
      journal = {\apj},
         year = 1995,
        month = sep,
       volume = {451},
        pages = {100},
          doi = {10.1086/176203},
archivePrefix = {arXiv},
       eprint = {astro-ph/9502077},
 primaryClass = {astro-ph},
       adsurl = {https://ui.adsabs.harvard.edu/abs/1995ApJ...451..100P}
}

@ARTICLE{Pettini2002,
       author = {{Pettini}, M. and {Ellison}, S.~L. and {Bergeron}, J. and {Petitjean}, P.},
        title = "{The abundances of nitrogen and oxygen in damped Lyman alpha systems}",
      journal = {\aap},
         year = 2002,
        month = aug,
       volume = {391},
        pages = {21-34},
          doi = {10.1051/0004-6361:20020809},
archivePrefix = {arXiv},
       eprint = {astro-ph/0205472},
 primaryClass = {astro-ph},
       adsurl = {https://ui.adsabs.harvard.edu/abs/2002A&A...391...21P}
}

@ARTICLE{Lu1998,
       author = {{Lu}, Limin and {Sargent}, Wallace L.~W. and {Barlow}, Thomas A.},
        title = "{The N/Si Abundance Ratio in 15 Damped Lyalpha Galaxies: Implications for the Origin of Nitrogen}",
      journal = {\aj},
         year = 1998,
        month = jan,
       volume = {115},
       number = {1},
        pages = {55-61},
          doi = {10.1086/300180},
archivePrefix = {arXiv},
       eprint = {astro-ph/9710266},
 primaryClass = {astro-ph},
       adsurl = {https://ui.adsabs.harvard.edu/abs/1998AJ....115...55L}
}

@article{Nandal2024a,
  title = {Explaining the High Nitrogen Abundances Observed in High-z Galaxies via Population {{III}} Stars of a Few Thousand Solar Masses},
  author = {Nandal, Devesh and Regan, John A. and Woods, Tyrone E. and Farrell, Eoin and Ekstr{\"o}m, Sylvia and Meynet, Georges},
  year = 2024,
  month = mar,
  journal = {Astronomy and Astrophysics},
  volume = {683},
  pages = {A156},
  publisher = {EDP},
  issn = {0004-6361},
  doi = {10.1051/0004-6361/202348035},
  urldate = {2025-11-06}
}

@ARTICLE{Pettini2008,
       author = {{Pettini}, Max and {Zych}, Berkeley J. and {Steidel}, Charles C. and {Chaffee}, Fred H.},
        title = "{C, N, O abundances in the most metal-poor damped Lyman alpha systems}",
      journal = {\mnras},
         year = 2008,
        month = apr,
       volume = {385},
       number = {4},
        pages = {2011-2024},
          doi = {10.1111/j.1365-2966.2008.12951.x},
archivePrefix = {arXiv},
       eprint = {0712.1829},
 primaryClass = {astro-ph},
       adsurl = {https://ui.adsabs.harvard.edu/abs/2008MNRAS.385.2011P}
}

@ARTICLE{Izotov1999,
       author = {{Izotov}, Yuri I. and {Thuan}, Trinh X.},
        title = "{Heavy-Element Abundances in Blue Compact Galaxies}",
      journal = {\apj},
         year = 1999,
        month = feb,
       volume = {511},
       number = {2},
        pages = {639-659},
          doi = {10.1086/306708},
archivePrefix = {arXiv},
       eprint = {astro-ph/9811387},
 primaryClass = {astro-ph},
       adsurl = {https://ui.adsabs.harvard.edu/abs/1999ApJ...511..639I}
}

@ARTICLE{Spite2005,
       author = {{Spite}, M. and {Cayrel}, R. and {Plez}, B. and {Hill}, V. and {Spite}, F. and {Depagne}, E. and {Fran{\c{c}}ois}, P. and {Bonifacio}, P. and {Barbuy}, B. and {Beers}, T. and {Andersen}, J. and {Molaro}, P. and {Nordstr{\"o}m}, B. and {Primas}, F.},
        title = "{First stars VI - Abundances of C, N, O, Li, and mixing in extremely metal-poor giants. Galactic evolution of the light elements}",
      journal = {\aap},
         year = 2005,
        month = feb,
       volume = {430},
        pages = {655-668},
          doi = {10.1051/0004-6361:20041274},
archivePrefix = {arXiv},
       eprint = {astro-ph/0409536},
 primaryClass = {astro-ph},
       adsurl = {https://ui.adsabs.harvard.edu/abs/2005A&A...430..655S}
}

@article{Nicholls2017,
  title = {Abundance Scaling in Stars, Nebulae and Galaxies},
  author = {Nicholls, David C. and Sutherland, Ralph S. and Dopita, Michael A. and Kewley, Lisa J. and Groves, Brent A.},
  year = 2017,
  month = apr,
  journal = {Monthly Notices of the Royal Astronomical Society},
  volume = {466},
  pages = {4403--4422},
  publisher = {OUP},
  issn = {0035-8711},
  doi = {10.1093/mnras/stw3235},
  urldate = {2024-12-19}
}

@article{Henry2000,
       author = {{Henry}, R.~B.~C. and {Edmunds}, M.~G. and {K{\"o}ppen}, J.},
        title = "{On the Cosmic Origins of Carbon and Nitrogen}",
      journal = {\apj},
         year = 2000,
        month = oct,
       volume = {541},
       number = {2},
        pages = {660-674},
          doi = {10.1086/309471},
archivePrefix = {arXiv},
       eprint = {astro-ph/0004299},
 primaryClass = {astro-ph},
       adsurl = {https://ui.adsabs.harvard.edu/abs/2000ApJ...541..660H}
}

@ARTICLE{Kobayashi2020,
       author = {{Kobayashi}, Chiaki and {Karakas}, Amanda I. and {Lugaro}, Maria},
        title = "{The Origin of Elements from Carbon to Uranium}",
      journal = {\apj},
         year = 2020,
        month = sep,
       volume = {900},
       number = {2},
          eid = {179},
        pages = {179},
          doi = {10.3847/1538-4357/abae65},
archivePrefix = {arXiv},
       eprint = {2008.04660},
 primaryClass = {astro-ph.GA},
       adsurl = {https://ui.adsabs.harvard.edu/abs/2020ApJ...900..179K}
}

@ARTICLE{Heintz2023,
       author = {{Heintz}, Kasper E. and {Brammer}, Gabriel B. and {Gim{\'e}nez-Arteaga}, Clara and {Strait}, Victoria B. and {Lagos}, Claudia del P. and {Vijayan}, Aswin P. and {Matthee}, Jorryt and {Watson}, Darach and {Mason}, Charlotte A. and {Hutter}, Anne and {Toft}, Sune and {Fynbo}, Johan P.~U. and {Oesch}, Pascal A.},
        title = "{Dilution of chemical enrichment in galaxies 600 Myr after the Big Bang}",
      journal = {Nature Astronomy},
         year = 2023,
        month = dec,
       volume = {7},
        pages = {1517-1524},
          doi = {10.1038/s41550-023-02078-7},
archivePrefix = {arXiv},
       eprint = {2212.02890},
 primaryClass = {astro-ph.GA},
       adsurl = {https://ui.adsabs.harvard.edu/abs/2023NatAs...7.1517H}
}

@ARTICLE{ArellanoCordova2024b,
       author = {{Arellano-C{\'o}rdova}, K.~Z. and {Cullen}, F. and {Carnall}, A.~C. and {Scholte}, D. and {Stanton}, T.~M. and {Kobayashi}, C. and {Martinez}, Z. and {Berg}, D.~A. and {Barrufet}, L. and {Begley}, R. and {Donnan}, C.~T. and {Dunlop}, J.~S. and {Hamadouche}, M.~L. and {McLeod}, D.~J. and {McLure}, R.~J. and {Rowlands}, K. and {Shapley}, A.~E.},
        title = "{The JWST EXCELS survey: direct estimates of C, N, and O abundances in two relatively metal-rich galaxies at z ≃ 5}",
      journal = {\mnras},
         year = 2025,
        month = jul,
       volume = {540},
       number = {4},
        pages = {2991-3007},
          doi = {10.1093/mnras/staf855},
archivePrefix = {arXiv},
       eprint = {2412.10557},
 primaryClass = {astro-ph.GA},
       adsurl = {https://ui.adsabs.harvard.edu/abs/2025MNRAS.540.2991A}
}

@article{Asplund2021,
  title = {The Chemical Make-up of the {{Sun}}: {{A}} 2020 Vision},
  shorttitle = {The Chemical Make-up of the {{Sun}}},
  author = {Asplund, M. and Amarsi, A. M. and Grevesse, N.},
  year = {2021},
  month = sep,
  journal = {Astronomy and Astrophysics},
  volume = {653},
  pages = {A141},
  issn = {0004-6361},
  doi = {10.1051/0004-6361/202140445},
  urldate = {2025-09-15}
}

@article{Bunker2023,
  title = {{{JADES NIRSpec Spectroscopy}} of {{GN-z11}}: {{Lyman-$\alpha$}} Emission and Possible Enhanced Nitrogen Abundance in a z = 10.60 Luminous Galaxy},
  shorttitle = {{{JADES NIRSpec Spectroscopy}} of {{GN-z11}}},
  author = {Bunker, Andrew J. and Saxena, Aayush and Cameron, Alex J. and Willott, Chris J. and {Curtis-Lake}, Emma and Jakobsen, Peter and Carniani, Stefano and Smit, Renske and Maiolino, Roberto and Witstok, Joris and Curti, Mirko and D'Eugenio, Francesco and Jones, Gareth C. and Ferruit, Pierre and Arribas, Santiago and Charlot, Stephane and Chevallard, Jacopo and Giardino, Giovanna and {de Graaff}, Anna and Looser, Tobias J. and L{\"u}tzgendorf, Nora and Maseda, Michael V. and Rawle, Tim and Rix, Hans-Walter and Del Pino, Bruno Rodr{\'i}guez and Alberts, Stacey and Egami, Eiichi and Eisenstein, Daniel J. and Endsley, Ryan and Hainline, Kevin and Hausen, Ryan and Johnson, Benjamin D. and Rieke, George and Rieke, Marcia and Robertson, Brant E. and Shivaei, Irene and Stark, Daniel P. and Sun, Fengwu and Tacchella, Sandro and Tang, Mengtao and Williams, Christina C. and Willmer, Christopher N. A. and Baker, William M. and Baum, Stefi and Bhatawdekar, Rachana and Bowler, Rebecca and Boyett, Kristan and Chen, Zuyi and Circosta, Chiara and Helton, Jakob M. and Ji, Zhiyuan and Kumari, Nimisha and Lyu, Jianwei and Nelson, Erica and Parlanti, Eleonora and Perna, Michele and Sandles, Lester and Scholtz, Jan and Suess, Katherine A. and Topping, Michael W. and {\"U}bler, Hannah and Wallace, Imaan E. B. and Whitler, Lily},
  year = {2023},
  month = sep,
  journal = {Astronomy and Astrophysics},
  volume = {677},
  pages = {A88},
  publisher = {EDP},
  issn = {0004-6361},
  doi = {10.1051/0004-6361/202346159},
  urldate = {2025-01-21}
}

@article{Cameron2023_GNz11,
  title = {Nitrogen Enhancements 440 {{Myr}} after the Big Bang: Supersolar {{N}}/{{O}}, a Tidal Disruption Event, or a Dense Stellar Cluster in {{GN-z11}}?},
  shorttitle = {Nitrogen Enhancements 440 {{Myr}} after the Big Bang},
  author = {Cameron, Alex J. and Katz, Harley and Rey, Martin P. and Saxena, Aayush},
  year = {2023},
  month = aug,
  journal = {Monthly Notices of the Royal Astronomical Society},
  volume = {523},
  pages = {3516--3525},
  publisher = {OUP},
  issn = {0035-8711},
  doi = {10.1093/mnras/stad1579},
  urldate = {2024-12-19}
}

@article{Castellano2024,
  title = {{{JWST NIRSpec Spectroscopy}} of the {{Remarkable Bright Galaxy GHZ2}}/{{GLASS-z12}} at {{Redshift}} 12.34},
  author = {Castellano, Marco and Napolitano, Lorenzo and Fontana, Adriano and {Roberts-Borsani}, Guido and Treu, Tommaso and Vanzella, Eros and Zavala, Jorge A. and Arrabal Haro, Pablo and Calabr{\`o}, Antonello and Llerena, Mario and Mascia, Sara and Merlin, Emiliano and Paris, Diego and Pentericci, Laura and Santini, Paola and Bakx, Tom J. L. C. and Bergamini, Pietro and Cupani, Guido and Dickinson, Mark and Filippenko, Alexei V. and Glazebrook, Karl and Grillo, Claudio and Kelly, Patrick L. and Malkan, Matthew A. and Mason, Charlotte A. and Morishita, Takahiro and Nanayakkara, Themiya and Rosati, Piero and Sani, Eleonora and Wang, Xin and Yoon, Ilsang},
  year = {2024},
  month = sep,
  journal = {The Astrophysical Journal},
  volume = {972},
  pages = {143},
  publisher = {IOP},
  issn = {0004-637X},
  doi = {10.3847/1538-4357/ad5f88},
  urldate = {2025-09-25}
}

@book{Clayton1983,
  title = {Principles of Stellar Evolution and Nucleosynthesis},
  author = {Clayton, Donald D.},
  year = {1983},
  month = jan,
  urldate = {2025-09-03},
  publisher = {Chicago: University of Chicago Press}
}

@article{MarquesChaves2024,
  title = {Extreme {{N-emitters}} at High Redshift: {{Possible}} Signatures of Supermassive Stars and Globular Cluster or Black Hole Formation in Action},
  shorttitle = {Extreme {{N-emitters}} at High Redshift},
  author = {{Marques-Chaves}, R. and Schaerer, D. and Kuruvanthodi, A. and Korber, D. and Prantzos, N. and Charbonnel, C. and Weibel, A. and Izotov, Y. I. and Messa, M. and Brammer, G. and {Dessauges-Zavadsky}, M. and Oesch, P.},
  year = {2024},
  month = jan,
  journal = {Astronomy and Astrophysics},
  volume = {681},
  pages = {A30},
  publisher = {EDP},
  issn = {0004-6361},
  doi = {10.1051/0004-6361/202347411},
  urldate = {2025-01-21}
}

@ARTICLE{Napolitano2025,
       author = {{Napolitano}, Lorenzo and {Castellano}, Marco and {Pentericci}, Laura and {Vignali}, Cristian and {Gilli}, Roberto and {Fontana}, Adriano and {Santini}, Paola and {Treu}, Tommaso and {Calabr{\`o}}, Antonello and {Llerena}, Mario and {Piconcelli}, Enrico and {Zappacosta}, Luca and {Mascia}, Sara and {Tripodi}, Roberta and {Arrabal Haro}, Pablo and {Bergamini}, Pietro and {Bakx}, Tom J.~L.~C. and {Dickinson}, Mark and {Glazebrook}, Karl and {Henry}, Alaina and {Leethochawalit}, Nicha and {Mazzolari}, Giovanni and {Merlin}, Emiliano and {Morishita}, Takahiro and {Nanayakkara}, Themiya and {Paris}, Diego and {Puccetti}, Simonetta and {Roberts-Borsani}, Guido and {Rojas Ruiz}, Sofia and {Rosati}, Piero and {Vanzella}, Eros and {Vito}, Fabio and {Vulcani}, Benedetta and {Wang}, Xin and {Yoon}, Ilsang and {Zavala}, Jorge A.},
        title = "{The Dual Nature of GHZ9: Coexisting Active Galactic Nuclei and Star Formation Activity in a Remote X-Ray Source at z = 10.145}",
      journal = {\apj},
         year = 2025,
        month = aug,
       volume = {989},
       number = {1},
          eid = {75},
        pages = {75},
          doi = {10.3847/1538-4357/ade706},
archivePrefix = {arXiv},
       eprint = {2410.18763},
 primaryClass = {astro-ph.GA},
       adsurl = {https://ui.adsabs.harvard.edu/abs/2025ApJ...989...75N}
}

@misc{NavarroCarrera2024,
  title = {The Interstellar Medium Conditions of a Strong {{Lya}} Emitter at z = 8.279 Revealed by {{JWST}}: A Robust {{LyC}} Leaker Candidate at the {{Epoch}} of {{Reionization}}},
  shorttitle = {The Interstellar Medium Conditions of a Strong {{Lya}} Emitter at z = 8.279 Revealed by {{JWST}}},
  author = {{Navarro-Carrera}, Rafael and Caputi, Karina I. and Iani, Edoardo and Rinaldi, Pierluigi and Kokorev, Vasily and Kerutt, Josephine},
  year = {2024},
  month = jul,
  publisher = {arXiv},
  doi = {10.48550/arXiv.2407.14201},
  urldate = {2025-09-25}
}

@article{PlanckCollaboration2020,
  title = {Planck 2018 Results. {{VI}}. {{Cosmological}} Parameters},
  author = {{Planck Collaboration} and Aghanim, N. and Akrami, Y. and Ashdown, M. and Aumont, J. and Baccigalupi, C. and Ballardini, M. and Banday, A. J. and Barreiro, R. B. and Bartolo, N. and Basak, S. and Battye, R. and Benabed, K. and Bernard, J. -P. and Bersanelli, M. and Bielewicz, P. and Bock, J. J. and Bond, J. R. and Borrill, J. and Bouchet, F. R. and Boulanger, F. and Bucher, M. and Burigana, C. and Butler, R. C. and Calabrese, E. and Cardoso, J. -F. and Carron, J. and Challinor, A. and Chiang, H. C. and Chluba, J. and Colombo, L. P. L. and Combet, C. and Contreras, D. and Crill, B. P. and Cuttaia, F. and {de Bernardis}, P. and {de Zotti}, G. and Delabrouille, J. and Delouis, J. -M. and Di Valentino, E. and Diego, J. M. and Dor{\'e}, O. and Douspis, M. and Ducout, A. and Dupac, X. and Dusini, S. and Efstathiou, G. and Elsner, F. and En{\ss}lin, T. A. and Eriksen, H. K. and Fantaye, Y. and Farhang, M. and Fergusson, J. and {Fernandez-Cobos}, R. and Finelli, F. and Forastieri, F. and Frailis, M. and Fraisse, A. A. and Franceschi, E. and Frolov, A. and Galeotta, S. and Galli, S. and Ganga, K. and {G{\'e}nova-Santos}, R. T. and Gerbino, M. and Ghosh, T. and {Gonz{\'a}lez-Nuevo}, J. and G{\'o}rski, K. M. and Gratton, S. and Gruppuso, A. and Gudmundsson, J. E. and Hamann, J. and Handley, W. and Hansen, F. K. and Herranz, D. and Hildebrandt, S. R. and Hivon, E. and Huang, Z. and Jaffe, A. H. and Jones, W. C. and Karakci, A. and Keih{\"a}nen, E. and Keskitalo, R. and Kiiveri, K. and Kim, J. and Kisner, T. S. and Knox, L. and Krachmalnicoff, N. and Kunz, M. and {Kurki-Suonio}, H. and Lagache, G. and Lamarre, J. -M. and Lasenby, A. and Lattanzi, M. and Lawrence, C. R. and Le Jeune, M. and Lemos, P. and Lesgourgues, J. and Levrier, F. and Lewis, A. and Liguori, M. and Lilje, P. B. and Lilley, M. and Lindholm, V. and {L{\'o}pez-Caniego}, M. and Lubin, P. M. and Ma, Y. -Z. and {Mac{\'i}as-P{\'e}rez}, J. F. and Maggio, G. and Maino, D. and Mandolesi, N. and Mangilli, A. and {Marcos-Caballero}, A. and Maris, M. and Martin, P. G. and Martinelli, M. and {Mart{\'i}nez-Gonz{\'a}lez}, E. and Matarrese, S. and Mauri, N. and McEwen, J. D. and Meinhold, P. R. and Melchiorri, A. and Mennella, A. and Migliaccio, M. and Millea, M. and Mitra, S. and {Miville-Desch{\^e}nes}, M. -A. and Molinari, D. and Montier, L. and Morgante, G. and Moss, A. and Natoli, P. and {N{\o}rgaard-Nielsen}, H. U. and Pagano, L. and Paoletti, D. and Partridge, B. and Patanchon, G. and Peiris, H. V. and Perrotta, F. and Pettorino, V. and Piacentini, F. and Polastri, L. and Polenta, G. and Puget, J. -L. and Rachen, J. P. and Reinecke, M. and Remazeilles, M. and Renzi, A. and Rocha, G. and Rosset, C. and Roudier, G. and {Rubi{\~n}o-Mart{\'i}n}, J. A. and {Ruiz-Granados}, B. and Salvati, L. and Sandri, M. and Savelainen, M. and Scott, D. and Shellard, E. P. S. and Sirignano, C. and Sirri, G. and Spencer, L. D. and Sunyaev, R. and {Suur-Uski}, A. -S. and Tauber, J. A. and Tavagnacco, D. and Tenti, M. and Toffolatti, L. and Tomasi, M. and Trombetti, T. and Valenziano, L. and Valiviita, J. and Van Tent, B. and Vibert, L. and Vielva, P. and Villa, F. and Vittorio, N. and Wandelt, B. D. and Wehus, I. K. and White, M. and White, S. D. M. and Zacchei, A. and Zonca, A.},
  year = {2020},
  month = sep,
  journal = {Astronomy and Astrophysics},
  volume = {641},
  pages = {A6},
  issn = {0004-6361},
  doi = {10.1051/0004-6361/201833910},
  urldate = {2025-09-17}
}

@article{Sanders2024,
  title = {Direct {{T}} E-Based {{Metallicities}} of z = 2--9 {{Galaxies}} with {{JWST}}/{{NIRSpec}}: {{Empirical Metallicity Calibrations Applicable}} from {{Reionization}} to {{Cosmic Noon}}},
  shorttitle = {Direct {{T}} E-Based {{Metallicities}} of z = 2--9 {{Galaxies}} with {{JWST}}/{{NIRSpec}}},
  author = {Sanders, Ryan L. and Shapley, Alice E. and Topping, Michael W. and Reddy, Naveen A. and Brammer, Gabriel B.},
  year = {2024},
  month = feb,
  journal = {The Astrophysical Journal},
  volume = {962},
  pages = {24},
  publisher = {IOP},
  issn = {0004-637X},
  doi = {10.3847/1538-4357/ad15fc},
  urldate = {2025-07-23}
}

@article{Schaerer2024,
  title = {Discovery of a New {{N-emitter}} in the Epoch of Reionization},
  author = {Schaerer, D. and {Marques-Chaves}, R. and Xiao, M. and Korber, D.},
  year = {2024},
  month = jul,
  journal = {Astronomy and Astrophysics},
  volume = {687},
  pages = {L11},
  publisher = {EDP},
  issn = {0004-6361},
  doi = {10.1051/0004-6361/202450721},
  urldate = {2025-09-25}
}

@article{Senchyna2024,
  title = {{{GN-z11}} in {{Context}}: {{Possible Signatures}} of {{Globular Cluster Precursors}} at {{Redshift}} 10},
  shorttitle = {{{GN-z11}} in {{Context}}},
  author = {Senchyna, Peter and Plat, Adele and Stark, Daniel P. and Rudie, Gwen C. and Berg, Danielle and Charlot, St{\'e}phane and James, Bethan L. and Mingozzi, Matilde},
  year = {2024},
  month = may,
  journal = {The Astrophysical Journal},
  volume = {966},
  pages = {92},
  publisher = {IOP},
  issn = {0004-637X},
  doi = {10.3847/1538-4357/ad235e},
  urldate = {2025-01-22}
}

@ARTICLE{Stanton2025,
       author = {{Stanton}, T.~M. and {Cullen}, F. and {Carnall}, A.~C. and {Scholte}, D. and {Arellano-C{\'o}rdova}, K.~Z. and {McLeod}, D.~J. and {Begley}, R. and {Donnan}, C.~T. and {Dunlop}, J.~S. and {Hamadouche}, M.~L. and {McLure}, R.~J. and {Shapley}, A.~E. and {Bondestam}, C. and {Stevenson}, S.},
        title = "{The JWST EXCELS survey: tracing the chemical enrichment pathways of high-redshift star-forming galaxies with O, Ar, and Ne abundances}",
      journal = {\mnras},
         year = 2025,
        month = feb,
       volume = {537},
       number = {2},
        pages = {1735-1748},
          doi = {10.1093/mnras/staf106},
archivePrefix = {arXiv},
       eprint = {2411.11837},
 primaryClass = {astro-ph.GA},
       adsurl = {https://ui.adsabs.harvard.edu/abs/2025MNRAS.537.1735S}
}




\appendix

\section{SED Fitting Model}
\label{app:appendix_sed_fitting}

In \S\ref{sec:sed_fitting} we briefly discussed the SED fitting model applied to our data. In this appendix, we expand on the details and show our model assumptions and optimised parameters in Table~\ref{tab:sed_models}.

\begin{table}
\centering
\renewcommand{\arraystretch}{1.2}
\caption{Stellar population parameters and their priors used for SED modelling with \texttt{bagpipes}.}

\begin{tabular}{l@{\hspace{4pt}}l@{\hspace{2pt}}l@{\hspace{2pt}}}
\toprule\toprule
Parameter & Range & Prior \\
\midrule
    SFR (age bin)\(^\ast\) & \(\Delta \log{\text{SFR}} \in [-10, 10]\) & \(\text{Student-}t(\nu=2, \sigma=0.3)\) \\

    Metallicity & \(Z/Z_{\odot} \in [0.001, 3]\) & \(\mathcal{U}(\log{Z/Z_{\odot}})\) \\
    Ionisation parameter & \(\log{U} \in [-4, 0]\) & \(\mathcal{N}(\log{U}; \mu=-2.5, \sigma=0.2)\) \\
   
    Dust Component\(^{\ast\ast}\), \(A_V\) & \([0,4]\) mag & \(\mathcal{N}(A_V; \mu=0.2, \sigma=0.2)\) \\
    
    Dust Variable Slope, \(n\) & [0.3, 2.5] & \(\mathcal{N}(n; \mu=0.7, \sigma=0.3)\) \\
   
    Velocity dispersion & \(\sigma \in [500, 5000]\,\text{km\,s}^{-1}\) & \(\mathcal{N}(\sigma; \mu=10^{3}, \sigma=200)\) \\
   
    \makecell[l]{Extra noise parameter\\(multiplicative)} & \(f_{\text{noise}} \in [1, 2]\) & \(\mathcal{U}(\log{f_{\text{noise}}})\) \\

    \makecell[l]{Intergalactic Medium\\
    Absorption\(^{\dagger}\)} & --- & --- \\
\bottomrule
\end{tabular}

\par\raggedright\footnotesize
\(^{\ast}\)Dust attenuation model from \cite{Charlot2000}.\\
\(^{\ast\ast}\)We use a non-parametric star formation history from \cite{Leja2019} with the continuity SFR prior calculated for each of the total of seven age bins that we define based on the spectroscopic redshift.\\
\(^\dagger\)Intergalactic Medium Absorption model from \cite{Inoue2014}.\\

\label{tab:sed_models}
\end{table}

One shortcoming of the SED model we observe is the insufficient strength of the nebular lines or unaccounted nebular attenuation. While \ha fluxes of most of the SED models are in excellent agreement with the data, some cannot reproduce the strong lines in the data, as seen in Figure~\ref{fig:SFR_Ha_comparison}, and the median is lower by 0.3~dex. This bias is likely explained by stronger dust attenuation in the nebular regions compared to the locations where the UV and optical continuum originates. To demonstrate this effect we colour the points by \(A_V\) magnitudes estimated from the Balmer decrement values (\S\ref{sec:data_spec_lines}).

\begin{figure}
\begin{center}
    \includegraphics[angle=0,width=1\columnwidth]{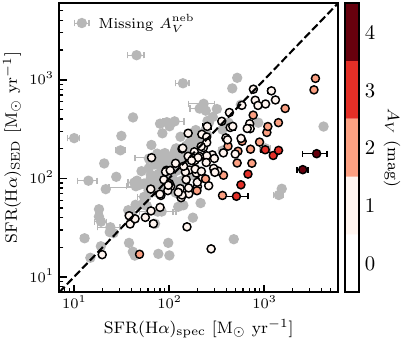}
\end{center}
\caption{Comparison between the star formation rates computed from \ha line fluxes and \ha from best-fit BPASS SED models.  The magnitude of dust attenuation \(A_V\) measured from the Balmer decrement \ha\,/\,\hb is shown as the colour of the data points.  The data without \(A_V\) measurements is shown in grey colour with no black outline.  This comparison reveals a slight bias of 0.3~dex in SFR, which is likely explained by the difference in dust attenuation measured from the Balmer decrement compared to the SED continuum.}
\label{fig:SFR_Ha_comparison}
\end{figure}

\section{NIRCam Images \& S\'{e}rsic profiles}
\label{app:nircam_images}

In Figure~\ref{fig:appendix_sersic_images}, we demonstrate 2-by-2 arcsecond snapshots of galaxies with the highest \NO (left panel) and lowest \NO (right panel), as well as results of our morphological modelling (\S\ref{sec:properties_morphology}). The residual images show that most sources are modelled very accurately, although when modelling point sources (e.g. 2478-RXCJ2248-3), there are significant residuals indicating that our ePSFs may not always accurately capture the PSF across a field. As such sources have upper limits on their radii, this does not affect our results. Note that sources centred within ten pixels of image edges are masked out and hence appear in some residual images as non-modelled sources.

In Figure~\ref{fig:appendix_size_mass} we compare the sizes of galaxies in our sample with the size-mass relation at \(5<z<14\) from \cite{Morishita2024_sizes}. We note that the relation is based on the rest-UV morphology from short-wavelength JWST/NIRCam bands (F115W, F150W, F200W and F277W), whereas our sizes are measured in F444W probing the rest-optical wavelengths (0.46--0.88 \micron). Therefore, our measurements are more restricted by the F444W PSF size which is 1.6--3.6 times greater than for the shorter-wavelength NIRCam filters. This results in most NOEG sizes being upper limits in our work.

\begin{figure*}
    \centering
    \includegraphics[width=1.01\columnwidth]{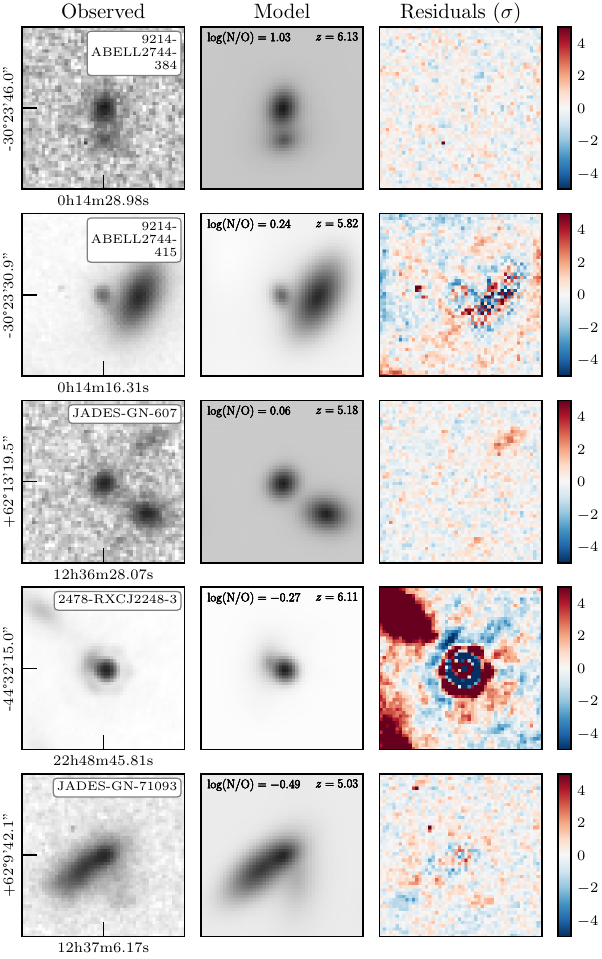}
    \hfill
    \includegraphics[width=1.01\columnwidth]{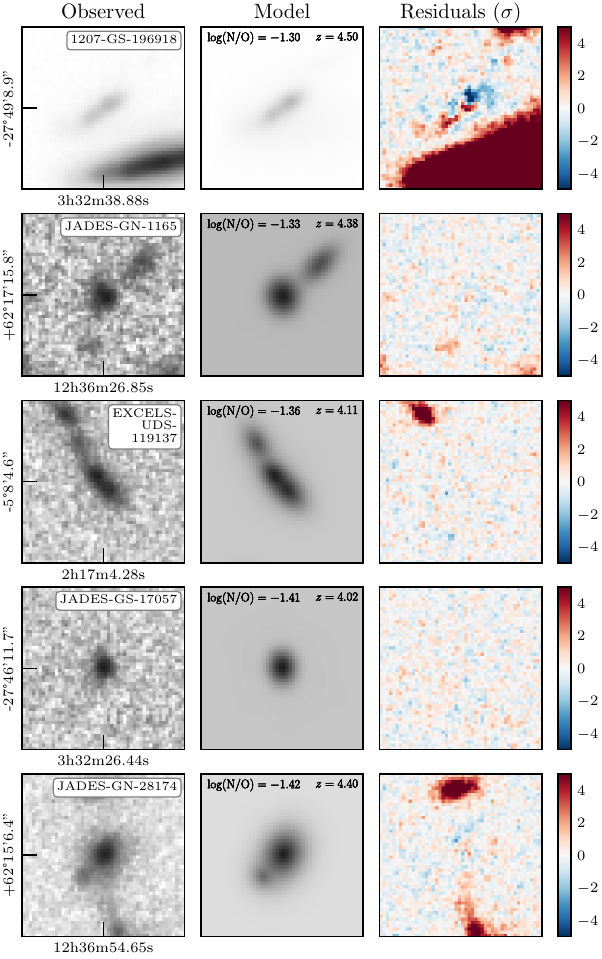}
    \caption{JWST/NIRCam images (F444W band) of 5 objects with the highest \NO (left) and 5 objects with the lowest \NO (right) in our sample. In both figures, we show the observations, the best-fit model and residuals next to each other. On average, the highest-\NO systems appear to be more compact and spheroidal, whereas lowest-\NO systems tend to be less compact and more elliptical. Most systems appear to have companions. Object 2478-RXCJ2248-3 (left column) is a multiply-lensed galaxy with an effective radius of around 20~pc \protect\citep{Claeyssens2025,Berg2026}. We note that sources centred within 10 pixels of an edge were masked out (see \S\ref{sec:properties_morphology}) and therefore show up in some of the residual images.}
    \label{fig:appendix_sersic_images}
\end{figure*}

\begin{figure}
    \centering
    \includegraphics[width=1.01\columnwidth]{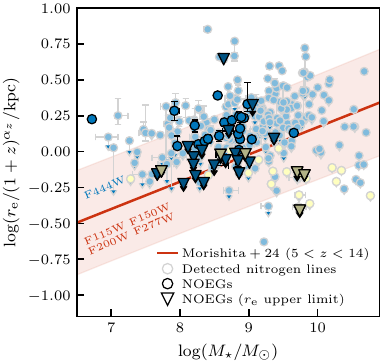}
    \caption{Comparison of our sample with the size-stellar mass relation for star-forming galaxies. The relation is taken from \protect\cite{Morishita2024_sizes} with \(\alpha_z=-0.44\pm0.21\) for rest-UV morphology of galaxies at redshift \(z=5-14\). Our AGN or LRD classifications are highlighted with a brighter colour (light-yellow)---the stellar masses are likely overestimated for these systems as AGN SED models were not included in the fit. Resolved NOEGs have similar typical sizes to nitrogen-detected galaxies at the same stellar masses, although more than half of NOEGs are unresolved and likely have smaller sizes, lying below the size-mass relation. We note that our total sample on average is above this relation by 0.2--0.3 dex, which corresponds to the differences between PSF sizes of the F444W used in our study and the F115W, F150W, F200W and F277W bands used to construct the relation, as the F444W PSF is approximately 1.5--3.0 times larger.}
    \label{fig:appendix_size_mass}
\end{figure}

\section{Atomic and collisional databases}
\label{app:pyneb_databses}

In this section, in Table~\ref{tab:atomic_data} we list all atomic databases with \texttt{pyneb} for our emissivity and abundance calculations.

\begin{table*}
\centering
\renewcommand{\arraystretch}{1.2}
\caption{Atomic data (transition probabilities and collision strengths) adopted in the \texttt{PyNeb\,1.1.30} calculations of nebular conditions and ionic abundances.}
\label{tab:atomic_data}
\begin{tabular}{l l l}
\toprule\toprule
Ion & Transition probabilities & Collision strengths \\
\midrule
    \op & \cite{Zeippen1982,Wiese1996} & \cite{Kisielius2009} \\
    \opp & \cite{Storey2000,FroeseFischer2004} & \cite{Palay2012,Aggarwal1999} \\
    \np & \cite{FroeseFischer2004} & \cite{Tayal2011} \\
    \npp & \cite{Galavis1998} & \cite{Blum1992} \\
    \nppp & \cite{Wiese1996} & \cite{Ramsbottom1994} \\
    \cpp & \cite{Glass1983,Nussbaumer1978,Wiese1996} & \cite{Berrington1985} \\
    \arpp & \cite{MunozBurgos2009} & \cite{MunozBurgos2009} \\
    \arppp & \cite{Rynkun2019} & \cite{RamsbottomBell1997} \\
    \nepp & \cite{Galavis1997} & \cite{McLaughlin2000} \\
\bottomrule
\end{tabular}
\end{table*}


\bsp	
\label{lastpage}
\end{document}